\documentclass[aps,prb,twocolumn,amsmath,amssymb,groupedaddress]{revtex4}

\usepackage{dcolumn}
\usepackage{graphicx,subfigure}
\usepackage{bm}
\usepackage{verbatim}
\usepackage{amsmath}
\usepackage{amssymb}
\usepackage[T1]{fontenc}
\usepackage{ae,aecompl}
\usepackage{appendix}
\usepackage{float}
\usepackage{color}
\usepackage[export]{adjustbox}
\usepackage{array}

\newcommand{\rv}{{\bf r}}

\newcommand{\be}{\begin{equation}}
\newcommand{\ee}{\end{equation}}
\newcommand{\bea}{\begin{eqnarray}}
\newcommand{\eea}{\end{eqnarray}}
\newcommand{\bse}{\begin{subequations}}
\newcommand{\ese}{\end{subequations}}

\begin{document}
\title{High-Sensitivity Photonic Crystal Biosensors using Topological Light Trapping}
\author{Zhengzheng Zhai and Sajeev John}
\affiliation{
Department of Physics, University of Toronto, Toronto, Ontario, Canada M5S 1A7}
\date{08/05/2025}
\email{zhengzheng.zhai@utoronto.ca}

\begin{abstract}
   Photonic crystals (PCs) with localized optical cavity modes arising from topological domain-wall line defects are simulated for optical biosensing by numerical solution of Maxwell's equations. These consist of a square lattice of square silicon blocks with a significant photonic band gap (PBG). Optical transmission through the PBG at specific frequencies occurs by defect-mediated optical tunneling. Biofluid flows perpendicular to light propagation, through a channel containing the PC, defined by silica side-walls and an underlying silica substrate. Replacing the silicon blocks with thin silicon strips throughout the domain-wall region, analyte binding coincides with regions of maximal field intensity. As a result, the sensitivity is improved by almost 16 times higher than the previous designs. We analyze optical mode hybridization of two nearby domain walls and its close relation to the transmission-levels and correlations in frequency shifts of nearby optical resonances in response to analyte-bindings. We illustrate three high-sensitivity chips each with three domain-wall defects, all of which can distinguish three analyte-bindings and their combinations completely in a single spectroscopic measurement. In a photonic crystal, consisting of silicon squares embedded in a water background and a 5-micron lattice spacing, the biosensor sensitivity to a thin analyte binding layer is nearly 3000 nm/RIU, and to the overall background biofluid is over 8000 nm/RIU.
\end{abstract}

\maketitle

\section{Introduction}

Photonic crystal (PC)-based biosensors\cite{Lee07,Chow04,Skivesen07,Vannahme13,Lee15,Zhao16} are wavelength-scale optical microchips\cite{Damborsky16} that can distinguish different concentrations of disease markers in a microscopic bio-fluid sample. This occurs through distinct spectral fingerprints in optical transmission through a photonic band gap (PBG), that reveal resonant frequency shifts and transmission-level modulations. Their integration into compact, chip-based platforms enables \textit{lab-on-a-chip diagnostics}, supporting in situ analysis without the need for extensive sample preparation or labeling procedures~\cite{Damborsky16}.

The functional principle of optical biosensing based on PBG materials was described in detail earlier \cite{Abdullah15, Shuai16}.  Liquid-infiltrated photonic crystals (PCs) are structures with a silicon dielectric pillars and interstitial spaces infiltrated by flowing bio-fluids. In the defect regions of the otherwise periodic photonic crystal, the silicon surface is functionalized through the anchoring of DNA aptamers that bind specific molecules (analyte) from the injected bio-fluid. Optical biosensing relies on the local refractive-index changes due to the thickness increments of the bound analyte layers. Frequency shifts and changes in the transmission peak levels, in response to the increments of analyte-layer thickness, provide a detailed fingerprint of different disease markers. Defects, created by any breaking of the periodicity in photonic crystals (PCs), support localized, resonant, optical modes within the band gap. There are two fundamental classes of defects, namely topological and substitutional. Topological defects involve a global disruption of periodicity and require an infinite rearrangement of material to remove. Substitutional defects involve only local disruptions of periodicity and often involve material shape changes within one or a few unit cells. Topological and substitutional defects can also be combined. Carefully chosen defect architectures can concentrate optical fields near the defect regions to enhance light-analyte interaction and sensitivity. The photonic band gap (PBG) prohibits light propagation over a broad range of frequencies. This free spectral range provides a ``clean slate" for engineering  optical defect modes that appear as transmission resonances, without other irrelevant and spurious transmission of light. A schematic of our two-dimensional (2D) PBG-based biosensor using topological domain-wall line defects is illustrated in Fig. \ref{fig: Schematic Biosensor}.

Two metrics are generally used to evaluate the performance of optical biosensors: (i) sensitivity -- the change of the output signal, resulting from infinitesimal change of the input, namely the resonant frequency shift due to the increment of the analyte-layer thickness; (ii) limit of detection -- the smallest variation required in the input, namely analyte thickness, to detect a change in the output. 

In our PBG-based biosensors, the sensitivity is defined as the rate of change of the resonant frequency with the increment of the analyte thickness for fixed analyte refractive index or the increment of analyte refractive index for fixed analyte thickness.
The limit-of-detection is defined as the minimum change in analyte coating thickness 
 (or refractive index) required to optically detect that a change occurred. For detection based on an optical transmission resonance, the limit-of-detection is usually determined by the quality factor (Q) of the resonance. The frequency shift of the resonance must be at least the full-width at half-maximum of the original resonance for unambiguous detection. Much effort has been made to achieve a good biosensor with both a high sensitivity and a low limit-of-detection\cite{Farida25}. Photonic crystal biosensors provide a unique opportunity to simultaneously achieve high sensitivity and low-limit-of-detection using a minimal volume of bio-fluid.

Other mechanisms for biosensing have also been proposed in the literature. For example, nanoplasmonic devices have demonstrated single-molecule sensitivity through extreme electromagnetic 
field enhancements in nanoscale hot spots\cite{Taylor2017,Wang2013,Mauriz2021,Radziuk2015}. However, the purpose of point-of-care diagnostics is to provide a nearly instantaneous (in a few minutes) outcome, without the need for more time-consuming testing in an external laboratory. The long timescale required for a single molecule to diffuse within a bio-fluid sample and find a specific nanometer scale hotspot makes the nanoplasmonic device less suitable for point-of-care medical diagnostics. In other words, it may require many hours (or longer) for the nanoscale “mouse” in a realistic fluid sample to find the nanoscale “mousetrap”. This might be overcome by having a high concentration of the disease-marker molecules in the sample, but then it would not be truly “single molecule detection” (since a large number of molecules was necessary for timely detection). Alternatively, the timescale problem could be overcome by a device hosting a large number of spatially separated nanoplasmonic hotspots, creating a large spectral density of nanoplasmonic resonances. Within this “noisy” background it may be difficult to identify the spectral shift of a single resonance due to the attachment of a single molecule to the single hotspot. Plasmonic resonances are intrinsically broad and lossy (low Q), and have limited reproducibility, significant photothermal heating, and restricted multiplexing capability\cite{Vollmer2008, Jin2025}.

Our lab-in-a-PC enables nearly instantaneous detection and diagnosis by providing a spatially broad “net”, occupying a sizeable volume of photonic crystal flow channel, for the capture of disease-marker molecules\cite{Abdullah15,Shuai16, Abdullah19,Dragan21}. Each  “line-defect” within the 2D PC extends along the length of the fluid flow direction. The resulting optical modes, offer much higher quality factors and narrower line-widths than nanoplasmonic devices, leading to high refractive-index figures of merit and quantitative, label-free detection. These high-Q photonic modes enable precise spectral readout and low limits-of-detection, while maintaining stability and low absorption losses. Our
larger mode volumes (enabling rapid detection) and weaker local field enhancements make unassisted single-molecule detection more challenging than for plasmonic hot-spots. Nevertheless, our photonic platform provides spectral precision, reproducibility, and rapid detection.

\begin{widetext}
\begin{center}
\begin{figure}[htbp]
 \centering
  \subfigure[]{
    \hspace{0in}\includegraphics[width=0.95\textwidth]{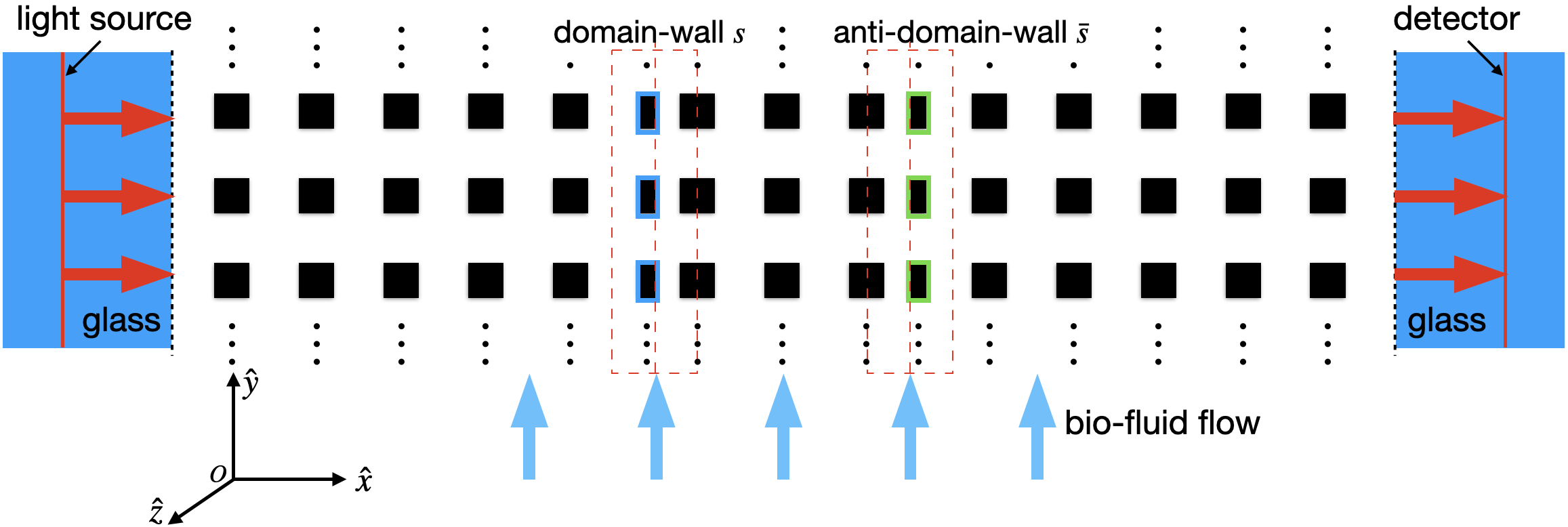}}
 \caption{Schematic two-dimensional (2D) PBG-based biosensor. A 2D PC is sandwiched by glass walls along the $\hat{x}$-direction and periodically repeated in the $\hat{y}$-direction (as indicated by dots). Interstitial spaces within the periodically arranged silicon dielectric blocks are infiltrated by bio-fluids flowing in the $\hat{y}$-direction. A normally incident light pulse, emitted by the light source (depicted as a red line) embedded in the left-side glass, propagates in the $\hat{x}$-direction through the chip and the transmitted light is collected by the detector (depicted as another red line) embedded in the right-side glass. Topological domain-wall line defects are used to support localized, resonant optical modes within the PBG. Two different types of analytes (colored as blue and green) are bound to the surfaces of the rectangular silicon blocks within the left- and right-defect region, respectively.}
\label{fig: Schematic Biosensor}
\end{figure}
\end{center}
\end{widetext}

Since its two-dimensional (2D) conceptual paradigm with 2D PC slabs was proposed \cite{Abdullah15},  a realistic three-dimensional (3D) design with relatively long nanopllars followed \cite{Abdullah19}. These previous PBG-based biosensors involve correlated spectral responses of the multiple resonances and complex spectral fingerprints, which undergo transmittance-level changes and correlated frequency-shifts in response to analyte-bindings. This is the result of strong interaction between spatially proximal optical resonances that are also close in frequency and exhibit mode hybridization. Recently, a realistic 3D architecture with short-height nanopillars was proposed \cite{Dragan21}, which makes these PCs amenable to fabrication by techniques such as nanoimprinting. This latter biosensing is based on mutually independent frequency-shifts of multiple resonances, well-separated in frequency. This frequency separation prevents mode hybridization, leading to considerably simpler spectral fingerprints.

These previous biosensor chips are based on a conventional line defect and surface truncation structures.These support an extended optical waveguide mode and surface guided modes, respectively. When these defect surfaces are functionalized, they can capture disease-markers along their entire length and therefore have a relatively broader ``net" compared with other point-defect biosensors. 

Recently, motivated by Su-Schrieffer-Heeger (SSH) model\cite{SSH1979}, a PC architecture with topological domain-wall defects formed at the interface of two identical but displaced photonic-crystals was proposed\cite{Yang24}. These topological defects support spatially localized boundary modes within the PBG, and can be functionalized for biosensing. This alternative design does not require period doubling along the direction of the waveguide in order to allow coupling to an external plane wave. However, the initial study\cite{Yang24} showed very low sensitivity due to weak overlap of the field with the analyte. Moreover, they exhibit complex spectral fingerprints due to unregulated mode hybridization. These features lead to non-ideal biosensor functionality. 
\\

In this paper, we demonstrate that suitably redesigned topological domain-wall defects can achieve unprecedented high sensitivity for biosensing and adjustable mode hybridization by modifying the dielectric distribution in the defect region. In Sec. II, we review the domain-wall defect formed at the interface of two PC's, and the nature of the resulting spatially trapped optical waveguide mode. We then replace the simple square dielectric blocks in the defect region by disconnected rectangular strips with carefully adjusted sizes to improve the sensitivity by a factor of almost 16 times compared to previous estimates \cite{Yang24}. In Sec. III, we analyze the coupling of two domain-wall defect modes and the resulting mode hybridization. This is generally required to have high transmission-levels but unavoidably leads to correlated spectral responses  observed in the transmission spectrum of the two resonances to separate analyte-bindings.  The trade-off between spectral independence (mode dehybrization) and high transmission peaks is analyzed. We present several chips with two domain walls and different levels of mode hybridization. We demonstrate how correlations in spectral responses can be suppressed by dehybridization. On the other hand, transmission-levels can be improved by enhancing the mode hybridization. Finally, We discuss three distinct chips each with three domain-wall defects in Sec IV that can distinguish all the eight analyte-binding cases, involving three distinct analytes in a given bio-fluid sample. The first two biosensor chips, one with high sensitivity but reduced transmission and the other with high transmission but reduced sensitivity, have three effectively dehybridized resonances that respond independently to analyte-bindings, but with relatively low transmission-levels. The third chip has three highly hybridized resonances and correlated spectral responses, but with high transmission-levels.

\begin{widetext}
\begin{center}
\begin{figure}[htbp]
  \subfigure[]{
   \hspace{0in} \includegraphics[width=0.4\textwidth]{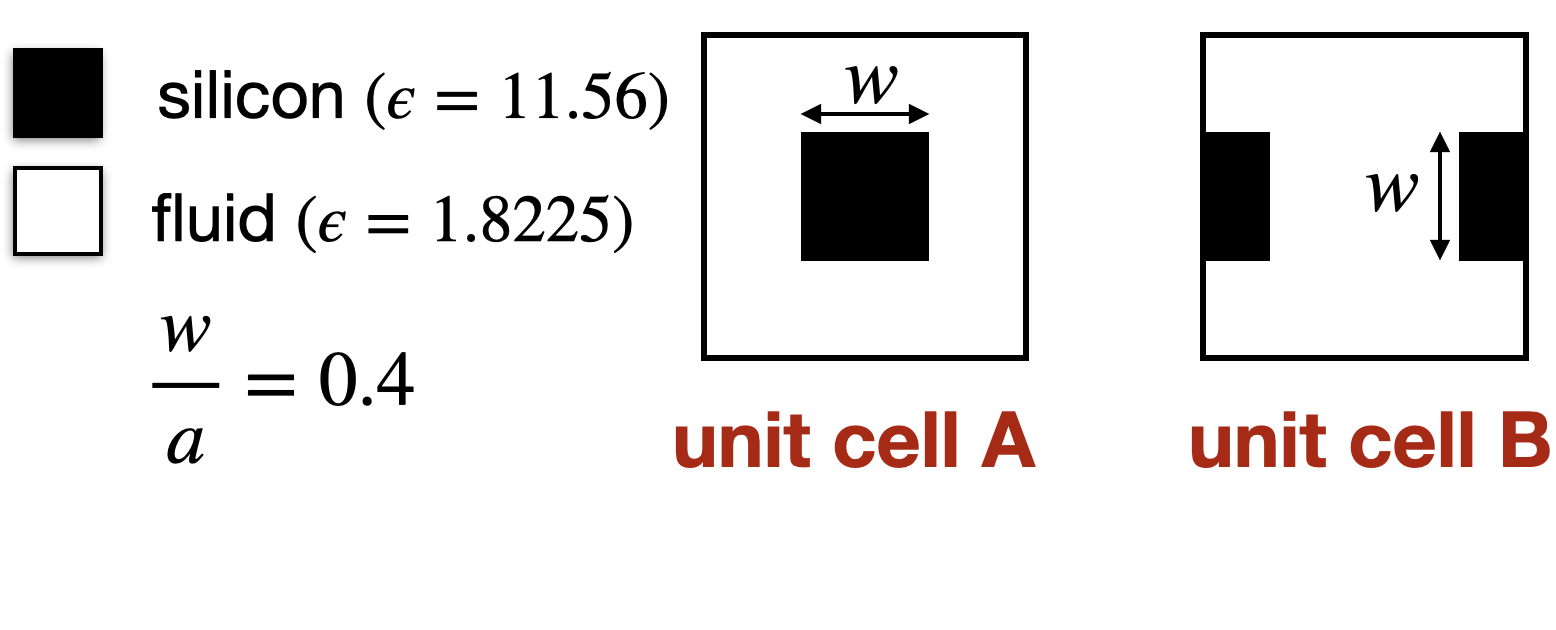}}
    \label{fig:unit cells}
  \hspace{1.5cm}
  \subfigure[]{
    \hspace{0in}\includegraphics[width=0.48\textwidth]{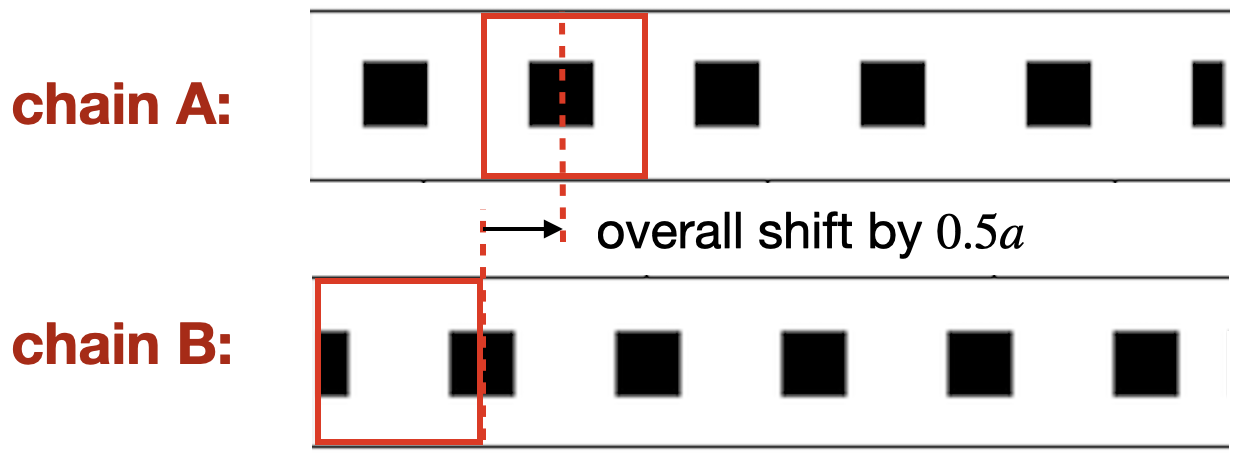}}
    \label{fig:SSH chains}
 \hspace{0cm}
  \subfigure[]{
   \hspace{0in} \includegraphics[width=0.95\textwidth,center]{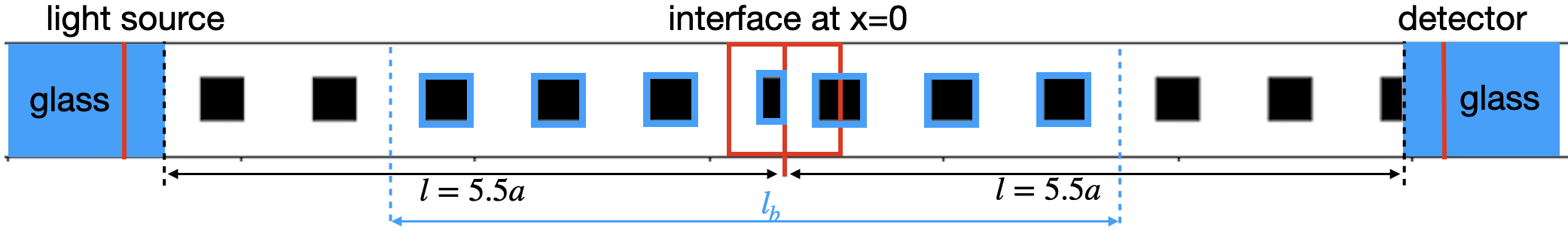}}
    \label{fig:doamin-wall}
  \subfigure[]{
   \hspace{0in} \includegraphics[width=0.54\textwidth]{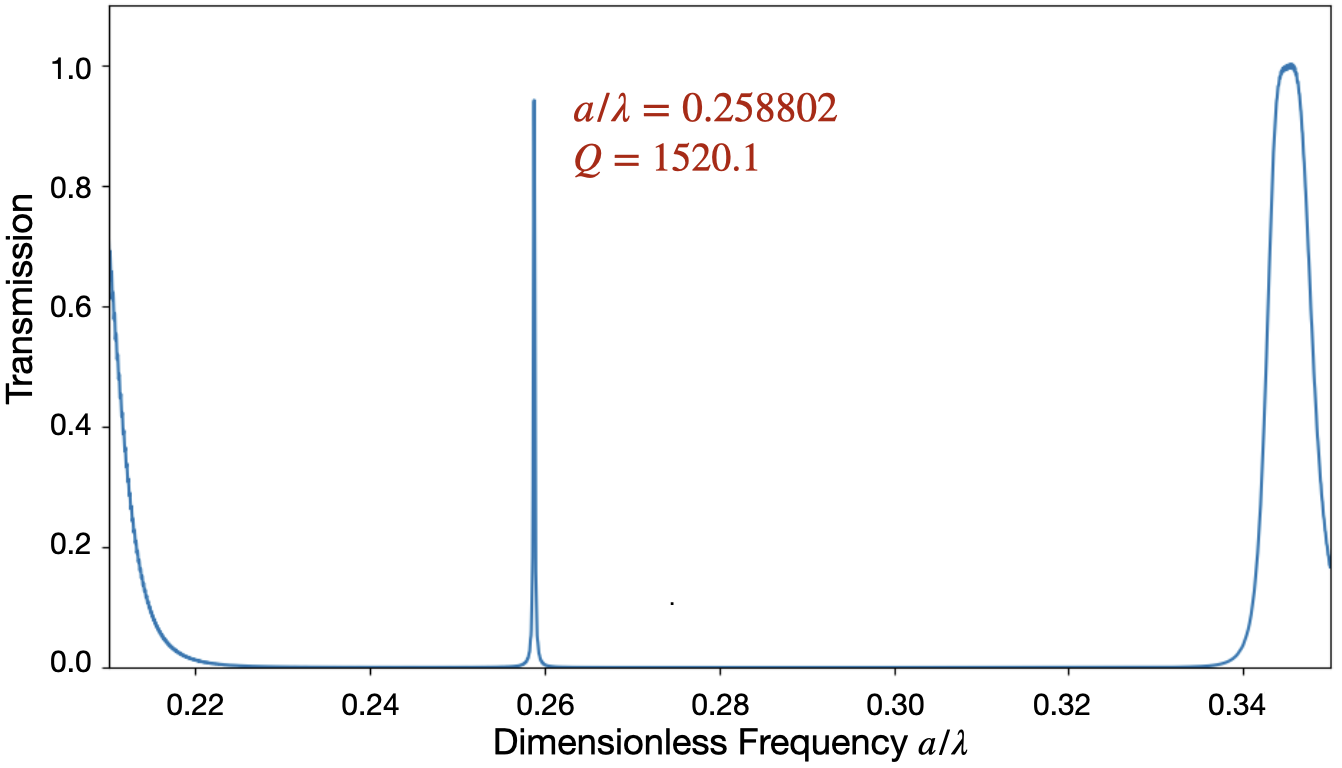}}
    \label{fig:defect mode}
  \subfigure[]{
   \hspace{0in} \includegraphics[width=0.43\textwidth]{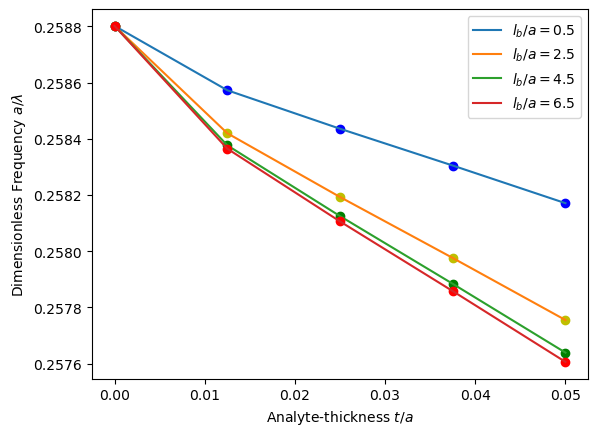}}
    \label{fig:original low sensitivity}
 \caption{Localized optical mode created by a domain-wall defect. (a) Unit cell containing a silicon square with side length $w/a=0.4$. (b) Two photonic crystal (PC) line segments, differing by an overall translation of half a unit cell. (c) A domain-wall defect formed at the junction of the two PC line segments. (d) This domain-wall defect supports a localized photonic mode with $a/\lambda=0.258802$ within the PBG, where $\lambda$ is the vacuum wavelength of light. (e) This topological defect mode can be used for biosensing. With only one rectangular block within the defect region whose surface is functionalized for analyte-binding, the resonant frequency shift is $0.000402$ $[2\pi c/a]$ for analyte-binding thickness $t/a=0.05$, corresponding to very low sensitivity. By increasing the number $l_b/a$ of dielectric blocks whose surfaces are functionalized for analyte-bindings, the sensitivity can be improved. The maximum sensitivity achieved by this enlargement ($l_b/a=6.5$) of analyte-binding region is $0.0012$ $[2\pi c/a]$ per analyte-thickness of $t/a=0.05$, but less than ideal. Here the dimensionless frequency $a/\lambda=(\omega a)/(2\pi c)$.}
\label{fig: localied photonic mode}
\end{figure}
\end{center}
\end{widetext}

\section{Light trapping by a domain wall}
The Su-Schrieffer-Heeger (SSH) model \cite{SSH1979} is a one-dimensional (1D) tight-binding model with alternating strong and weak bonds, first proposed to describe the electron propagation in some conducting polymers like polyacetylene. This sublattice bond structure gives rise to two degenerate ground states, and therefore a topological domain-wall soliton exists at the interface of these two degenerate structures. The localized electronic state associated with the domain wall occurs at the center of the band gap.

In a photonic crystal, the alternating higher and lower dielectric constants provides an effective means to create domain-wall defects that localize light within a photonic band gap. As shown in Fig. \ref{fig: localied photonic mode}, a topological domain-wall defect at the interface of two identical PC's is formed by translating one PC by half a unit cell relative to the other. Unlike a conventional subsitutional impurity defect, such a defect cannot be removed by any local modification, but only by a global operation involving infinite rearrangement of the structure.

 In our 2D PC, the square-lattice unit cell consists of a square dielectric (silicon) block with a high refractive index $n=3.4$  ($\varepsilon=11.56$) immersed in a fluid of $n=1.35$  ($\varepsilon=1.8225$). We choose the optimal ratio of the width $w$ of the silicon block to the lattice constant $a$ to be $w/a=0.40$, in order to maximize the PBG. Two identical PC chains with length $l/a=5.5$, differing by an overall shift in the direction of the chain by half a unit cell, form a topological defect at the interface of these two chains as shown in Fig. \ref{fig: localied photonic mode}. The transmission spectrum is calculated after the silicon-in-fluid PC is encased in a flow channel with walls of glass ($n=1.5$, $\varepsilon=2.25$). This structure is periodically repeated in the $\hat{y}$-direction. A normally incident Gaussian pulse is emitted by the light source (depicted as a red line) embedded in the left-side glass. It propagates in the $\hat{x}$-direction through the chip and the transmitted light is collected by the detector (depicted as another red line) embedded in the right-side glass. The electric field ${\bf E}(\rv,t)$ is polarized in the $\hat{z}$ direction, perpendicular to the plane of the 2D PC, i.e. only transverse-magnetic (TM) polarized light is considered. The biofluid flows in the $\hat{y}$ direction. We apply periodic boundary conditions in the $\hat{y}$-direction and absorbing boundary conditions in the $\hat{x}$-direction to avoid spurious light reflection from the two ends. We then perform an FDTD simulation of total optical transmission through the chip using MEEP\cite{MEEP}. The spatial resolution implemented for the FDTD calculations is 100 mesh points per lattice constant. As shown in Fig. \ref{fig: localied photonic mode}(d), this topological domain-wall defect supports a localized resonance with resonant frequency $\omega=0.258802$ $[2\pi a/c]$ (equivalently $a/\lambda=0.258802$) and quality factor $Q=1520.1$ within the PBG. 
 
 Note: Here we use a typical value $n=1.35$  ($\varepsilon=1.8225$) for the refractive-index, corresponding to human blood serum. The choice of specimen
may shift the exact numerical values of the resonant frequencies, but the overall detection scheme will remain the
same. For example, a fluid specimen with a higher refractive index will cause all the resonant frequencies to shift
slightly lower. It is a simple matter to re-calibrate the device to account for such a change in specimen. For all analytes (disease biomarkers) that bind to the functionalized silicon surfaces of the optical biosensor, a representative refractive index of $n=1.45$ ($\varepsilon=2.1025$) is adopted. This analyte index applies to all figures where spectral shift as a function of analyte layer thickness is depicted. If an alternative analyte with a higher or lower refractive index (from a different specimen) is chosen, the corresponding spectral shifts will accordingly be higher or lower.
\begin{widetext}
\begin{center}
\begin{figure}[htbp]
 \centering
  \subfigure[]{ \includegraphics[width=0.98\textwidth]{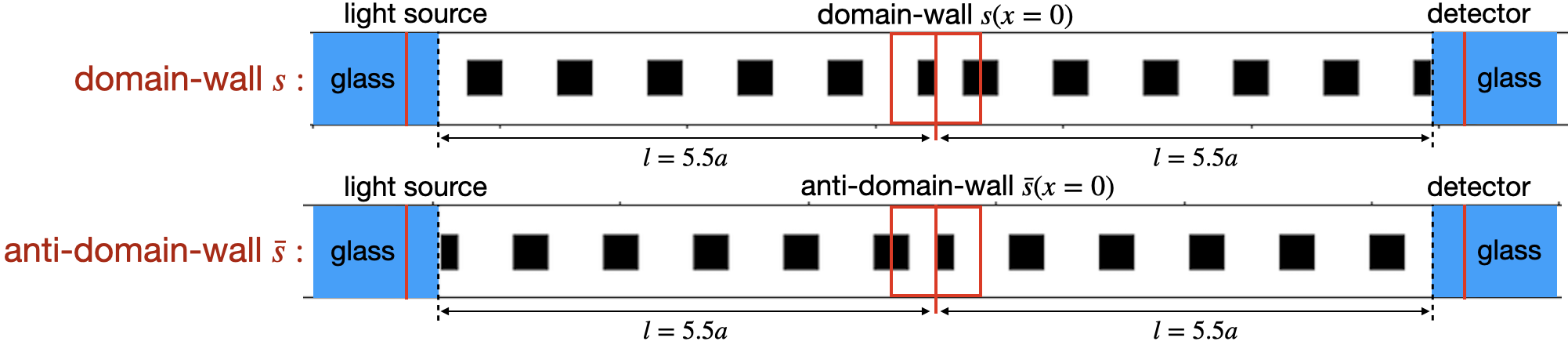}}
  \label{fig: DW and anti-DW}
  \hspace{0.5cm}
 \subfigure[]{\includegraphics[width=0.88\textwidth]{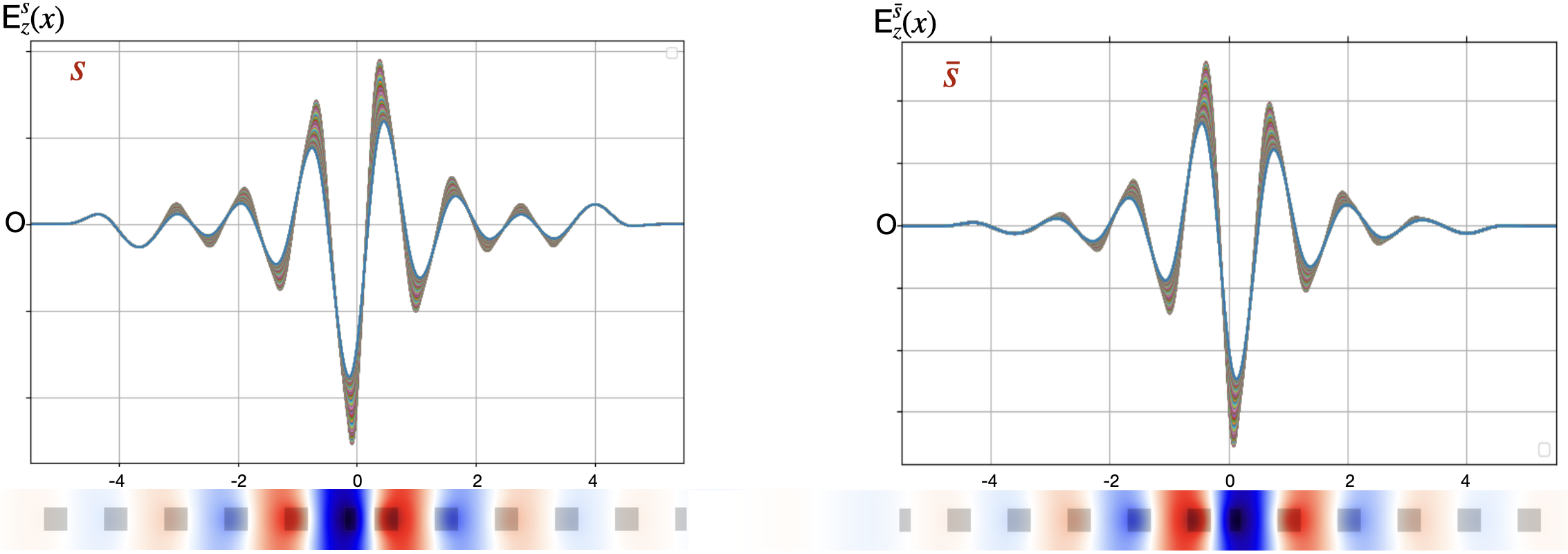}}
 \label{fig:DW-antiDW field profiles}
 \caption{(a) Photonic crystal (PC) chains with a domain-wall and an anti-domain-wall. The anti-domain-wall is a mirror image of the domain-wall configuration with respect to the middle $yoz$ plane, located at the junction $x=0$. (b) Electric field profiles for the domain-wall mode and the anti-domain-wall mode are mirror images of one another.}
\label{fig: DW and anti-DW}
\end{figure}
\end{center}
\end{widetext}

\subsection{Domain-wall and Anti-domain-wall}
 There are two distinct topological defects that can be formed with these two PC chains, shown by Fig. \ref{fig: DW and anti-DW}(a): with chain A to the left and chain B to the right, or its mirror image -- with chain B to the left and chain A to the right. If we call the first defect configuration a domain-wall, denoted by $s$, then the other configuration may be called an anti-domain-wall, denoted by $\bar{s}$. The refractive index distribution  of the anti-domain-wall is simply the spatially reflected one of the domain-wall, $\epsilon^{\bar{s}}(x,y)=\epsilon^s(-x,y)$. Accordingly, the electric field profile ${\bf E}^{\bar{s}}({\bf r})$ of the anti-domain-wall mode is the reflection of the domain-wall mode (at any given time): 
\begin{equation}
    {\bf E}^{\bar{s}}({\bf r})={\bf E}^{s}\left(-x,y\right).
\end{equation}

However, ${\bf E}^{s}\left(x,y\right)\neq \pm {\bf E}^{s}\left(-x,y\right)$, since the chip geometry is not reflection symmetric with respect to the $yoz$ mirror plane. The actual electric field profiles for the domain wall and anti-domain wall are shown in Fig. \ref{fig: DW and anti-DW}(b).

\subsection{Quality factor enhancement}
Domain-wall defect modes were introduced for optical biosensing in Ref. 13. Like earlier designs using line defects and surface defects\cite{Abdullah15,Shuai16, Abdullah19,Dragan21}, domain-wall modes are strongly localized in the $x$-direction (of light propagation), but spatially extended along the optical waveguide in the $y$-direction (direction of biofluid flow). These waveguide modes allow analyte-binding all along the fluid flow path, providing a broader ``net" to capture the disease-markers. On the other hand, the topological defect design avoids the ``period doubling" requirement, required in previous (substitutional defect) designs\cite{Abdullah15, Shuai16, Abdullah19, Dragan21} for the incident wave to couple to the waveguide modes. 
 
 The topological light-trapping mode for the structure depicted in Fig. \ref{fig: localied photonic mode}(c) has a quality factor of only 1520.1. The full-width at half-maximum (FWHM) of this resonance is $\Delta \omega^{\text{FWHM}}=\omega/Q=1.7 \times 10^{-4} [2\pi c/a]$.

\begin{widetext}
\begin{center}
\begin{figure}[htbp]
 \centering
  \subfigure[]{
   \hspace{0in} \includegraphics[width=0.9\textwidth]{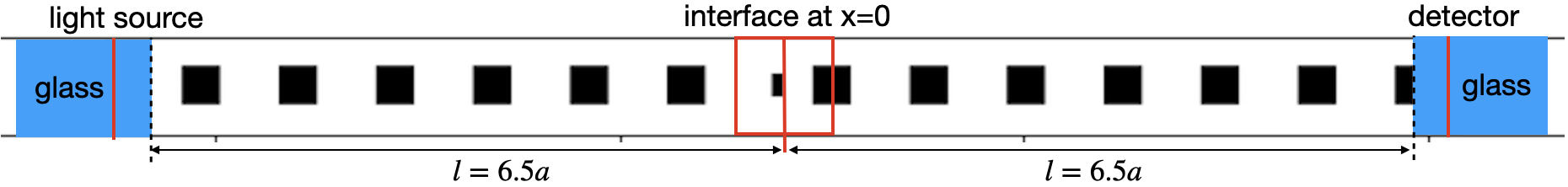}}
    \label{fig: domain-wall_l=6 chip}
  \hspace{0.5cm}
  \subfigure[]{
    \hspace{0in}\includegraphics[width=0.55\textwidth]{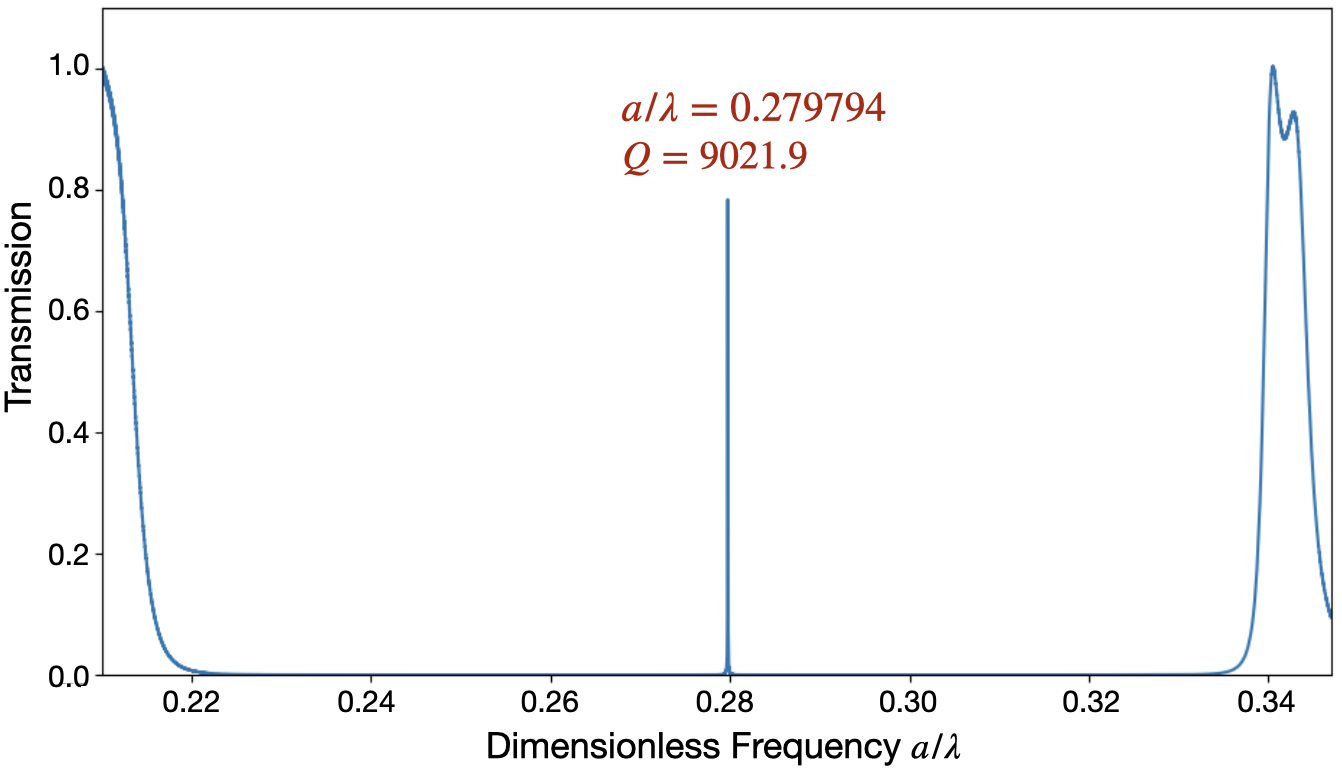}}
    \label{fig: doamin-wall_l=6_trans}
 \caption{The domain-wall optical quality factor is improved by reducing the size of the rectangular dielectric block in the defect region from $(0.2a \times 0.4a)$ to $(0.12a \times 0.24a)$. This optical resonance is moved to $a/\lambda=0.279794$, near the middle of the band gap, and the quality factor is improved from $1520.1$ to $2299.7$. By increasing the number of unit cells on each side of the defect from $l/a=5.5$ to $l/a=6.5$, the quality factor is further improved to $9021.9$. (a) The chip with length $(2l+a)=13a$ and defect dielectric block of dimensions $(0.12a \times 0.24a)$; (b) The resonance at $\omega=0.279794$ $[2\pi a/c]$, near the middle of the band gap, with quality factor $Q=9021.9$.}
\label{fig:improve quality factor}
\end{figure}
\end{center}
\end{widetext}

 The quality factor can be improved by two independent methods. The first involves moving the resonance towards the middle of the PBG. This decreases the spatial extent of the localized optical mode and its coupling to glass encasing structure. By reducing the size of the rectangular dielectric block in the defect region from $(0.2a \times 0.4a)$ to $(0.12a \times 0.24a)$, the optical resonance moves to $a/\lambda=0.279794$, almost in the middle of the band gap. The quality factor is improved from $1520.1$ to $2299.7$. The second method to decrease the coupling of the localized mode to the glass encasing is by widening the entire flow channel. By additionally increasing the number of unit cells on each side of the defect from $l/a=5.5$ to $l/a=6.5$, the quality factor is further improved to $9021.9$, as shown in Fig. \ref{fig:improve quality factor}. The FWHM, which is inversely proportional to the quality factor, is reduced to $\Delta \omega^{\text{FWHM}}=\omega/Q=3.1 \times 10^{-5} [2\pi c/a]$. 
 
 The quality factor, in fact, rises exponentially with increasing $l/a$ until it saturates due to weak absorption of light in a real medium. \cite{PC-book08} The structures we illustrate have widths (in the light propagation direction) ranging from 11 to 13 lattice constants. This corresponds to about 3-4 optical wavelengths. On such short length scales, absorption is generally negligible. A very small residual absorption would place a limit on the highest quality factor we can obtain for any of our optical resonances.  In the absence of absorption, the quality factors increase exponentially with the width of photonic crystal chip (because the optical mode amplitude decreases exponentially with distance from the center of each defect).  If the residual absorption is characterized by an effective absorption length scale, the quality factor of the optical transmission resonance will reach its maximum value (saturate) once the chip width equals the absorption length. That is to say, there is no advantage in making a photonic crystal chip that is wider than the absorption length scale. 
 
From a computational perspective, enlarging the chip width increases the numerical simulation time greatly. Therefore, we stop at a reasonable flow channel half-width ($l/a=5.5$ for most of our simulations).

\subsection{Sensitivity enhancement}
Analyte-binding on surfaces of only one square block within the defect region, as in Ref. 13, provides a resonant frequency shift of only $0.0006$ $[2\pi c/a]$ per analyte-thickness $t=0.1a$. With such a low sensitivity, the limit-of-detection of this defect mode, placed in the center of a PC that is 11-unit-cells in width as shown in Figure 2, is a modest $\delta t^{\text{lim}}/a=0.028$, which is the thinnest analyte coating thickness required to shift the resonance by one FWHM. For the  unfragmented domain-wall defect of Figure 4, placed in the center of a photonic crystal that is 13-unit-cells in width, the sensitivity is improved to a frequency shift of $0.000665$ $[2\pi c/a]$ per analyte-thickness $t=0.05a$ and the limit-of-detection is improved to $\delta t^{\text{lim}}/a=0.0023$.  Since the quality factor increases exponentially with the number of unit cells of the PC, a corresponding exponential decrease in $\delta t^{\text{lim}}/a$ will occur. 

\begin{widetext}
\begin{center}
\begin{figure}[htbp]
 \centering
  \subfigure[]{\includegraphics[width=0.33\textwidth]{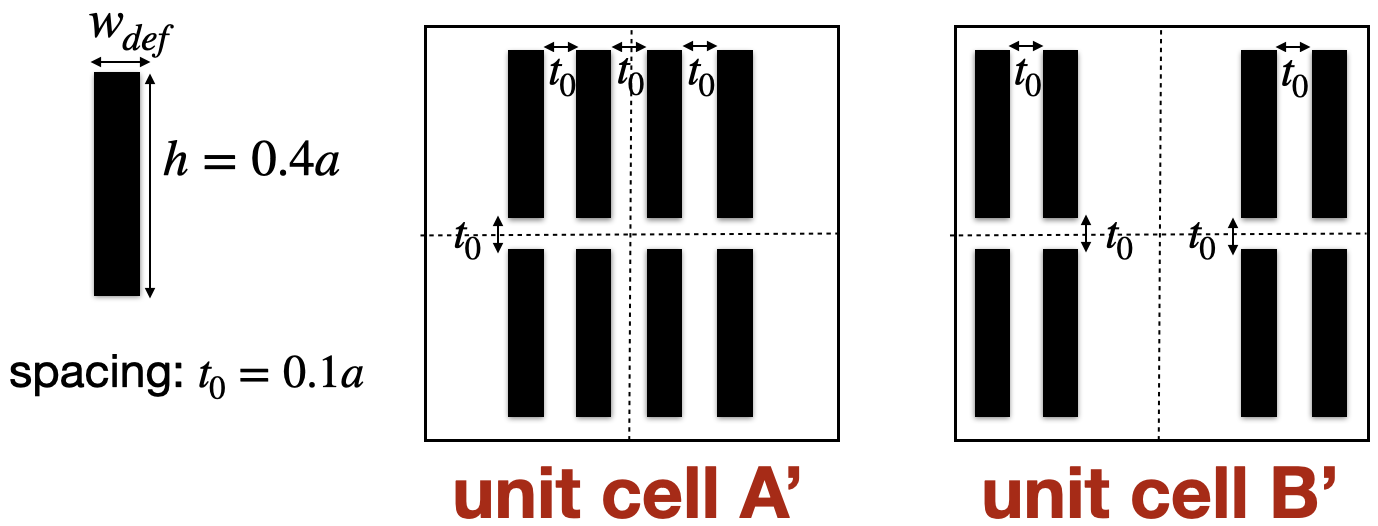}}
    \label{fig: rectangular striips unit cells}
  \hspace{2cm}
  \subfigure[]{\includegraphics[width=0.25\textwidth]{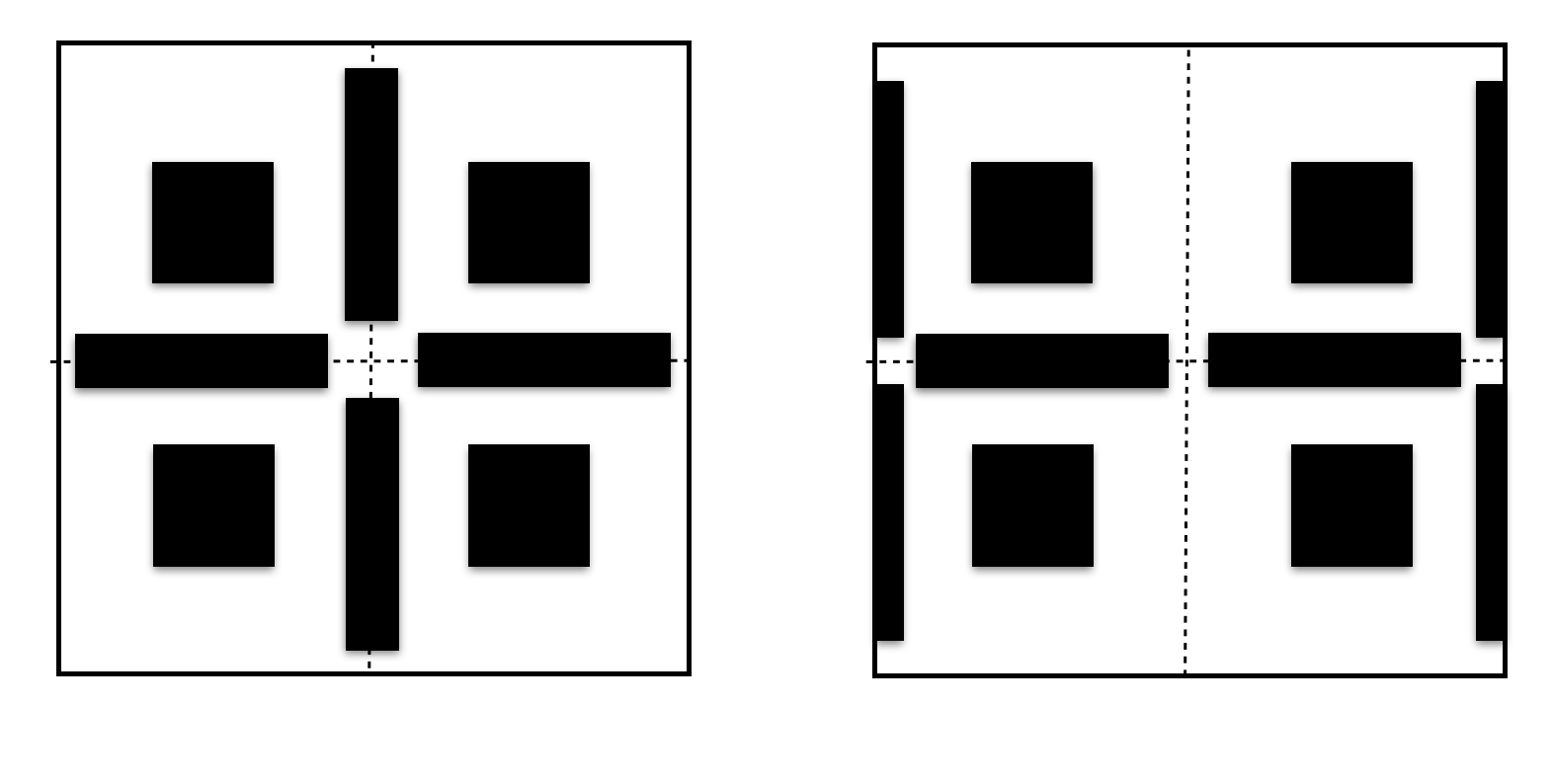}}
    \label{fig: plus-sign}
  \subfigure[]{\includegraphics[width=0.98\textwidth]{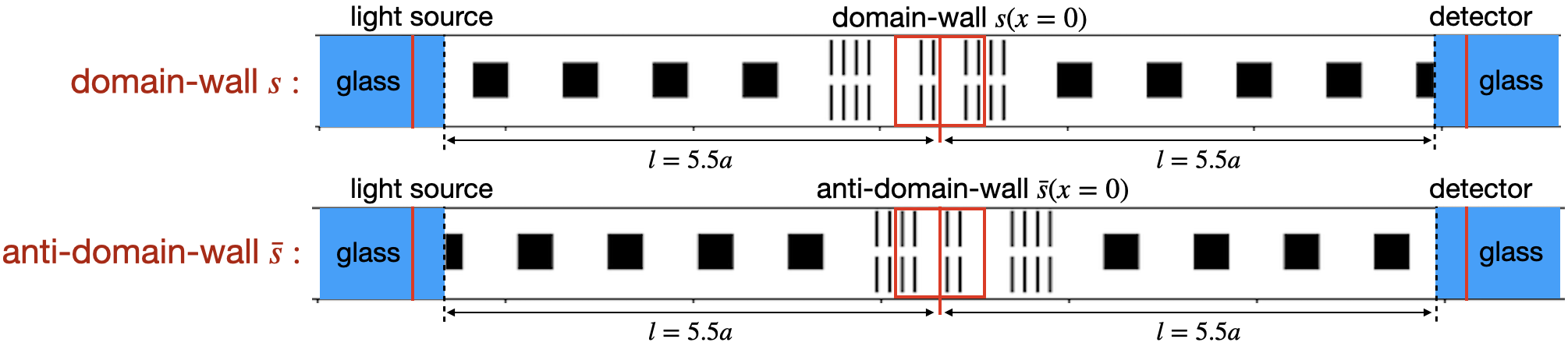}}
    \label{fig: revised s-s_bar} 
    \hspace{2cm}
      \subfigure[]{\includegraphics[width=0.58\textwidth]{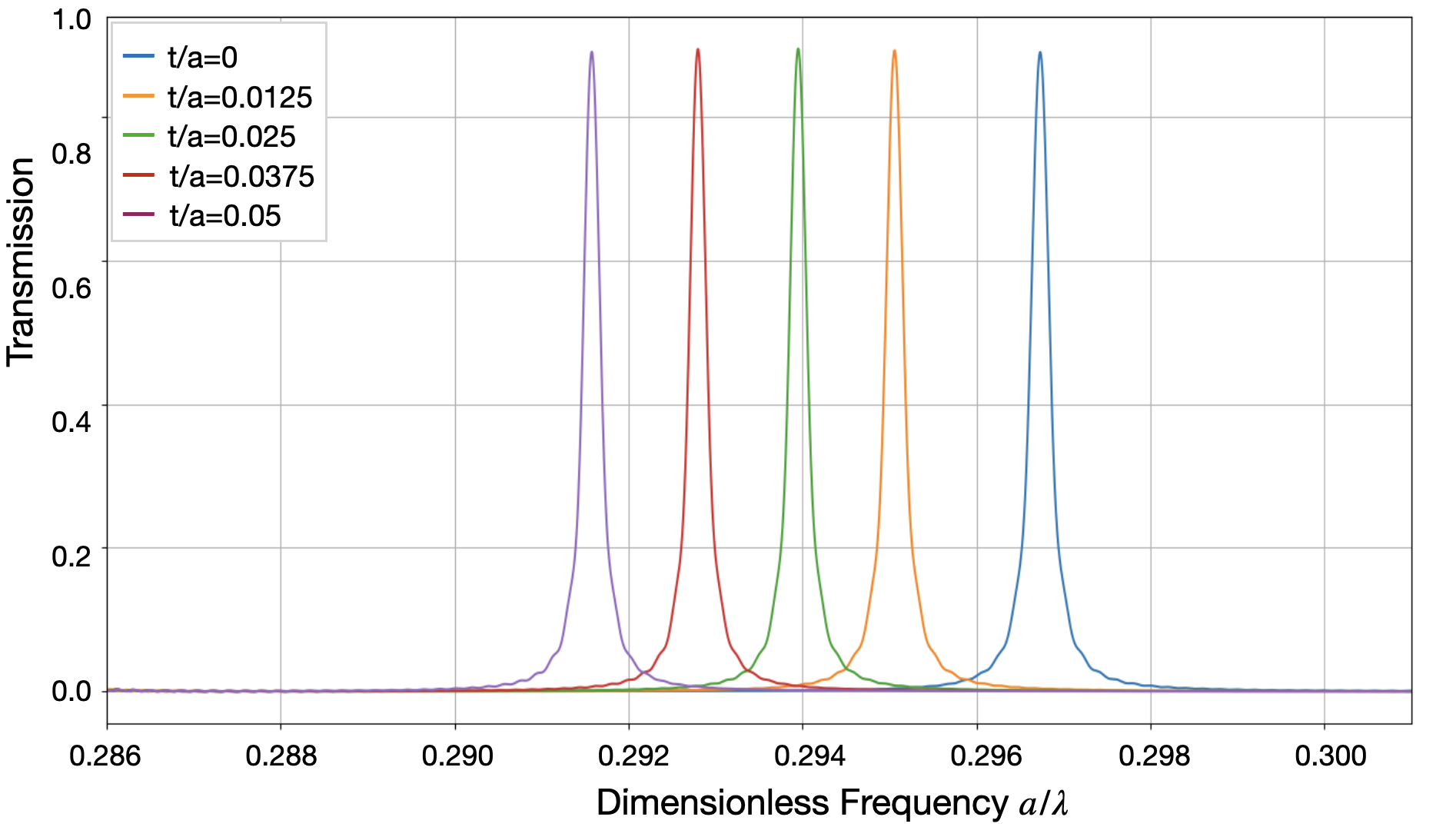}}
    \label{fig: bindings} 
     \subfigure[]{\includegraphics[width=0.45\textwidth]{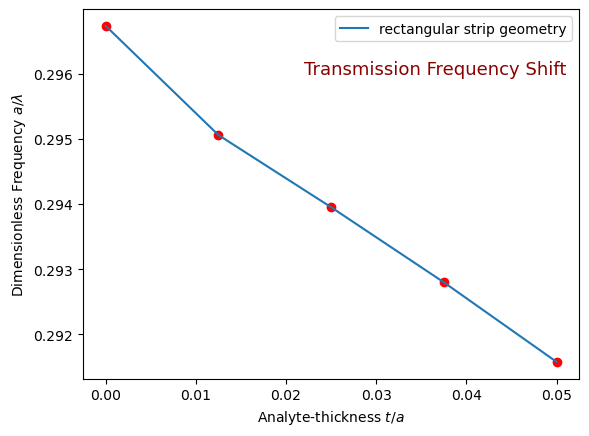}}
    \label{fig: improved sensitivity}
    \hspace{1cm}
     \subfigure[]{\includegraphics[width=0.45\textwidth]{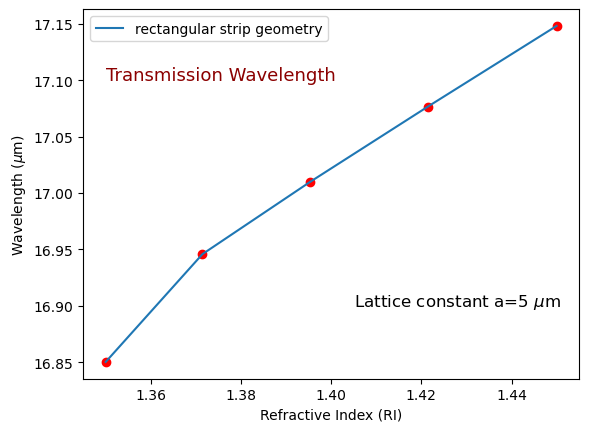}}
    \label{fig: Sensitivity nm/RIU}
 \caption{Sensitivity enhancement by improved geometry in the defect region such that the analytes bind near strong optical fields. (a) Unit cells with fragmented dielectric with eight rectangular strips, symmetrically distributed. The length of the defect region is $l_b=2.5a$. The length of each rectangular strip is $h=0.4a$ and the spatial separation between neighboring strips is $t_0=0.1a$. (b) Geometry with comparable sensitivity, but impeded biofluid flow. (c) Domain-wall and anti-domain-wall configuration with strip geometry in the defect region. (d) Red-shifts of the defect mode, in response to analyte-thickness increments, as observed in the transmission spectrum. (e) The highest sensitivity is achieved with strip width $w_{def}=0.03a$. In this case, the transmission resonance peak shifts in frequency by $\Delta \omega=0.0052$ $[2\pi c/a]$ per analyte-thickness $t=0.05a$. In this plot, the sensitivity is defined as the frequency shift per analyte-layer thickness change, for fixed refractive index of $n=1.35$. The analyte-thickness is measured in units of the PC lattice constant $a$. The frequency, $\omega$, and its shift, $\Delta \omega$, are measured in units of $[2\pi c/a]$. This corresponds to the dimensionless frequency, $a/\lambda$, depicted on the vertical axis, where $\lambda$ is the vacuum wavelength of light. (f) Sensitivity measured by the wavelength shift in units of nanometers per refractive index unit (RIU). With fixed analyte-thickness $t/a=0.05$, choosing $a=5 \mu$m, the sensitivity is $2979.1$ nm/RIU.}
\label{fig:improve sensitivity}
\end{figure}
\end{center}
\end{widetext}

There are likewise two approaches to improve the sensitivity. One is to enlarge the analyte binding region by functionalizing more dielectric surfaces in the vicinity of the localized optical mode. As shown in Fig. \ref{fig: localied photonic mode}(e), as the binding region enlarges, the sensitivity, defined as the slope of the frequency-shift with respect to the analyte-thickness increment, improves slowly. The sensitivity saturates when the binding region length, $l_b$, approaches the localization length of the defect mode, e.g. $l_b=6.5a$ as shown in Fig. \ref{fig: localied photonic mode}(c). The highest sensitivity achieved in this way is $0.0012$ $[2\pi c/a]$ per analyte-thickness $t=0.05a$. This is four times the sensitivity reported in Ref. 13. A second and very effective approach to increase the sensitivity is by forcing analyte binding into regions where the optical field concentration is strongest.

For this purpose, we replace a single silicon square  with eight fragmented strips in one unit cell, as shown in the Fig. \ref{fig:improve sensitivity}(a). This simple structure facilitates the flow of biofluid parallel to the strips and achieves a very high sensitivity. (Other geometries, such as the one shown in Fig. \ref{fig:improve sensitivity}(b), can also achieve high sensitivity, but impede the flow of biofluid for the required analyte bindings.) The size of every rectangular strip and the distribution of the eight strips within every unit cell is as shown in Fig. \ref{fig:improve sensitivity}(a): the width of every rectangular strip is $w_{def}=0.03a$ and its length is $h=0.4a$. The gap through which the biofluid can flow between neighboring strips is set to be $t_0=0.1a$. The eight dielectric strips are distributed symmetrically with respect to the center of the unit cell. We extend this geometry to occupy $2.5$ unit cells, and functionalize all the surfaces of these rectangular strips for analyte bindings. We call this the defect region or binding region. This design enables more analyte to bind to where the optical field strength is strong. With this change of geometry, the sensitivity is greatly improved from $0.0012$ $[2\pi c/a]$ per $t=0.1a$ to $0.0052$ $[2\pi c/a]$ per $t=0.05a$. This is more than 16 times the sensitivity reported in Ref. 13. Our sensitivity is also much higher than any other previous PBG-based biosensors\cite{Abdullah15,Shuai16, Abdullah19,Dragan21}.  Enlarging this defect region even further, replacing more square blocks by these rectangular strips can improve the sensitivity further, but will introduce extraneous modes within the PBG. This  complicates the transmission spectrum, making it difficult to spectroscopically distinguish different analyte binding configurations. Therefore, we limit the defect and analyte-binding region to a length of 2.5 unit cells ($l_b=2.5a$).

An alternative sensitivity measure used commonly by experimental groups is the rate of wavelength change (in units of nanometers) with change in refractive index, for fixed analyte thickness. This sensitivity is expressed in nm/RIU, where RIU refers to a refractive index unit. We consider a sequence of refractive indices $\{1.35, 1.3715, 1.3953, 1.4215 ,1.45\}$ for a hypothetical analyte that coats the silicon rectangular strips with a fixed thickness of $t/a=0.05$, corresponding to a total refractive-index variation of $\Delta n=0.1$ RIU. We consider a specific unit-cell size $a=5 \mu$m. The resonant wavelength shift per refractive index unit (RIU) is shown in Fig. \ref{fig:improve sensitivity}(f), revealing a value of $2979.1$ nm/RIU. 

A third sensitivity measure is the rate of wavelength shift with overall change in the background fluid refractive index. The sensitivity in this measure is $8255.1$ nm/RIU, with lattice constant $a=5\mu$m.

It is easy to verify that this sensitivity (measured in units nm/RIU) is linearly proportional to the unit-cell size $a$. We provide a specific value of the lattice constant $a=5 \mu$m as an illustration since it would make chip fabrication relatively inexpensive and the defect regions could be functionalized (by anchoring DNA aptamers) using simple inkjet printing methods. A lattice constant of $5 \mu$m is well-suited to an initial proof-of-principle experiment. The corresponding wavelength of the defect mode is about $18\mu$m. If we want to operate at a (telecommunication) wavelength of $1.5 \mu$m, we will need a lattice constant of only $420$ nm. For large lattice constants (and correspondingly long wavelengths), the changes induced by a fixed, thin analyte layer thickness would be weaker than for a smaller lattice constant (and correspondingly shorter wavelength). For example, the chip in Fig. \ref{fig:improve sensitivity} with $a=5 \mu$m would require a 10 nm analyte coating thickness for detection, in contrast to a 1 nm analyte coating thickness that would be detectable using $a=0.5 \mu$m (at a wavelength of $1.8 \mu$m). The other obvious advantage of using smaller lattice constants and shorter wavelengths is the need for a much-reduced volume of fluid specimen.  Our illustration involving the 18-micron IR wavelength is simply to provide an easier starting point for experimental testing.

For our enhanced sensitivity chip (in Fig. \ref{fig:improve sensitivity}) with fragmented rectangular strip construction, the limit-of-detection is greatly improved even for the narrower photonic crystal chip of 11-unit-cells in width. With such a modified domain-wall defect with resonant frequency $\omega =0.296732 [2\pi a/c]$ and quality factor $Q=1312.8$, $\Delta \omega^{\text{FWHM}}=2.3 \times 10^{-4} [2\pi a/c]$, and the limit-of-detection is $\delta t^{\text{lim}}/a=0.002$ in terms of analyte coating thickness. If the photonic crystal has a lattice constant of $a=1$ micron, a two-nanometer coating of analyte (of refractive index 1.45) would be detected. This minimal detectable analyte thickness can be further improved by increasing the number of unit cells of the photonic crystal as discussed above. For the 11-unit-cell-wide photonic crystal, the minimum detectable refractive index change of the background fluid is $\delta n^{\text{lim}}=7.35 \times 10^{-4} $RIU. Again, this can be improved by increasing the number of photonic crystal unit cells in the light propagation direction.

Note: In our design, we have fixed the values of the length of each rectangular strip and their nearest-neighbor spatial separation, i.e. $h=0.4a$ and $t_0=0.1a$ always, and only the width of every rectangular strip $w_{def}$ is varied. This means that for an individual defect mode, its resonant frequency and mode profile are controlled by $w_{def}$ solely. In the above discussion, we have chosen the optimal value $w_{def}=0.03a$ which maximizes the sensitivity. With a larger strip width, the sensitivity is slightly lowered.

\section{Two topological defects: mode hybridization and Spectral independence}

For biosensors capable of multi-parametric detection, which can distinguish different concentrations of multiple disease markers in a single measurement, multiple defects are required. This gives rise to mode hybridization, which complicates the interpretation of transmission spectra due to correlated spectral responses from different analyte-bindings.  We first study a chip with two defects to analyze mode hybridization and its relation to transmission-levels. We analyze correlations in frequency shifts of the two separate resonances, and how to achieve spectral independence. 

As the PBG excludes all other modes with nearby frequencies, the transmission occurs exclusively via the two defect modes. The Maxwell operator within the two-defect mode subspace, is expressible as:
\begin{equation}
    \hat{\Theta} \approx \begin{pmatrix} \omega^2_L & \kappa \\  \kappa & \omega^2_R \end{pmatrix}.
\label{2-defect operator}
\end{equation}
This is expressed in the basis of the left-defect mode $|\text L\rangle$ and the right-defect mode $|\text R\rangle$ with resonant frequencies $\omega_L$ and $\omega_R$ respectively. Here, $\kappa$ is the direct coupling strength between these two modes, which depends on the spatial separation $\text D$ between these two defects and their mode overlap. With our definition, $\kappa$ has the same units as the square of the resonant frequency of optical mode $\omega^2$. (Here we use the Dirac ket notation and denote the resonant mode as $|\text{label}\rangle$, where the ``label” within the ket notation serves to specify which mode the ket-state refers to.) The two resonances, denoted as $|\text L\rangle'$ and $|\text R\rangle'$, are eigenmodes of $\hat{\Theta}$ with resonant frequencies $\omega_L'$ and $\omega_R'$ respectively. They are linear combinations of $|\text L\rangle$ and $|\text R \rangle$. We refer to this mixing as hybridization of the left-defect mode $|\text L\rangle$ with the right-defect mode $|\text R\rangle$. The mixing fraction depends on the factor $\kappa/\Delta \omega^2$, where the frequency difference $\Delta \omega^2 \equiv|\omega_R^2-\omega_L^2|$. The direct transmission amplitude of each individual defect mode depends sensitively on its relative amplitudes at the left and right surfaces. High transmission occurs if the exponentially decaying amplitudes are equal. Strong asymmetry arises if the defect is located off-center. For two off-center defects, high transmission may occur by optical tunneling between the two defect modes. The tunneling effectiveness is determined by the parameter $\kappa$ and the frequency separation $\Delta \omega^2$, which together describe the hybridization. 

As we functionalize these two defect modes for optical biosensing, the analyte-binding will alter the resonant frequency difference $\Delta \omega$ and the coupling strength $\kappa$ between the two defect modes. Mode hybridization invariably leads to correlated spectral responses of the two resonances with respect to analyte-binding, i.e. the transmission-level change and red-shift of one resonance will induce a correlated transmission-level change and red-shift of the other resonance. This makes it more difficult to distinguish different analyte-bindings in the transmission spectrum. However, with weak mode hybridization, the correlation in spectral responses is small and these two transmission peaks respond nearly independently to analyte-binding. However, spectral independence unavoidably reduces the transmission-levels. In other words, there is a trade-off between the spectral independence and high transmission-levels.

We can change the spatial separation, D, between defects and the strip widths within the defect regions to adjust their coupling strength $\kappa$ and the mode hybridization $\kappa/\Delta \omega^2$. Different choices of domain-wall and anti-domain-wall combinations, $lr=\{ss, s\bar{s},\bar{s}s,\bar{s}\bar{s}\}$, also make a slight difference to the coupling strength $\kappa$.
The spatial separation between a domain-wall and an anti-domain-wall $(s-\bar{s}, \text{ or } \bar{s}-s)$ must be integer multiples of unit cells. However, spatial separation between two domain-walls $(s-s)$ or two anti-domain-walls $(\bar{s}-\bar{s})$ must be half integers. Also, the mode profile of a domain-wall is slightly different from that of an anti-domain-wall with the same strip width, as shown in Sec. II A. In the following, we consider a two-defect chip consisting of a domain-wall on the left and an anti-domain-wall on the right ($s-\bar{s}$). There are three binding cases for a chip with two defects: $\text L$-binding, with one type of analyte bound to the left-defect region; $\text R$-binding with another type of analyte bound to the right-defect region; and $\text {LR}$-binding, with both types of analyte bound to the left- and right-defect region respectively. A properly designed chip must distinguish these different binding profiles through their transmission spectra. 

\begin{widetext}
\begin{center}
\begin{figure}[htbp]
 \centering
  \subfigure[]{
   \hspace{0in} \includegraphics[width=0.95\textwidth]{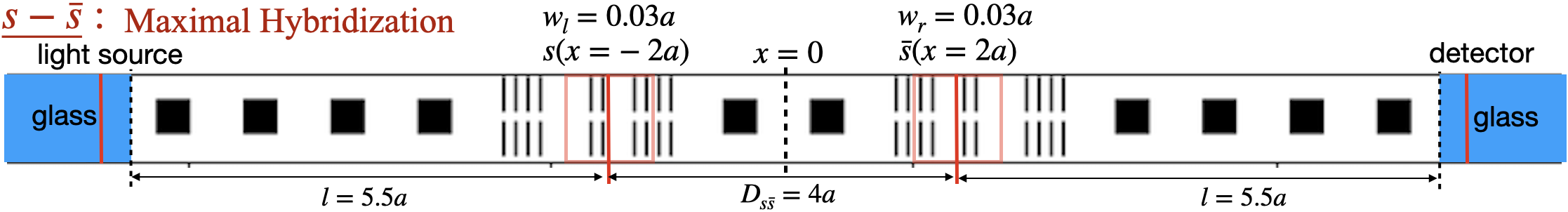}}
    \label{fig: DW-antiDW pair chip}
  \hspace{0cm}
  \subfigure[]{
    \hspace{0in}\includegraphics[width=0.85\textwidth]{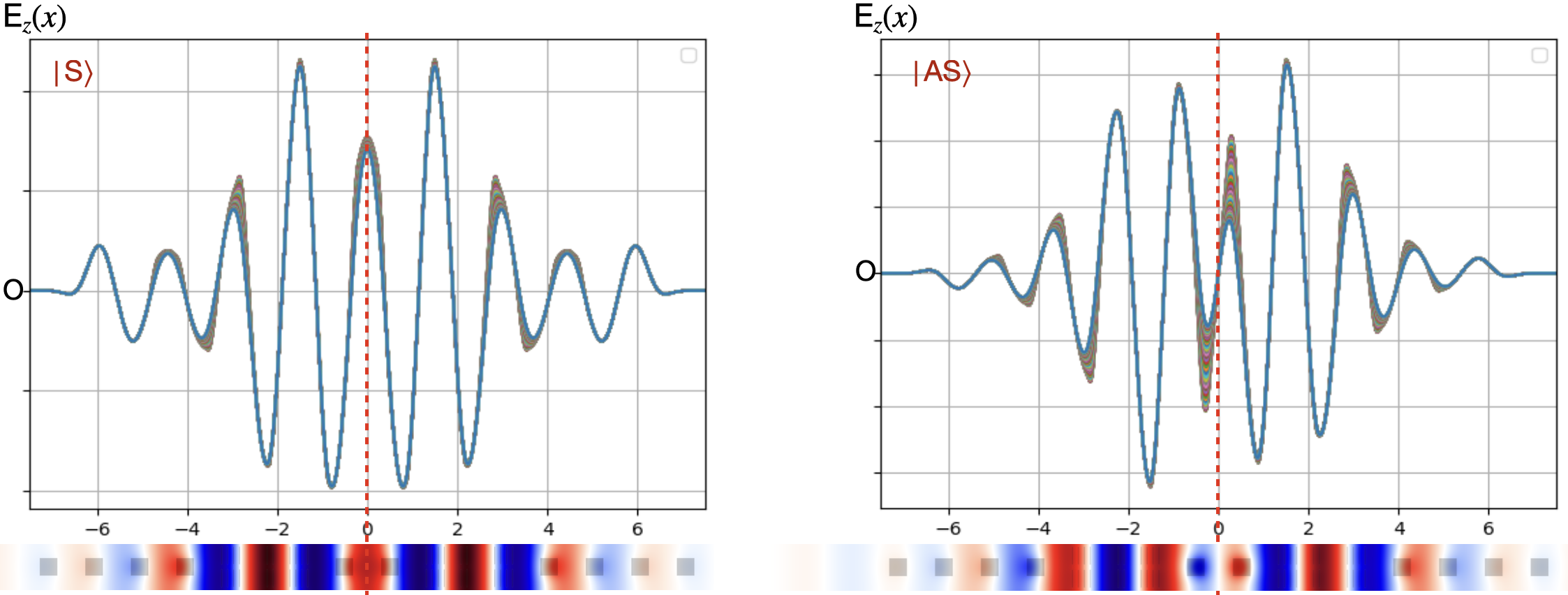}}
 \caption{ (a) Two-defect chip, consisting of a left domain-wall and a right anti-domain-wall with the same strip width $w_l=w_r=0.03a$, is symmetric under reflection about the middle $yoz$ plane.  (b) Eigenmode $|\text S\rangle$, with resonant frequency $\omega_S=0.292081$ $[2\pi c/a]$, has a symmetric electric field profile, and  eigenmode $|\text{AS}\rangle$, with resonant frequency $\omega_{AS}=0.302565$ $[2\pi c/a]$, is antisymmetric.}
\label{fig: DW and antiDW pair}
\end{figure}
\end{center}
\end{widetext}

\subsection{Maximal Hybridization}

We first investigate the case with a left domain-wall and a right anti-domain-wall with the same strip width $w_l=w_r=0.03a$, as shown in Fig. \ref{fig: DW and antiDW pair}(a). Here, the left- and right-defect modes are degenerate and have exactly the same resonant frequency $\omega_L=\omega_R=\omega$, $\Delta \omega^2=0$. The modes are maximally hybridized into symmetric and anti-symmetric combinations:
\begin{equation}
\begin{aligned}
    |\text S\rangle &= \frac{1}{\sqrt{2}}\left( |\text L\rangle+|\text R\rangle\right),\\
    |\text{AS}\rangle &= \frac{1}{\sqrt{2}}\left( |\text L\rangle-|\text R\rangle\right). 
\end{aligned}
\label{eqn: S and AS}
\end{equation}
with eigenvalues $\omega^2_S=\omega^2+\kappa  ,\omega^2_{AS}=\omega^2-\kappa$ respectively. Due to the reflection symmetry $\varepsilon (x,y)=\varepsilon(-x,y)$ of the chip, the field profiles of these two hybridized modes $|\text S\rangle$ and $|\text {AS}\rangle$ have a definite parity. As shown in Fig. \ref{fig: DW and antiDW pair}(b), the electric field for the $|\text S\rangle$ is symmetric, and that for the $|\text{AS}\rangle$ is antisymmetric with respect to the middle mirror plane:
\begin{equation}
    {\bf E}_{\text S}(x,y)={\bf E}_{\text S}(-x,y), \hspace{0.2cm} {\bf E}_{\text{AS}}(x,y)=-{\bf E}_{\text{AS}}(-x,y).
\end{equation}

\begin{widetext}
\begin{center}
\begin{figure}[htbp]
 \centering
  \subfigure[]{
   \hspace{0in} \includegraphics[width=0.47\textwidth]{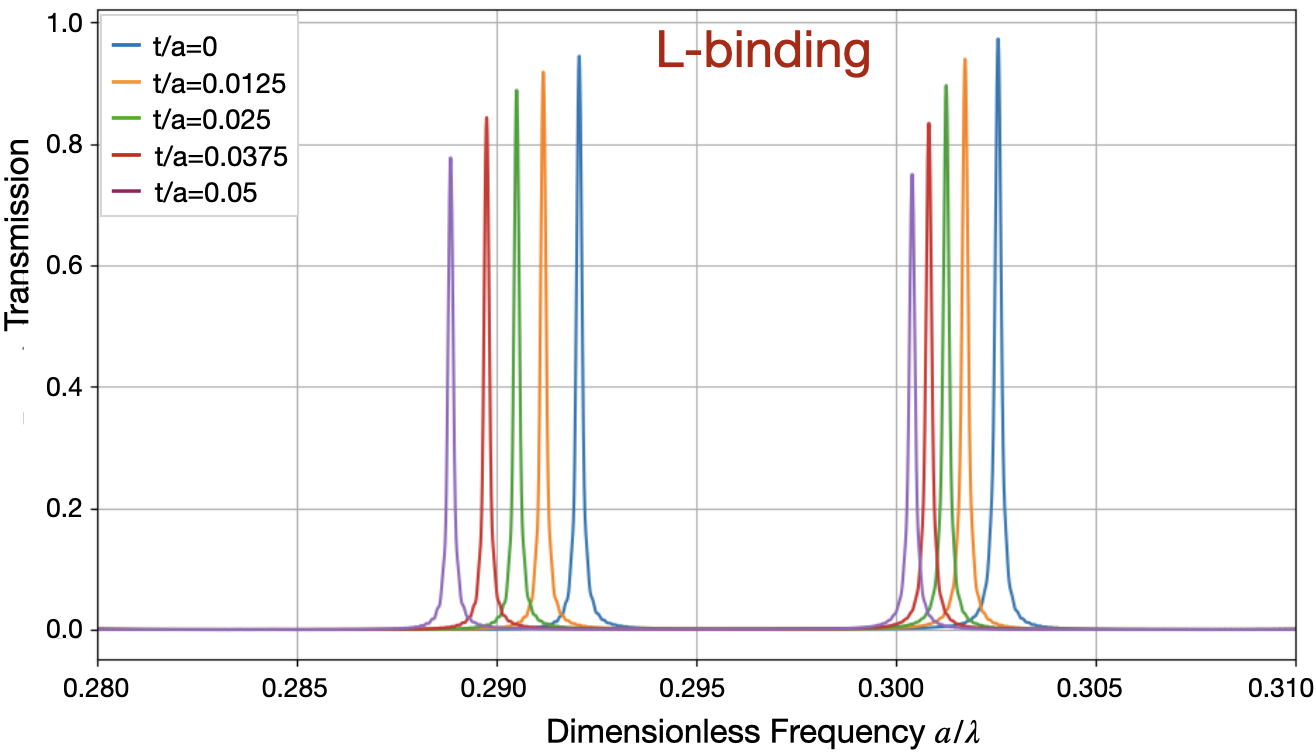}}
    \label{fig: S-AS_003-003_L-bindings}   
   \hspace{0.4cm}
  \subfigure[]{
    \hspace{0in}\includegraphics[width=0.47\textwidth]{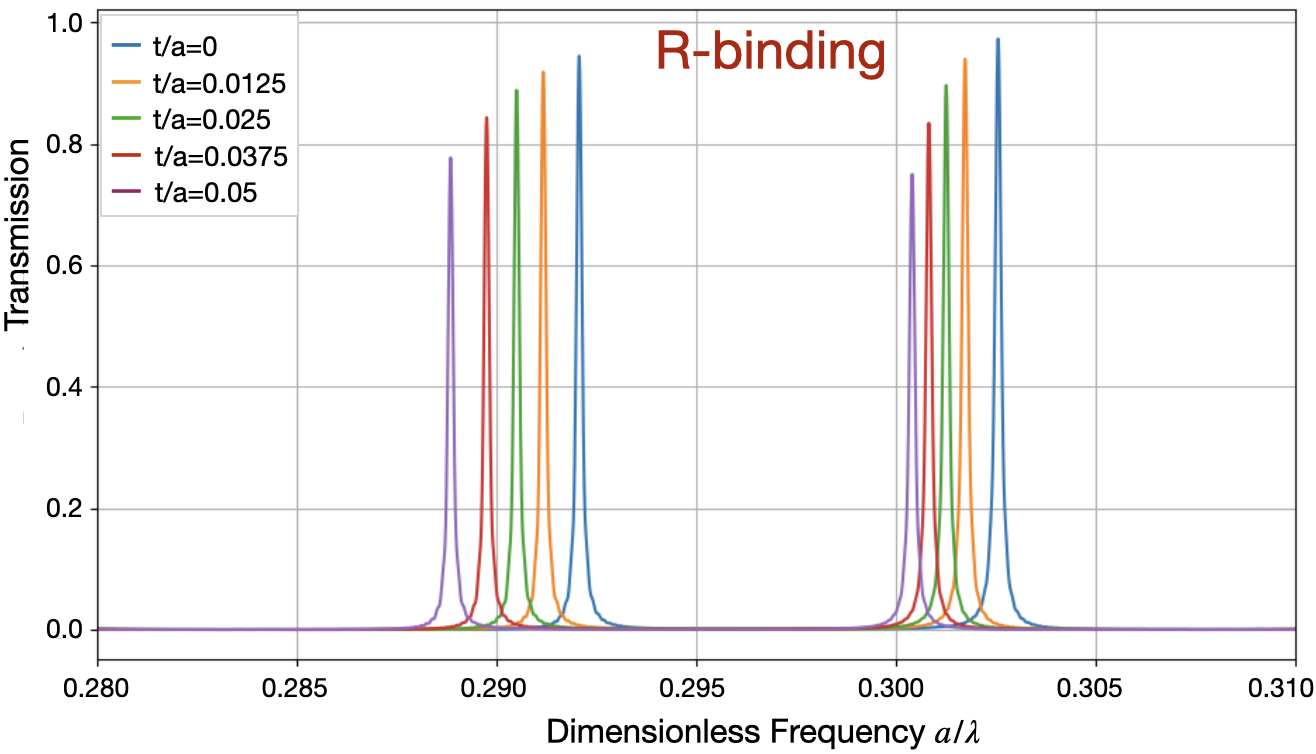}}
    \label{fig: S-AS_003-003_R-bindings}
  \subfigure[]{
    \hspace{0in}\includegraphics[width=0.47\textwidth]{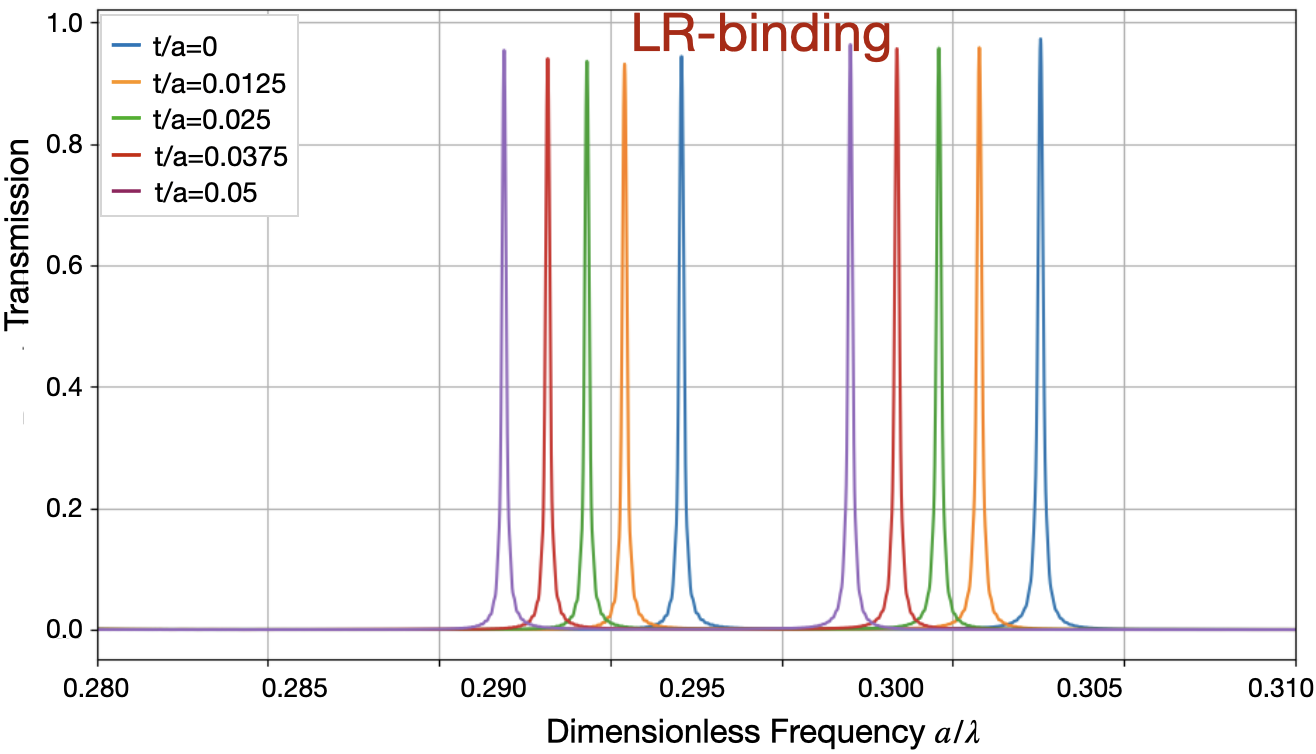}}
    \label{fig: S-AS_003-003_LR-bindings} 
   \subfigure[]{
   \hspace{0in} \includegraphics[width=0.5\textwidth]{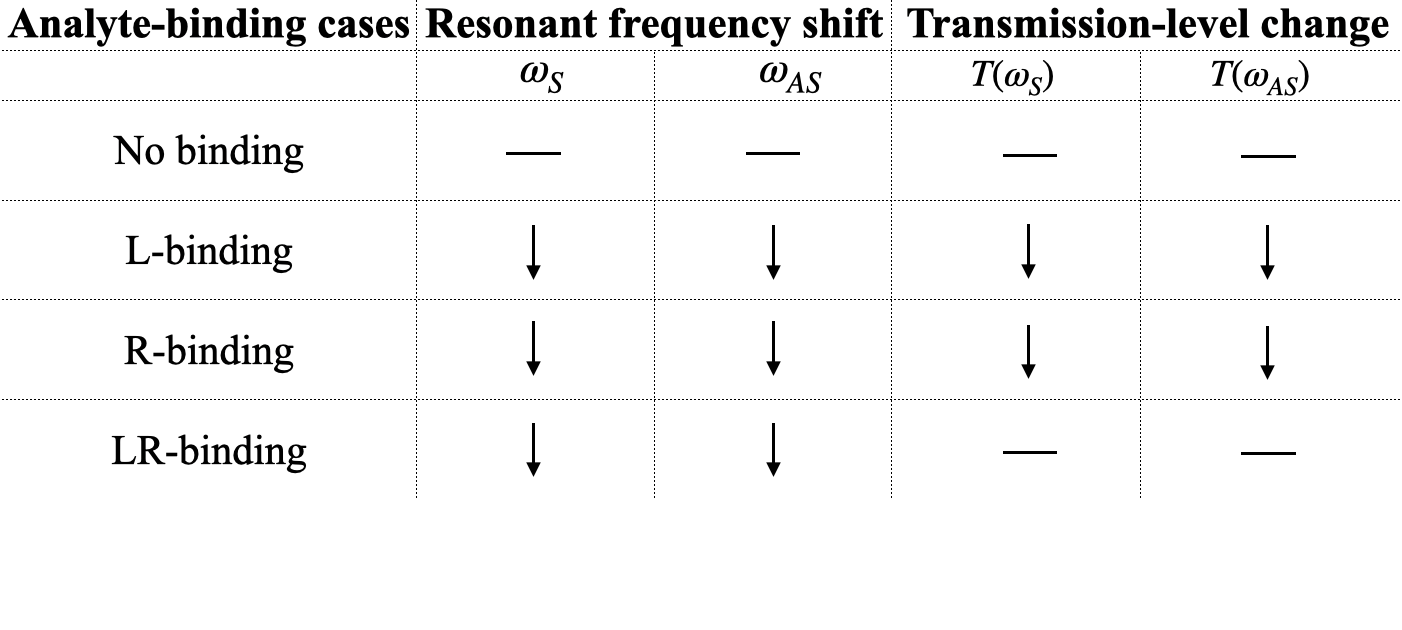}}
    \label{fig: S-AS_003-003_Truth table}
 \caption{ Transmission spectra for different analyte-binding cases for the symmetric two-defect chip with  domain-wall and anti-domain-wall, with the same strip width $w_r=w_l=0.03a$. (a) L-binding. (b) R-binding. (c) LR-binding. (d) Truth table for all the different analyte-binding cases. The spectral responses of two resonances are highly correlated: the two transmission peaks red-shift together and their transmission-levels change together with respect to analyte-binding. The L-binding and R-binding are indistinguishable due to the mirror symmetry.}
\label{fig: S-AS bindings}
\end{figure}
\end{center}
\end{widetext}

Due to this mirror symmetry, the $\text L$-binding and $\text R$-binding with the same analyte-thickness cannot be distinguished, since they lead to the same variation of mode hybridization. Fig. \ref{fig: S-AS bindings}(a)-(c) show the transmission spectra for different analyte-binding cases. The $\text L$-binding and $\text R$-binding cannot be distinguished, and their transmission peaks are highly correlated, i.e. the resonant frequency shift and transmission-level change of the $|\text S\rangle$ mode will induce a correlated frequency shift and transmission-level change of the $|\text{AS}\rangle$ mode, and vice versa. These resonant frequency shifts and transmission-level changes for all the binding cases are shown in the Truth table (Fig. \ref{fig: S-AS bindings}(d)). Consider L-binding as a pedagogical example. The binding of analyte makes the effective dielectric constant within the left-defect region larger, leading to a red-shift in resonant frequency. Due to the mode hybridization, both the resonances $|\text S\rangle$ and $|\text {AS}\rangle$ red-shift in unison. The lower frequency $|\text S\rangle$ resonance shifts more because its mode profile has stronger overlap with bound analyte than that of $|\text{AS}\rangle$ on average. As more analyte binds, the resonant frequency difference $\Delta \omega$ becomes larger, leading to weaker hybridization ($\kappa/\Delta \omega^2$) between the left-defect mode $|\text L\rangle$ and right-defect mode $|\text R\rangle$. As a result, the hybridized mode amplitudes between the right and left surfaces become more imbalanced and the transmission levels decline. The weaker hybridization leads to some small degree of spectral independence between the two transmission peaks. For the maximally hybridized chip, the transmission-levels of these two resonances are close to unity, but their spectral responses are excessively correlated.

\subsection{Effective Dehybridization}

We now illustrate a two-defect chip with $\kappa/\Delta \omega^2 \ll 1$. Here, the more weakly hybridized left-like $|\text L\rangle'$ and right-like $|\text R\rangle'$ are focused mostly on a single defect:
\begin{equation}
    |\text L\rangle' \approx |\text L\rangle, \hspace{0.5cm} |\text R\rangle' \approx |\text R\rangle.
\end{equation}
In this case, the $\text L$-binding will have almost no influence on the resonance $|\text R\rangle'$, and vice versa, leading to clear spectral independence of the transmission peaks with respect to analyte-binding.
However, complete suppression of mode hybridization can make transmission levels unacceptably low. An important mitigating factor is that as mode hybridization decreases, spectral correlation typically declines more rapidly than the decline in transmission. By choosing the ideal mode hybridization, $\kappa/\Delta \omega^2$, sufficient spectral independence can be retained while maintaining acceptable transmission. In this case, we refer to the modes as being effectively dehybridized. 

\begin{widetext}
\begin{center}
\begin{figure}[htbp]
 \centering
 \subfigure[]{
   \hspace{0in} \includegraphics[width=0.95\textwidth]{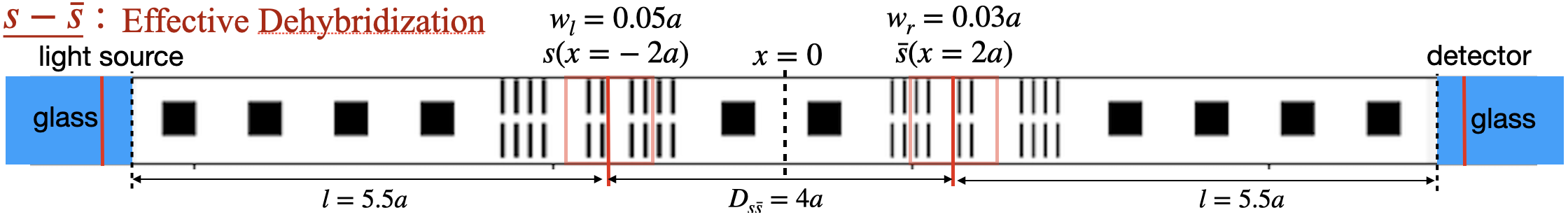}}
    \label{fig: 2-defect dehybridized chip}
  \subfigure[]{
   \hspace{0in} \includegraphics[width=0.85\textwidth]{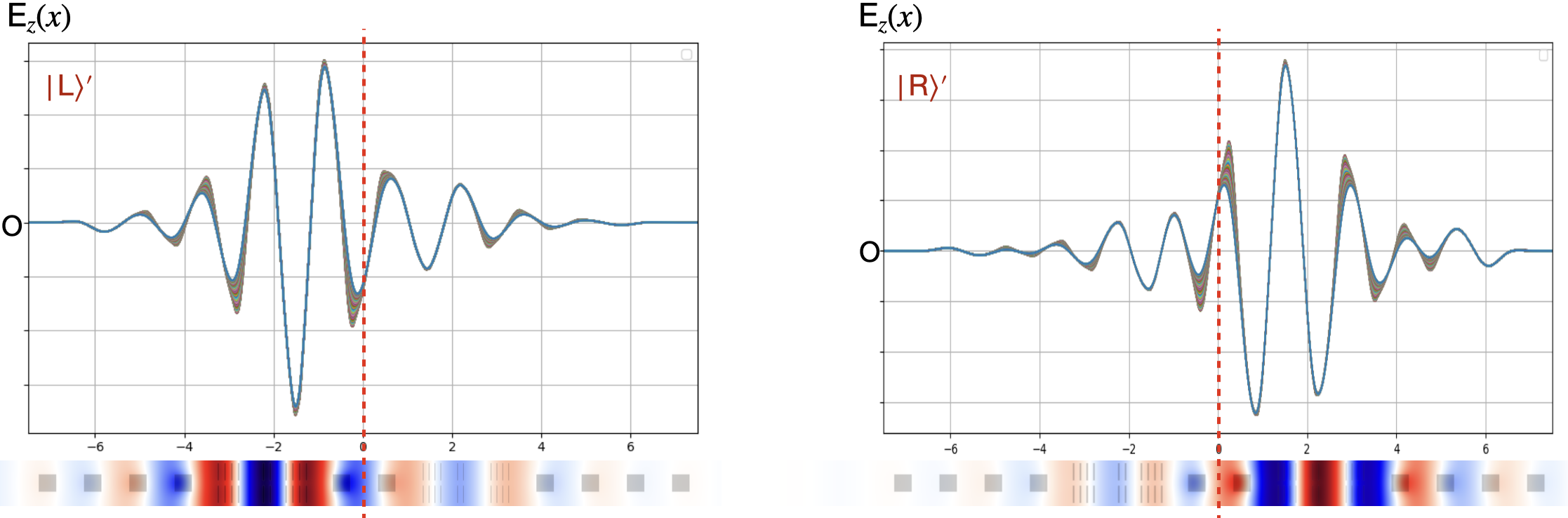}}
    \label{fig: 2 dehybridized mode profiles}
  \caption{Two-defect chip with two effectively dehybridized resonant modes. (a) The strip width $w_l=0.05a$ in the left domain-wall ($s$) region, whereas in the right anti-domain-wall ($\bar{s}$) region $w_r=0.03a$. This yields a resonant frequency difference $\Delta \omega=\omega_R-\omega_L=0.023102$ $[2\pi c/a]$.  (b) The two resonant modes are effectively dehybridized into a left-like mode with resonant frequency $\omega'_L=0.272954$ $[2\pi c/a]$, and a right-like mode with resonant frequency $\omega'_R=0.298088$ $[2\pi c/a]$.}
\label{fig: 2 dehybridized modes}
\end{figure}
\end{center}
\end{widetext}    

With different strip widths for the left and right defects, we introduce a resonance frequency difference $\Delta \omega \neq 0$, that suppresses mode hybridization. For example, choosing strip widths $w_l=0.05a$ and $w_r=0.03a$ within the left- and right-defect region respectively, the resonant frequency difference $\Delta \omega=0.023102$ $\left[2\pi c/a\right]$. This effectively dehybridizes these resonant modes into the left and right ones with a spatial separation $\text D=4a$ for this $s-\bar{s}$ chip. By making the frequency difference even larger, the spatial separation can be reduced further, but with a reduction in sensitivity (due to large strip width). For the chip shown in Fig. \ref{fig: 2 dehybridized modes}(a), the two effectively dehybridized resonances occur at frequency $\omega'_L=0.272954$ $[2\pi c/a]$ and $\omega'_R=0.298088$ $[2\pi c/a]$. Their electric field profiles are shown in Fig. \ref{fig: 2 dehybridized modes}(b). These two field profiles concentrate within the left- and right-defect region, with only a small overlap occurring through their tails. This enables detectable transmissions and nearly undetectable correlation in frequency shifts. 
\begin{widetext}
\begin{center}
\begin{figure}[htbp]
 \centering
  \subfigure[]{\includegraphics[width=0.47\textwidth]{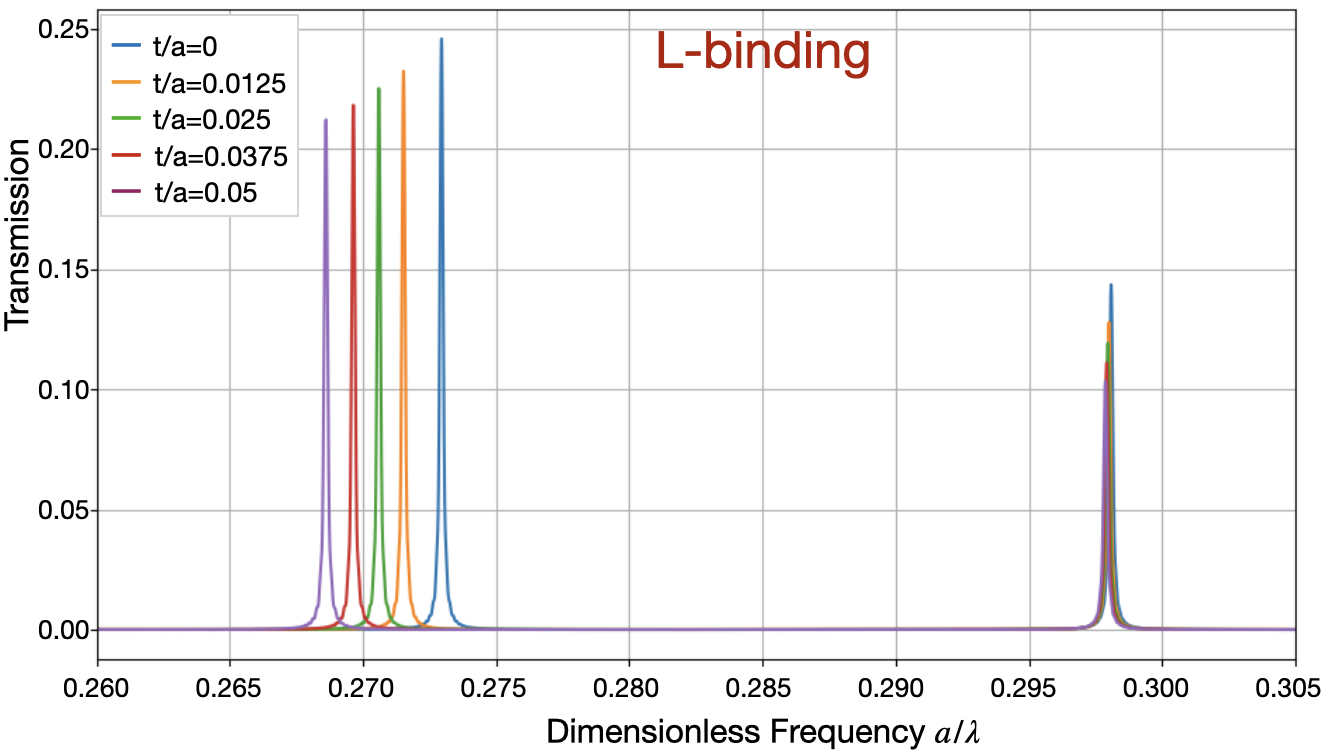}}
    \label{fig: 2 dehybrized L-bindings}
  \hspace{0.4cm}
  \subfigure[]{\includegraphics[width=0.47\textwidth]{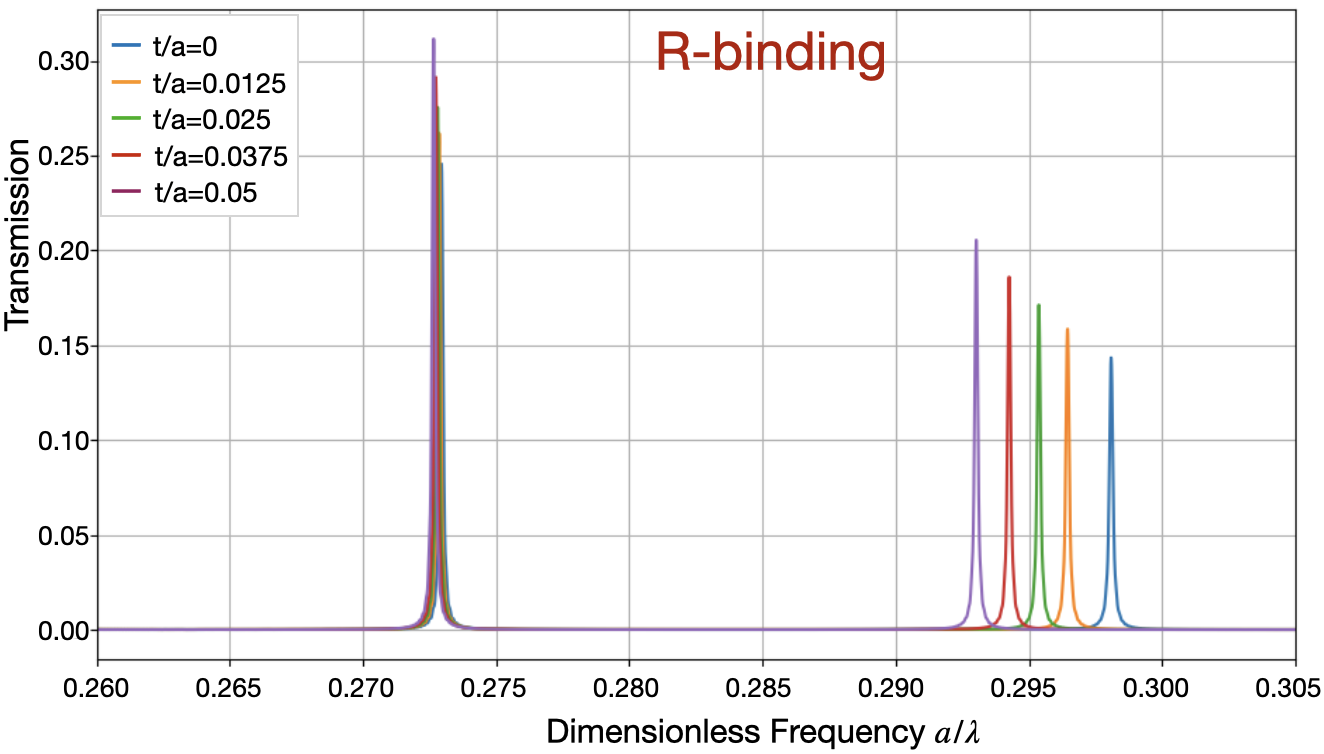}}
    \label{fig:2 dehybrized R-bindings}
     \subfigure[]{\includegraphics[width=0.47\textwidth]{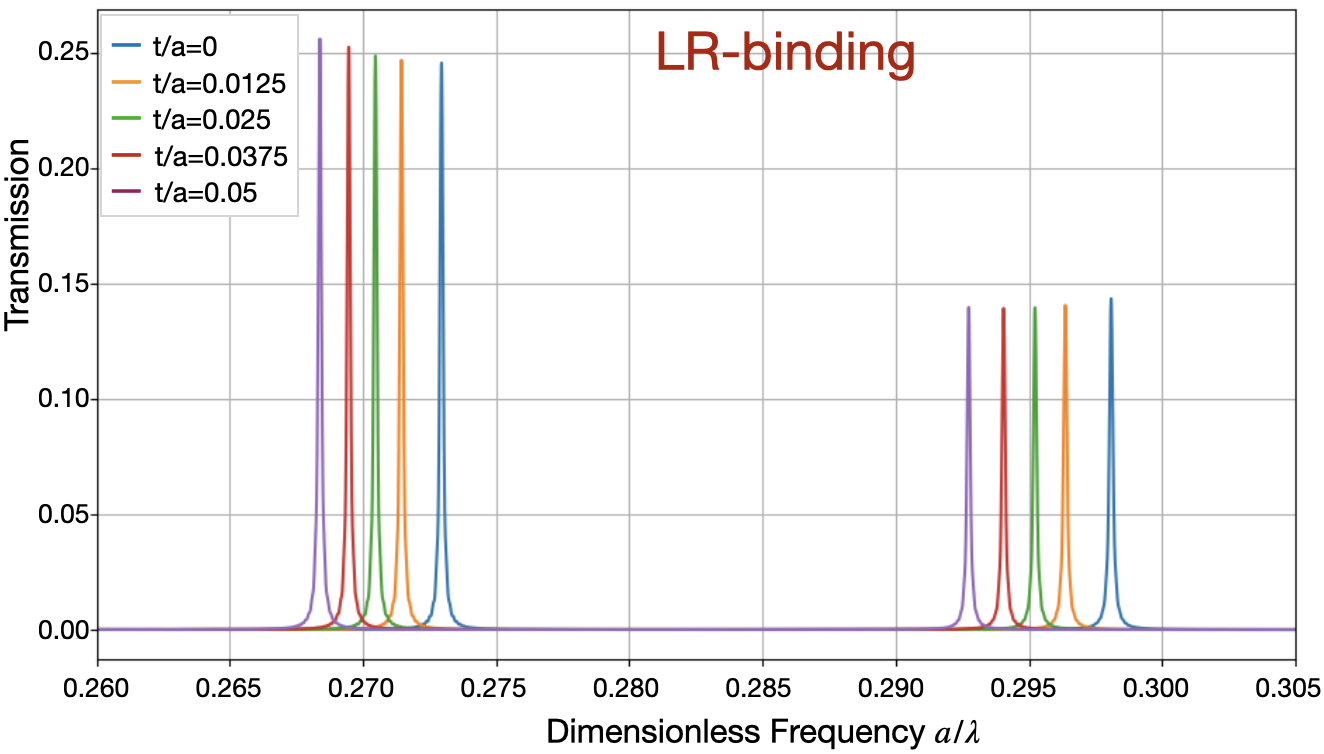}}
    \label{fig:2 dehybrized LR-bindings}
   \subfigure[]{ \includegraphics[width=0.5\textwidth]{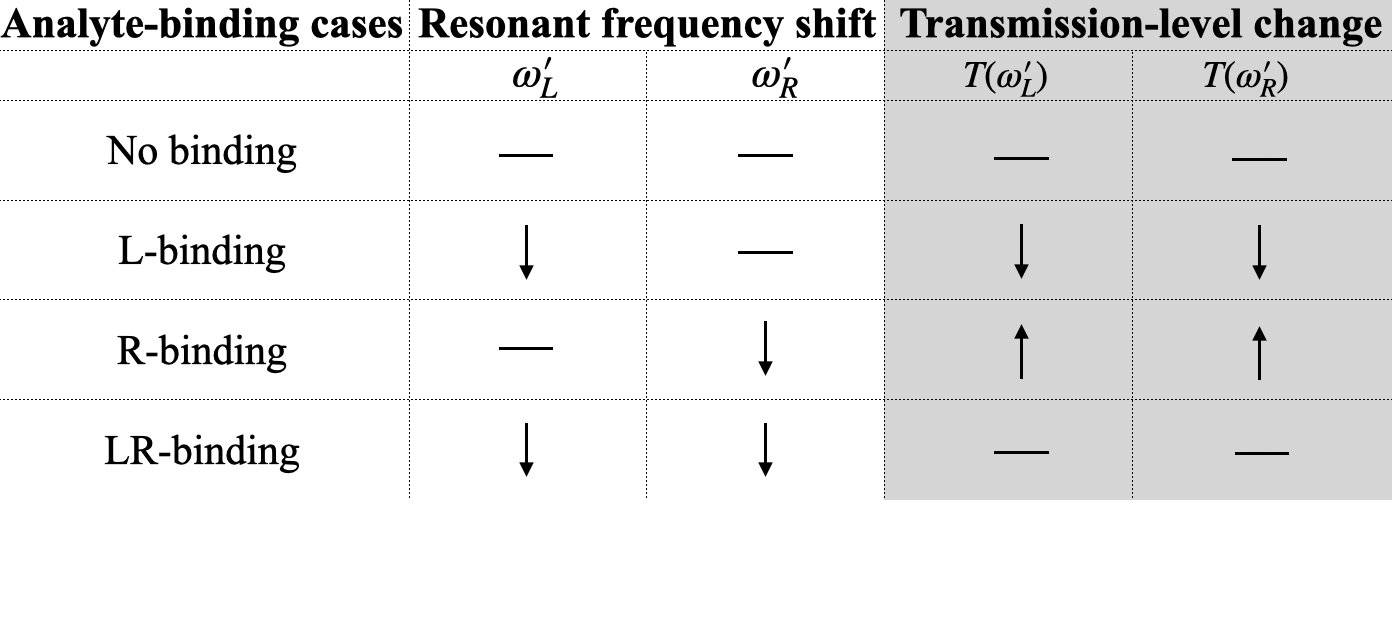}}
    \label{fig: S-AS_003-005_Truth table}
 \caption{ Transmission spectra for different analyte-binding cases for the two-defect chip supporting two effectively dehybridized modes. (a) L-binding. (b) R-binding. (c) LR-binding. The left-like resonance and the right-like resonance respond independently in frequency-shifts to analyte-binding. (d) Truth table for all the different analyte-binding cases reveals complete distinguishability by frequency shifts only. The observed transmission-level changes are due to mode hybridization changes.}
\label{fig: dehybridized modes bindings}
\end{figure}
\end{center}
\end{widetext}

 Fig. \ref{fig: dehybridized modes bindings}(a)-(c) show the completely distinguishable transmission spectra for all the different analyte-binding cases. For example, the $\text L$-binding mainly red-shifts the $|\text L\rangle'$ resonance, while the induced correlated frequency-shift of the $|\text R\rangle'$ resonance is nearly undetectable. Therefore, the three binding cases are distinguishable by frequency shifts only. The weak transmission-level changes are also listed in the Truth Table (Fig. \ref{fig: dehybridized modes bindings}(d)). They are not necessary for distinguishing analyte-binding cases but are indicative of the level of hybridization.

\begin{widetext}
\begin{center}
\begin{figure}[htbp]
 \centering
  \subfigure[]{
   \hspace{0in} \includegraphics[width=0.95\textwidth]{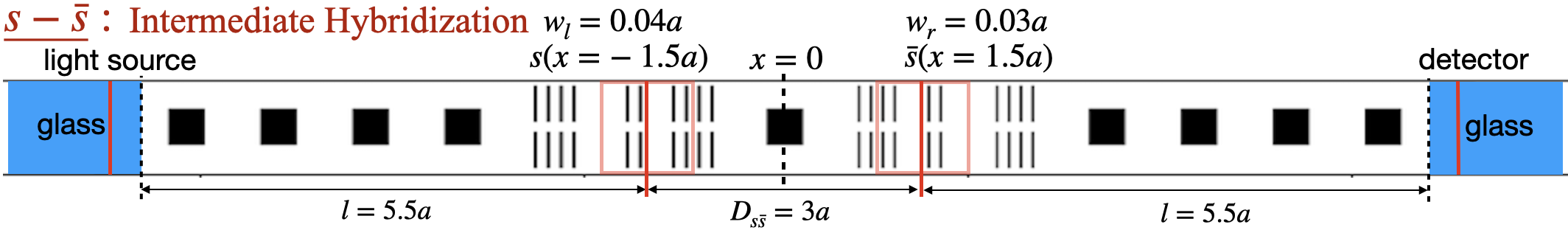}}
    \label{fig: 2-defect dehybridized chip}
  \subfigure[]{
   \hspace{0in} \includegraphics[width=0.47\textwidth]{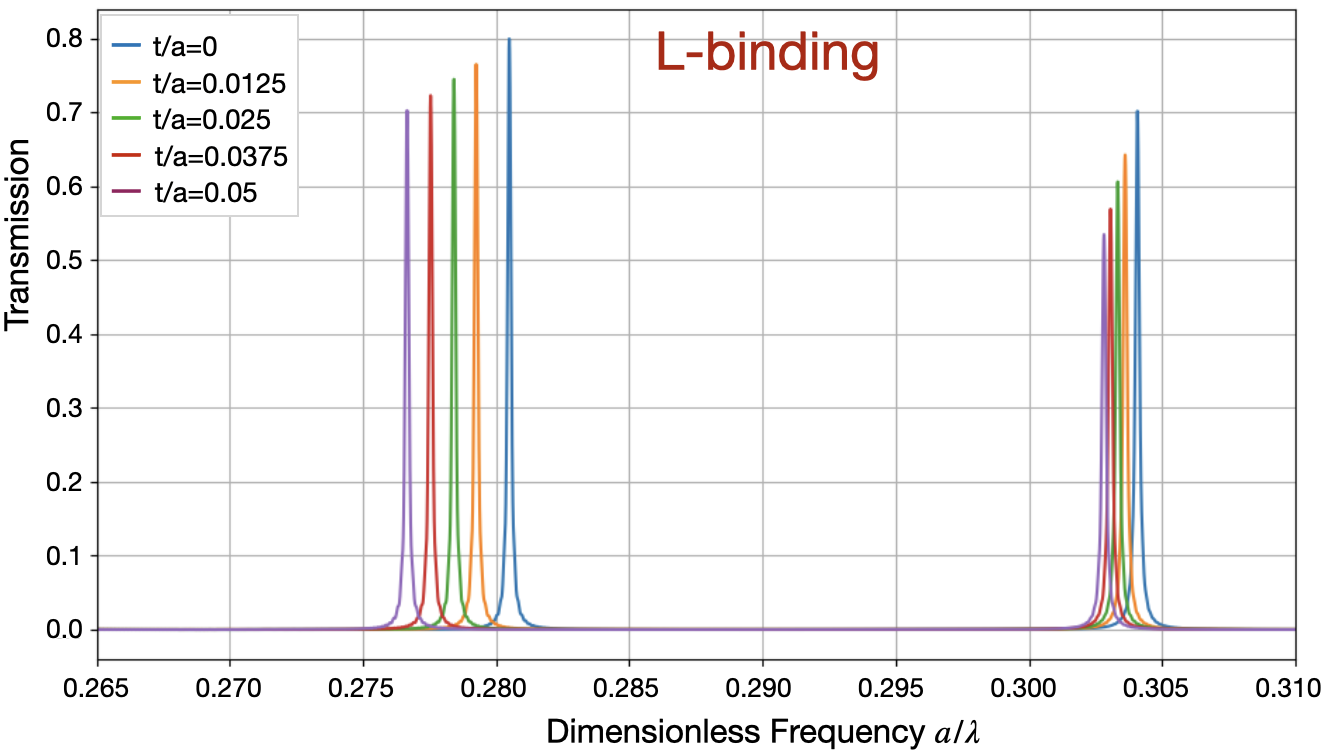}}
    \label{fig: 2 hyridized L-bindings}
    \hspace{0.4cm}
  \subfigure[]{
    \hspace{0in}\includegraphics[width=0.47\textwidth]{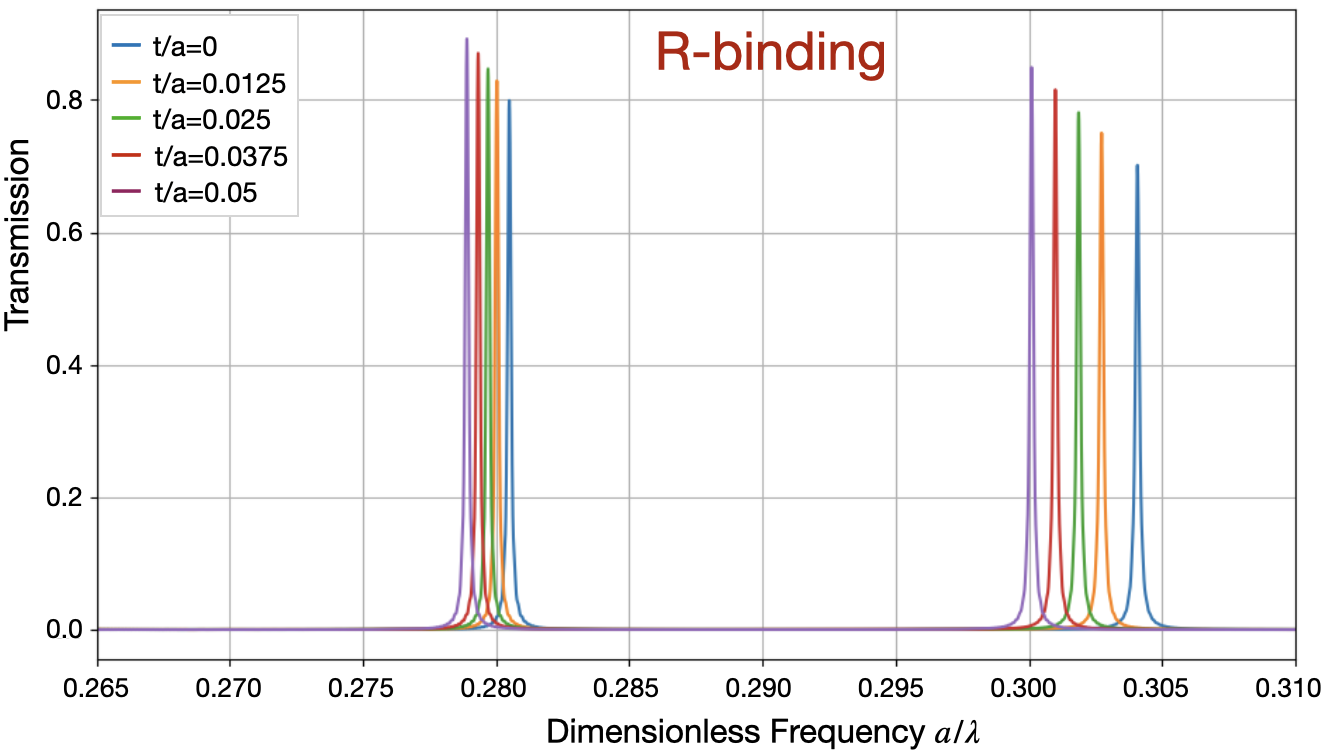}}
    \label{fig: 2 hybrized R-bindings}
     \subfigure[]{
    \hspace{0in}\includegraphics[width=0.47\textwidth]{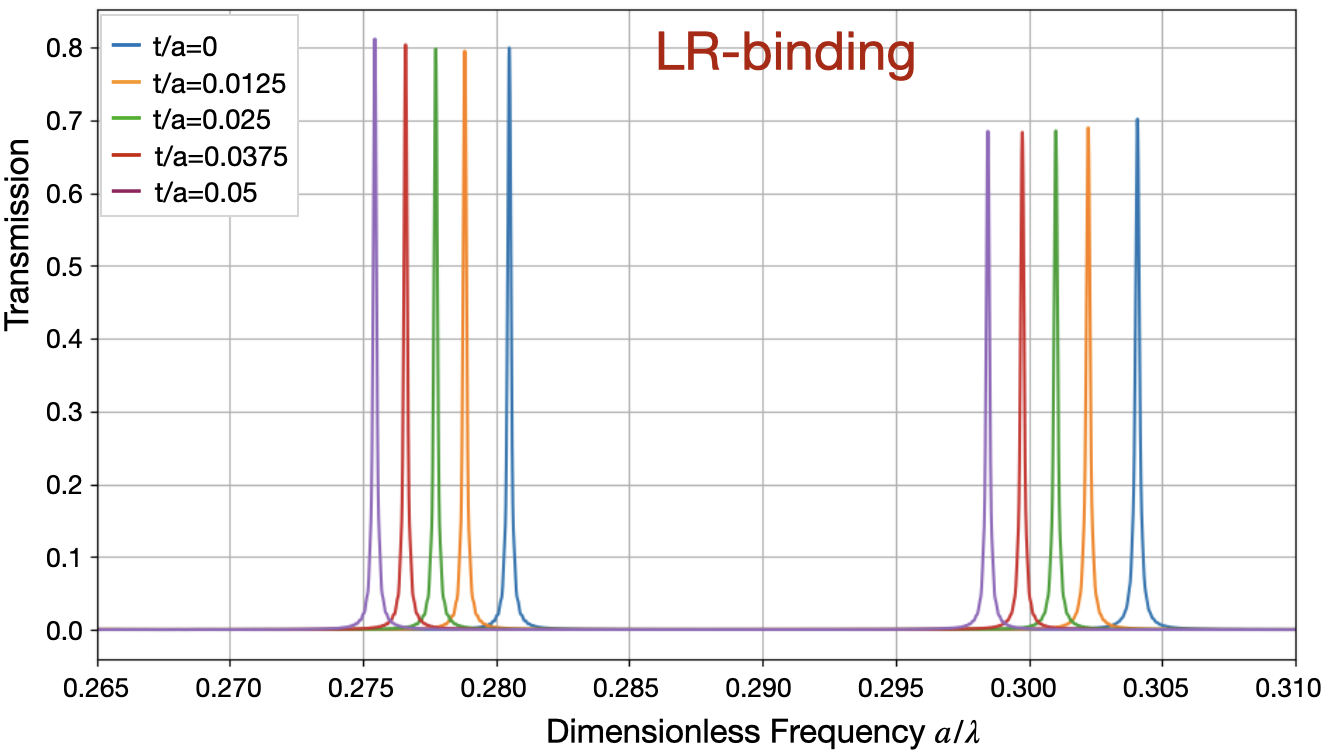}}
    \label{fig: 2 hybrized LR-bindings}
   \subfigure[]{
   \hspace{0in} \includegraphics[width=0.5\textwidth]{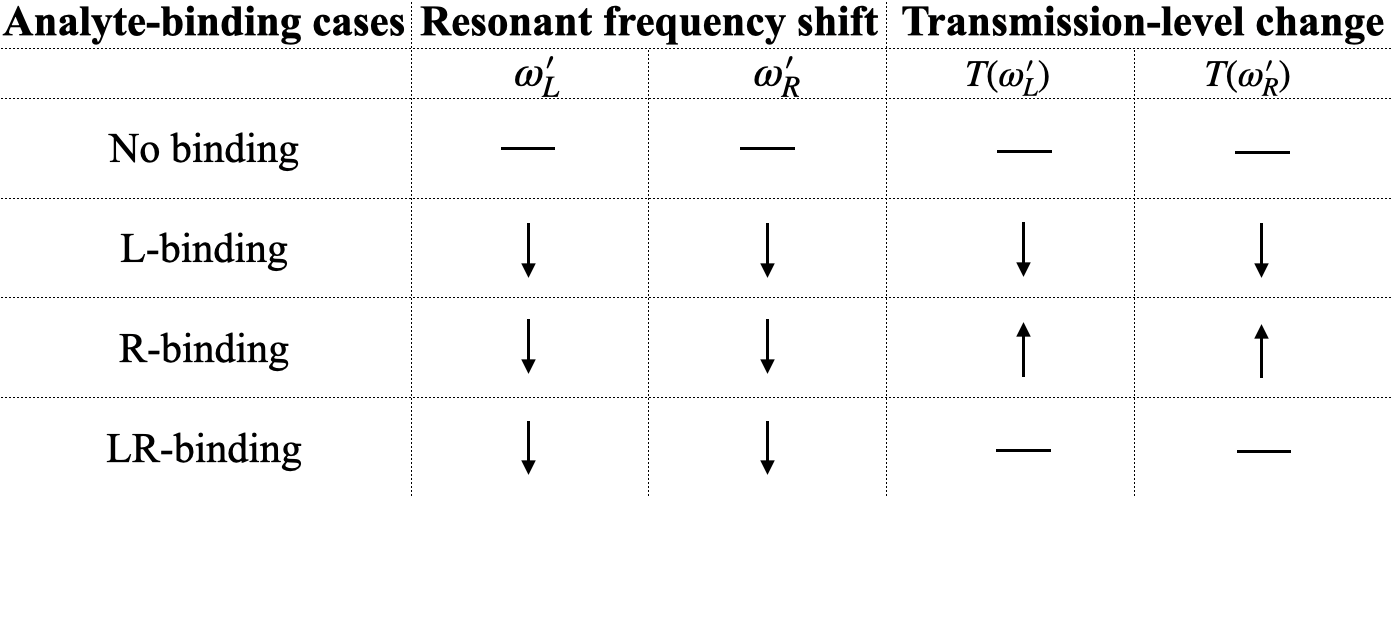}}
    \label{fig: S-AS_003-004_Truth table}
 \caption{ Transmission spectra for different analyte-binding cases for the two-defect chip with intermediate hybridization of modes. (a) L-binding. (b) R-binding. (c) LR-binding. (d) Truth table for all the different analyte-binding cases reveals distinguishability with respect to both frequency shifts and transmission-level changes. With increased hybridization, the transmission-levels are improved, but the spectral responses of the two resonances become more correlated.}
\label{fig: S-AS improved transmission}
\end{figure}
\end{center}
\end{widetext}

\subsection{Intermediate Hybridization}
 In this section, we increase the transmission levels by increasing the mode hybridization at the cost of more correlated spectral responses. We increase $\kappa/\Delta \omega^2$ by reducing the frequency difference $\Delta \omega$ and reducing the spatial separation D to raise the transmission peak heights.

By changing the strip width within the left-defect region to $w_l=0.04a$, slightly different from that within the right-defect region $w_r=0.03a$, the frequency difference is reduced to $\Delta \omega =0.012049$ $\left[2\pi c/a \right]$. We also reduce the spatial separation from $\text D=4a$ to $\text D=3a$. This improves the transmission coefficients to $0.7 \sim 0.8$, as shown in Fig. \ref{fig: S-AS improved transmission}. The resonant frequencies of these two hybridized modes are $\omega_L'=0.280499$ $[2\pi c/a]$ and $\omega_R'=0.304074$ $[2\pi c/a]$.

We functionalize this chip for biosensing and numerically simulate its transmission spectra for the various analyte-binding cases, as shown in Fig. \ref{fig: S-AS improved transmission}(b)-(d). Due to increased mode hybridization, the spectral responses of the two resonances become more correlated. Nevertheless, different binding cases are clearly distinguishable especially if we monitor the transmission-level changes, as shown in the Truth Table (Fig. \ref{fig: S-AS improved transmission}(e)). The frequency-shifts by themselves are also distinct. For example, with L-binding, the left-like resonance $|\text L\rangle'$ will red-shift, inducing a correlated red-shift of the right-like resonance $|\text R\rangle'$. Since the field strength of the left-like mode is higher than that of the right-like mode in the left-defect region, the left-like resonance $|\text L\rangle'$ will red-shift more. The resonant frequency difference $\Delta \omega$ becomes larger with L-binding. This reduces hybridization and makes the transmission-levels of both resonances decline. Similar analysis applies to the other binding cases.

In summary, there are two functional choices of two-defect chip structures. The first chip has effectively dehybridized resonances and mutually independent spectral responses to analyte-bindings, but low transmission-levels. The second chip has high transmission-levels and correlated spectral responses to analyte-binding. For the first chip, we need only to detect the frequency-shifts to distinguish all the binding cases. For the second, the transmission-level changes are also indicative. 
\\
\\

\section{Three topological defects}
Our most efficacious chip contains three topological defects with different strip widths $w_l$, $w_m$ and $w_r$ within the left-, middle- and right-defect region, respectively. This detects and distinguishes three different analytes (disease markers) individually and their various combinations through a single spectroscopic measurement. 

In Section III, effective dehybridization was achieved using the fact that spectral correlation decreases more rapidly than transmission levels. Effective dehybridization occurs when spectral correlation in frequency shifts is nearly undetectable but transmission levels remain acceptable. However, correlation detectability also depends on the sensitivity of individual domain-wall defects. We fix the spatial separation between each defect and the nearest surface to be $l/a=5.5$. By adjusting the strip width within the defect region, the quality factors and the limit-of-detection do not change significantly. However, changes in sensitivity can play a role in effective dehybridization. A larger strip width can reduce sensitivity to the level that spectral correlation is barely detectable. 

Altogether, there is a trade-off among three biosensing features: (1) effective dehybridization (spectral independence); (2) high transmission; and (3) high sensitivity. A given chip can typically achieve only two of the three goals. In the following, we describe three chips each with three domain-wall defects: The first chip offers independent spectral responses and high sensitivity, but low transmissions. The second chip provides spectral independence and higher transmissions, but low sensitivity. The third chip reveals high transmissions and high sensitivity, but highly correlated spectral responses.

With three domain-wall defects, there is no longer spatial reflection symmetry. Consequently, R-binding is always distinguishable from L-binding.
The Maxwell operator within the subspace of defect modes, written in the basis of individual defect modes ${|\text L\rangle, |\text M\rangle, |\text R\rangle}$ with resonant frequency $\omega_L, \omega_M, \omega_R$ respectively, takes the form:
\begin{equation}
    \hat{\Theta} \approx \begin{pmatrix} \omega^2_L & \kappa_{lm} & \kappa_{lr} \\ \kappa_{lm} & \omega^2_M & \kappa_{rm} \\ \kappa_{lr} & \kappa_{rm} & \omega^2_R \end{pmatrix}.
\end{equation}
Here, $\kappa_{lm}$, $\kappa_{rm}$, $\kappa_{lr}$ are the direct coupling strengths between the left ($l$), middle ($m$), and right ($r$) defect modes, as indicated by the subscripts. Now the left- and right-defect are further away from the chip center. Moreover, the direct coupling between the left- and right-defect mode is extremely weak, $\kappa_{lr}\ll 1$, due to their large separation. The strip widths within the left- and right-defect regions are chosen different such that the resonant frequencies of their modes are well-separated. This also reduces direct coupling between the left- and right-defect modes. The dominant transmission trajectory is through the indirect coupling of the $|\text L\rangle$ mode and the $|\text R\rangle$ mode via the intermediate $|\text M\rangle$ mode. This second-order coupling is usually much smaller than the first-order direct couplings $\kappa_{lm}$ and $\kappa_{rm}$. Therefore the transmission-levels for the $|\text L\rangle'$ and $|\text R\rangle'$ resonances are low, and much harder to enhance than in the two-defect case that exhibited direct mode hybridization. 
 
For the three-defect chip, we consider hybridization between the left- and middle-defect modes, $\kappa_{lm}/\Delta\omega_{lm}^2$, and the hybridization between the middle- and right-defect modes, $\kappa_{rm}/\Delta\omega_{rm}^2$, where $\Delta \omega^2_{lm}=|\omega_M^2-\omega_L^2|$ and $\Delta \omega^2_{rm}=|\omega_R^2-\omega_M^2|$. To enhance spectral independence of the three resonances, we have to suppress both mode hybridizations, while to improve the transmission-levels of the left-like $|\text L\rangle'$ and right-like $|\text R\rangle'$ resonances, we have to do the opposite. 

\begin{widetext}
\begin{center}
\begin{figure}[htbp]
 \centering
  \subfigure[]{\includegraphics[width=0.95\textwidth]{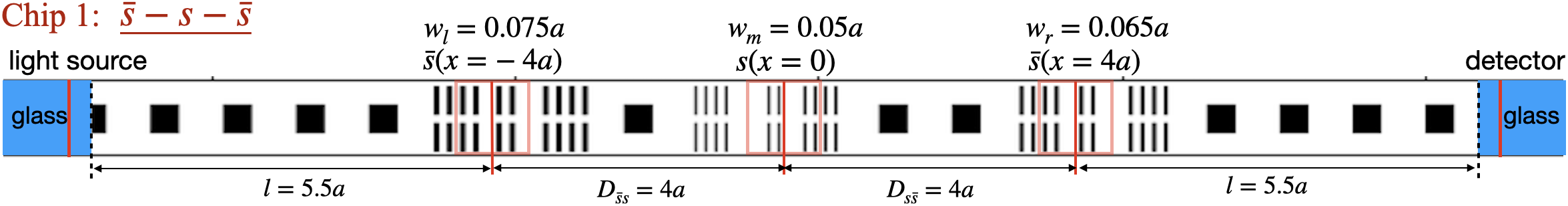}}
    \label{fig: 3 dehybridized modes chip}
  \subfigure[]{\includegraphics[width=0.43\textwidth]{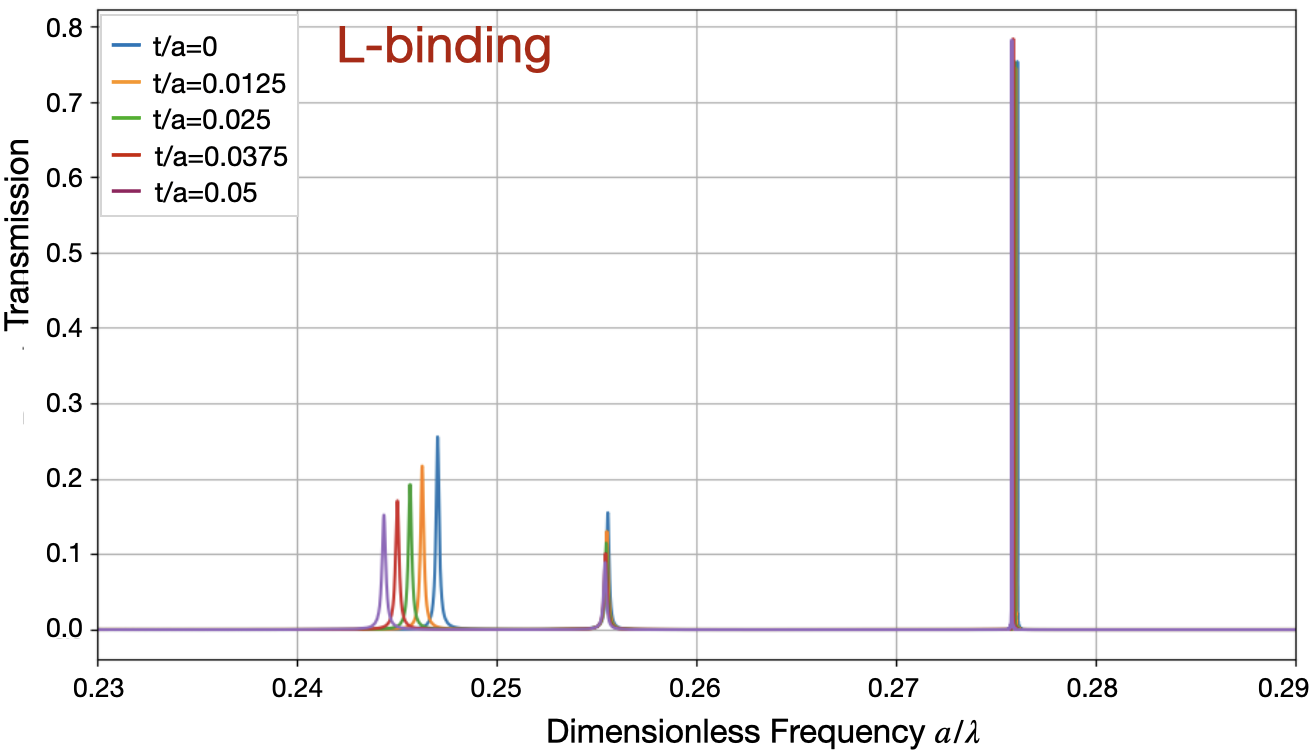}}
    \label{fig: 3 dehybrized L-bindings}
    \hspace{0.4cm}
     \subfigure[]{\includegraphics[width=0.43\textwidth]{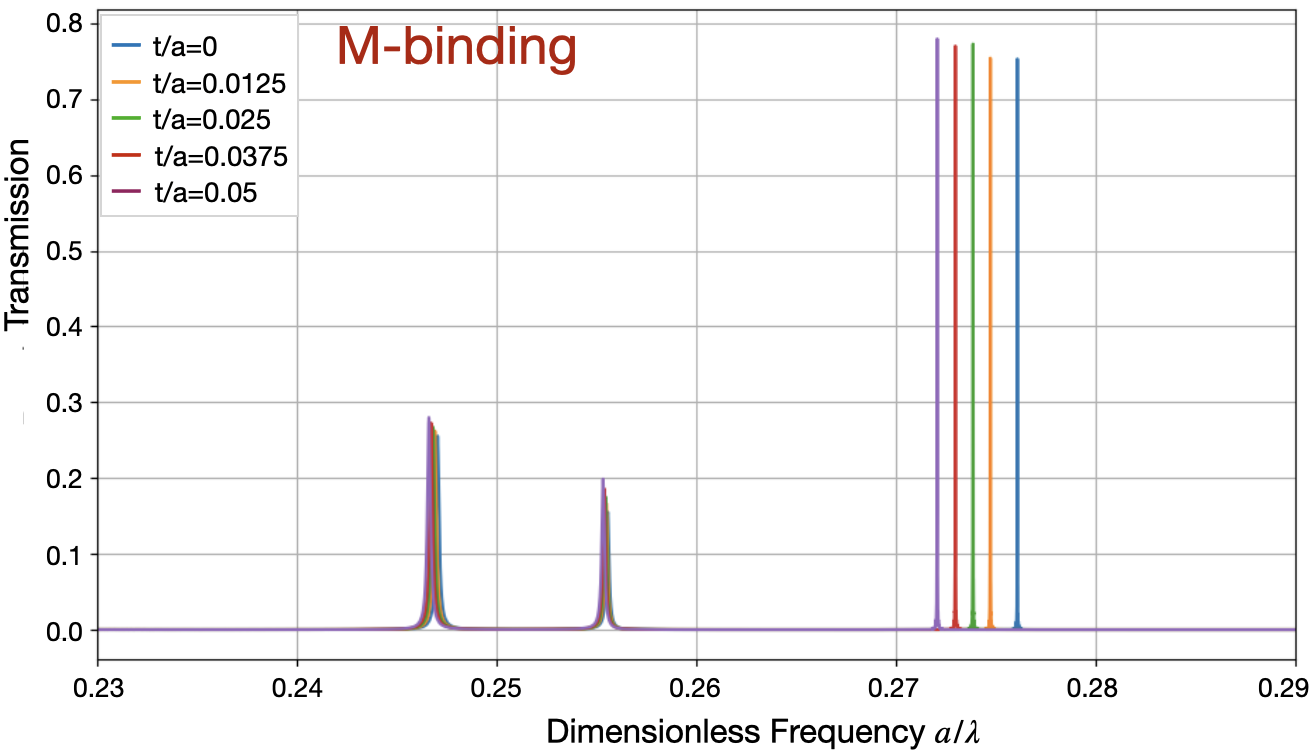}}
    \label{fig: 3 dehybrized M-bindings}
   \subfigure[]{ \includegraphics[width=0.43\textwidth]{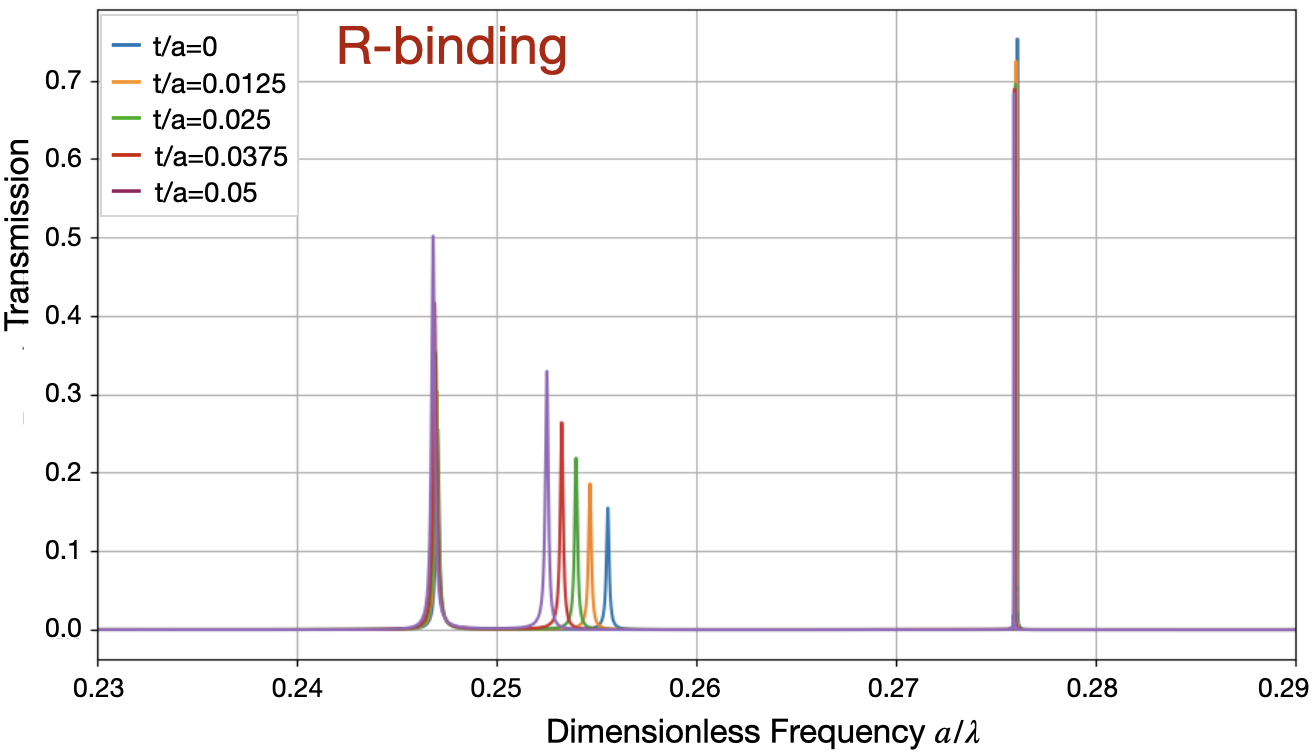}}
    \label{fig: 3 dehybrized R-bindings}
     \hspace{0.4cm}
  \subfigure[]{\includegraphics[width=0.43\textwidth]{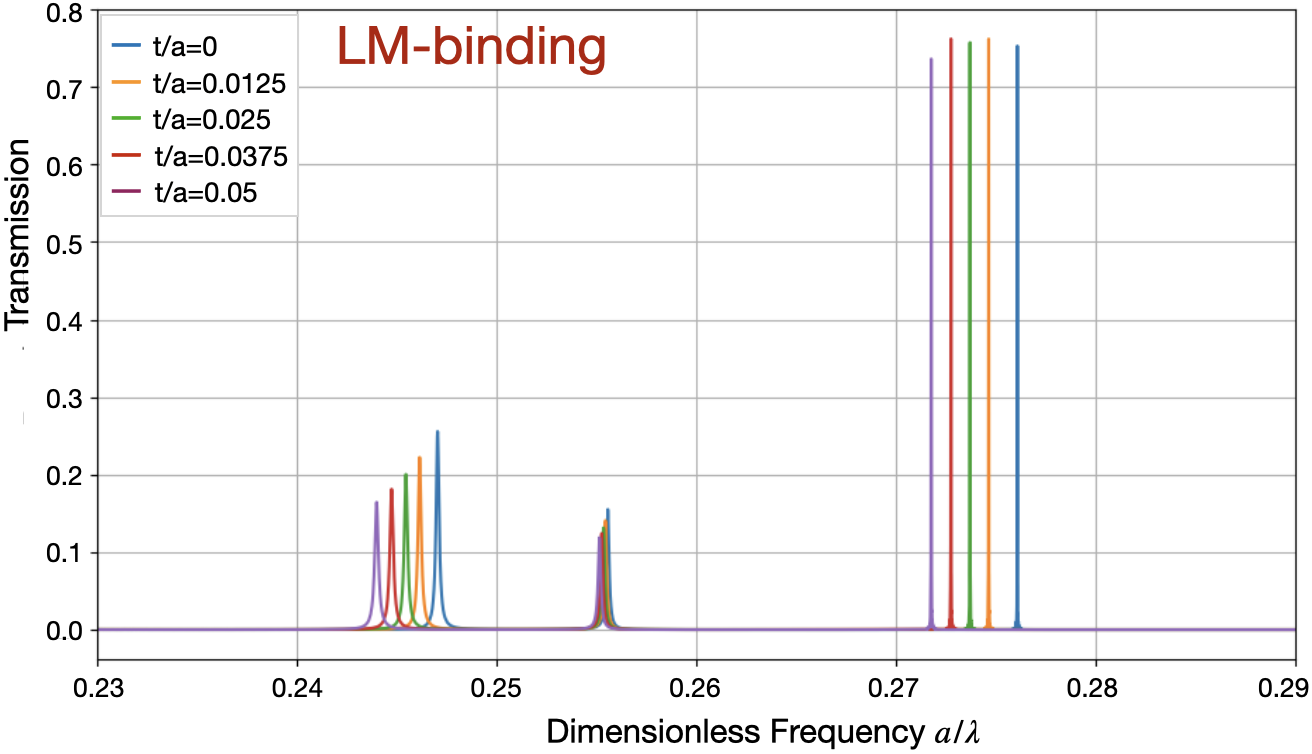}}
    \label{fig: 3 dehybrized LM-bindings}
     \subfigure[]{\includegraphics[width=0.43\textwidth]{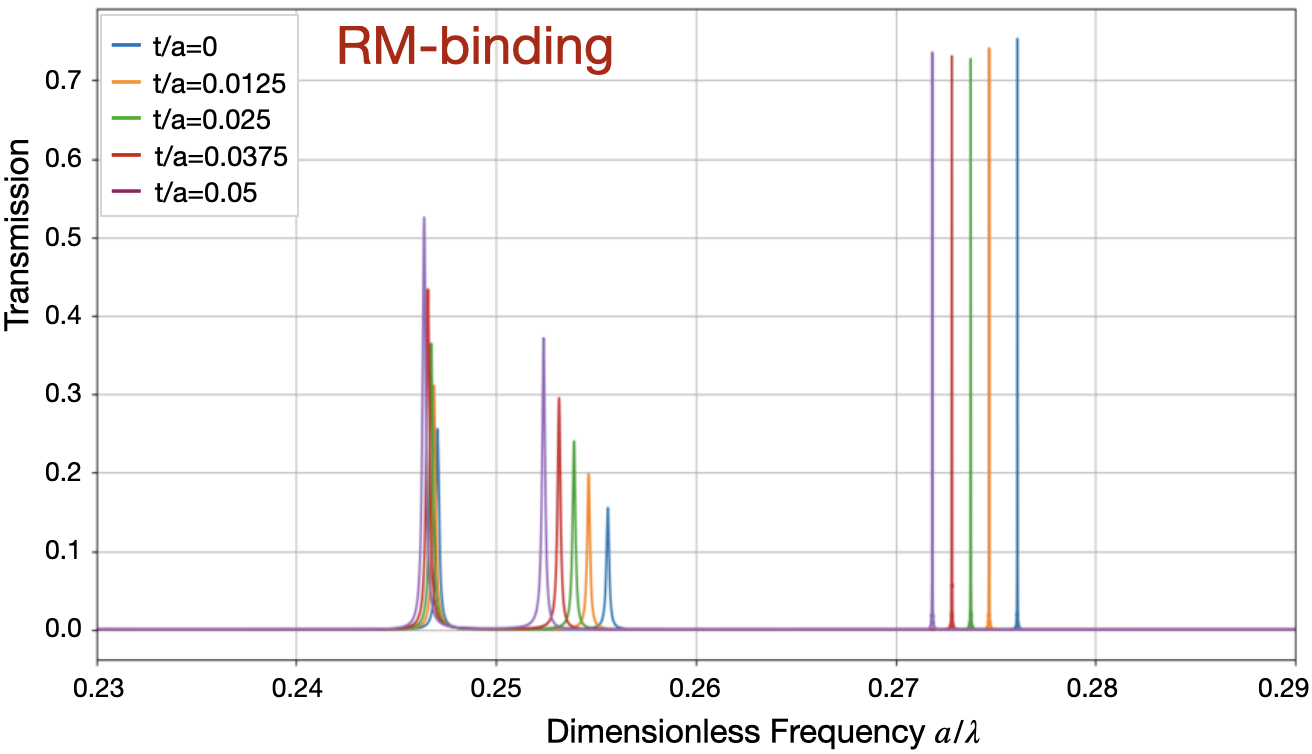}}
    \label{fig: 3 dehybrized RM-bindings}
    \hspace{0.4cm}
   \subfigure[]{ \includegraphics[width=0.43\textwidth]{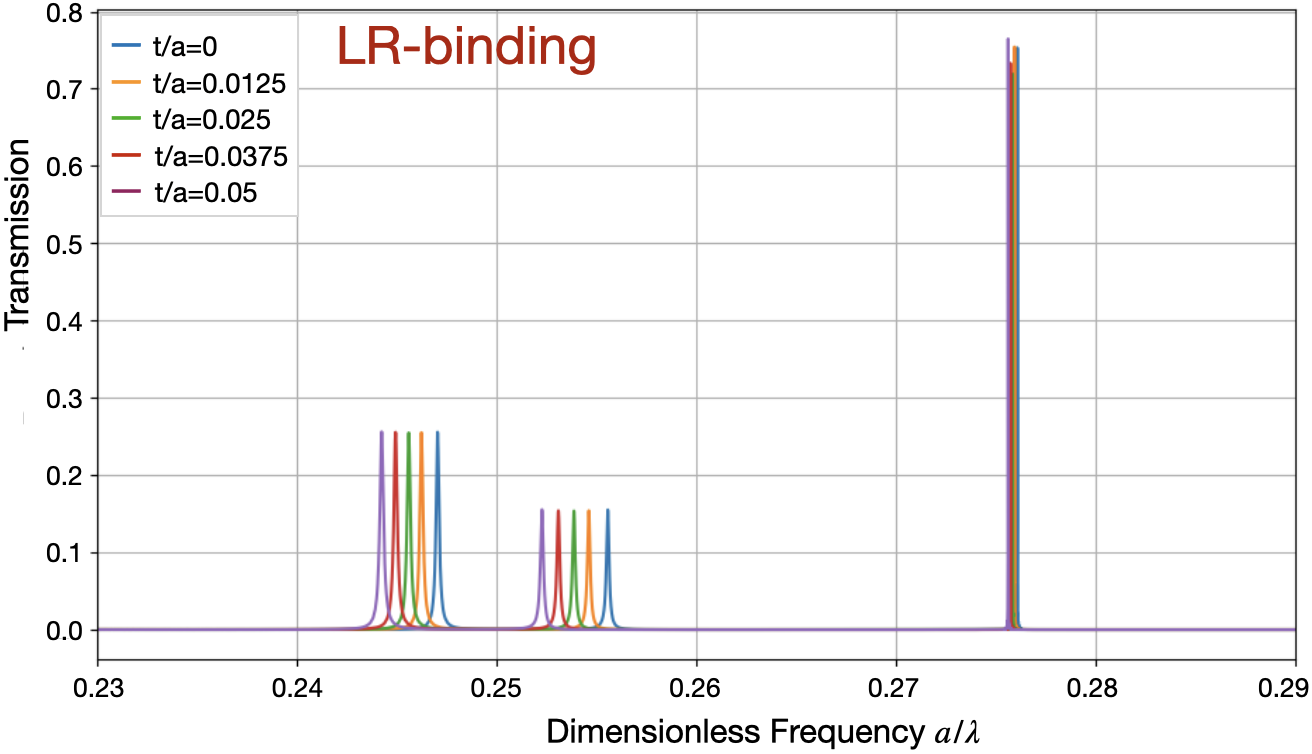}}
    \label{fig: 3 dehybrized LR-bindings}
   \subfigure[]{\includegraphics[width=0.43\textwidth]{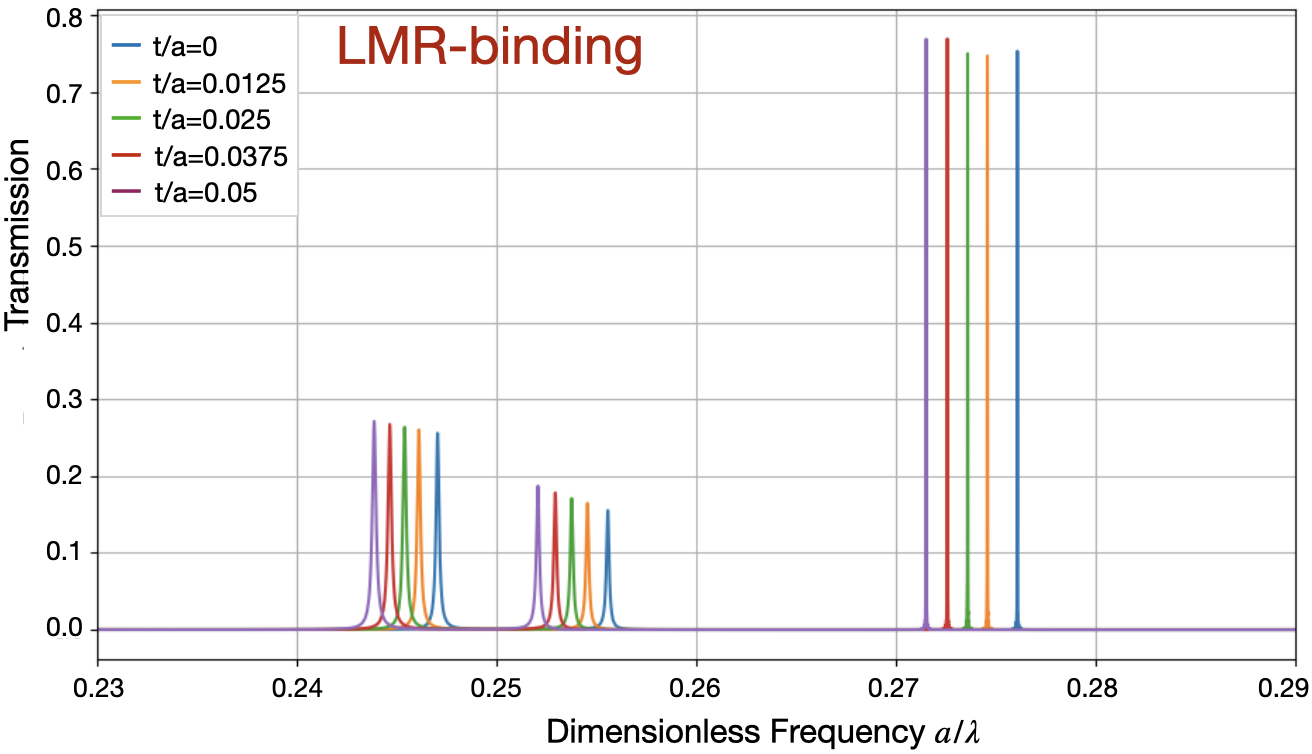}}
    \label{fig: 3 dehybridized LMR-bindings}
    \subfigure[]{\includegraphics[width=0.5\textwidth]{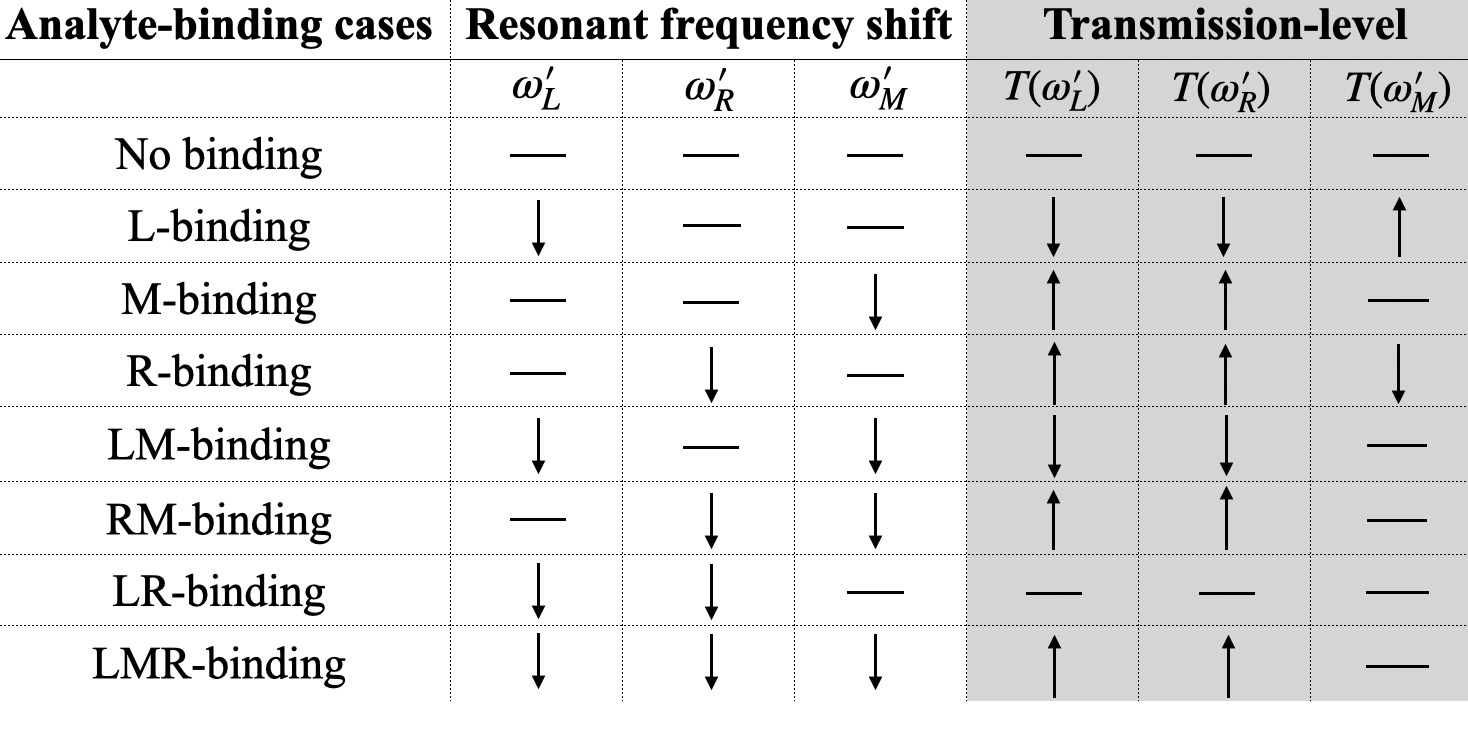}}
    \label{fig: 3-dehybridized Truth Table}
 \caption{Three-defect chip ($s-\bar{s}-\bar{s}$) with three effectively dehybridized resonances. (a) Strip widths within the left anti-domain-wall, middle domain-wall and right anti-domain-wall region are $w_l=0.075a$, $w_m=0.05a$ and $w_r=0.065a$ respectively, leading to well-separated resonance frequencies $\omega'_L=0.247076$ $[2\pi c/a]$, $\omega'_R=0.255595$ $[2\pi c/a]$ and $\omega'_M=0.276090$ $[2\pi c/a]$. Transmission spectra reveal spectral independence for different analyte-binding cases. (b) L-binding. (c) M-binding. (d) R-binding. (e) LM-binding. (f) RM-binding. (g) LR-binding. (h) LMR-binding. (i) Truth table for all the different analyte-binding cases. They are completely distinguishable with respect to frequency shifts alone. This chip achieves effective dehybridization and high sensitivity, but the transmission-levels of $|\text L\rangle'$ and $|\text R\rangle'$ resonances ($0.16\sim 0.26$) are relatively modest.}
\label{fig: 3 dehybridized modes bindings}
\end{figure}
\end{center}
\end{widetext}

\subsection{Chip 1: Spectrally independent resonances with high sensitivity but reduced transmission}
With three defects, it is problematic to achieve spectral independence with high enough transmission by only adjusting mode hybridizations. For illustration, a three-defect chip that achieves spectral independence using large frequency differences between modes is given in Appendix A. This has extremely low transmission levels.
However, by simultaneously increasing mode hybridizations and slightly lowering the sensitivity, we can achieve acceptable transmission with three effectively dehybridized resonances.
 For this purpose, we change the strip width within the middle-defect region to be $w_m=0.05a$, within the left-defect region to be $w_l=0.075a$ and within the right-defect region to be $w_r=0.065a$. In this case, $\Delta \omega_{lm}=0.02469$ $[2\pi c/a]$ and $\Delta \omega_{rm}=0.017257$ $[2\pi c/a]$, enabling adequate direct mode hybridization.

This chip with revised defect strip widths, depicted in Fig. \ref{fig: 3 dehybridized modes bindings}(a), supports three effectively dehybridized resonances with frequencies $\omega'_L=0.247076$ $[2\pi c/a]$, $\omega'_R=0.255595$ $[2\pi c/a]$ and $\omega'_M=0.276090$ $[2\pi c/a]$. The transmission-levels of $|\text L\rangle'$ and $|\text R\rangle'$ resonances, given by the simulation results, are $0.16 \sim 0.26$. The correlation in frequency shifts increases more slowly than the rise of the transmission-levels with mode hybridizations. Combined with slightly decreased sensitivity, the overall spectral correlation of these three resonances is almost undetectable.

We functionalize this chip for biosensing and depict transmission spectra for all the eight binding cases, in Fig. \ref{fig: 3 dehybridized modes bindings}(b)-(h). The eight binding cases are completely distinct with respect to frequency shifts alone. The observed transmission-level changes that are still detectable due to the existence of weak mode hybridizations, are also listed in the Truth Table (Fig. \ref{fig: 3 dehybridized modes bindings}(i)).

With this choice of larger strip width, the sensitivity is lowered only a small amount. For example, with the strip width $w_{def}=0.05a$, instead of the original $w_{def}=0.03a$, the sensitivity is lowered to $0.004451$ $[2\pi c/a]$ frequency shift per analyte-thickness $t=0.5a$, only $13.7\%$ lower than the maximal sensitivity.

\begin{widetext}
\begin{center}
\begin{figure}[htbp]
 \centering
  \subfigure[]{\includegraphics[width=0.95\textwidth]{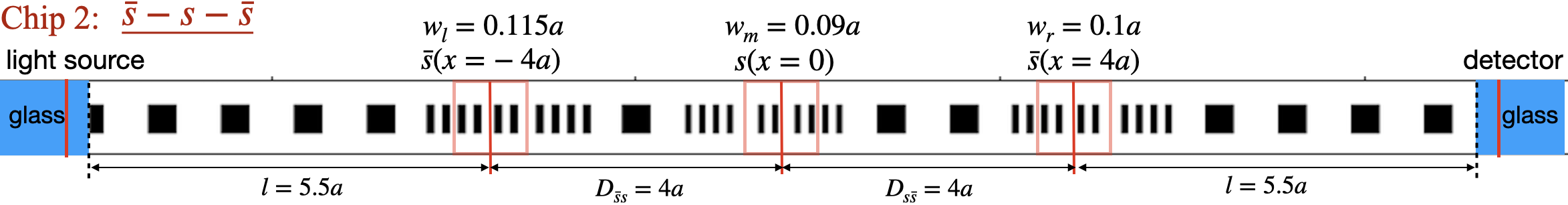}}
    \label{fig: 3 dehybridized low-sensitivity chip}
  \subfigure[]{\includegraphics[width=0.43\textwidth]{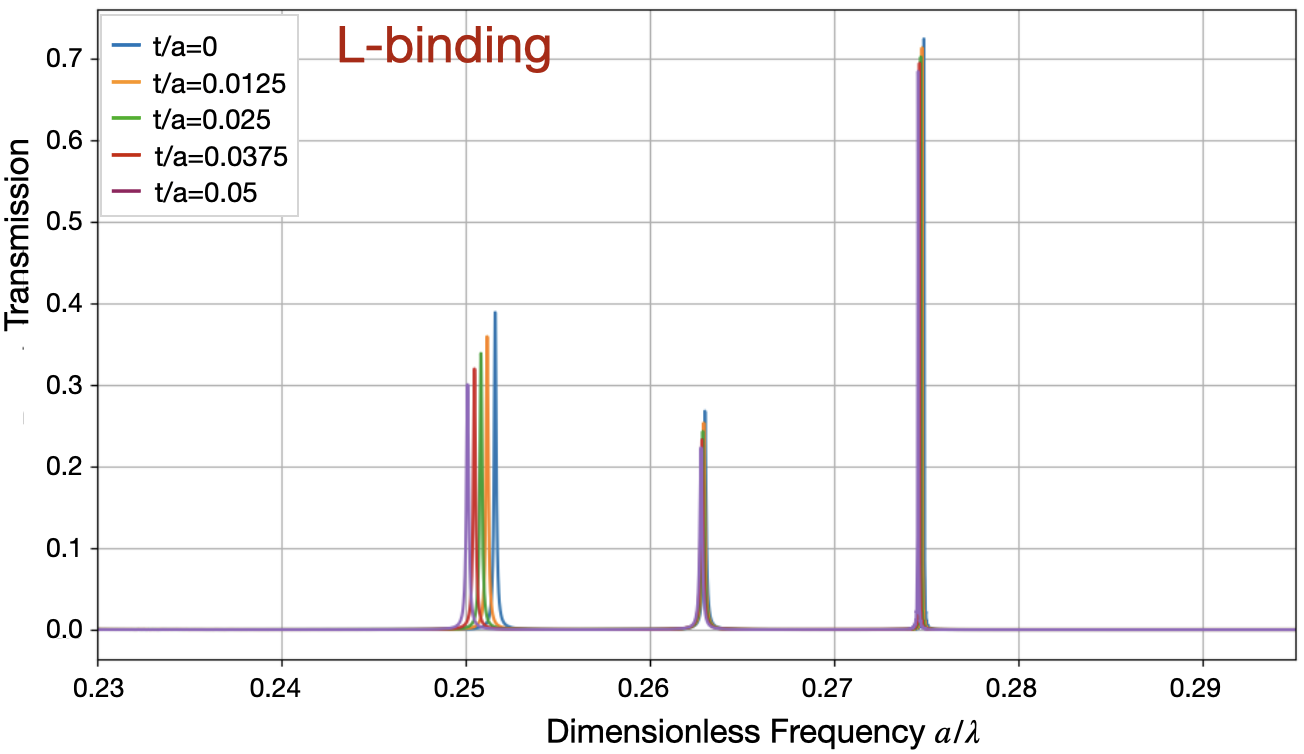}}
    \hspace{0.4cm}
     \subfigure[]{\includegraphics[width=0.43\textwidth]{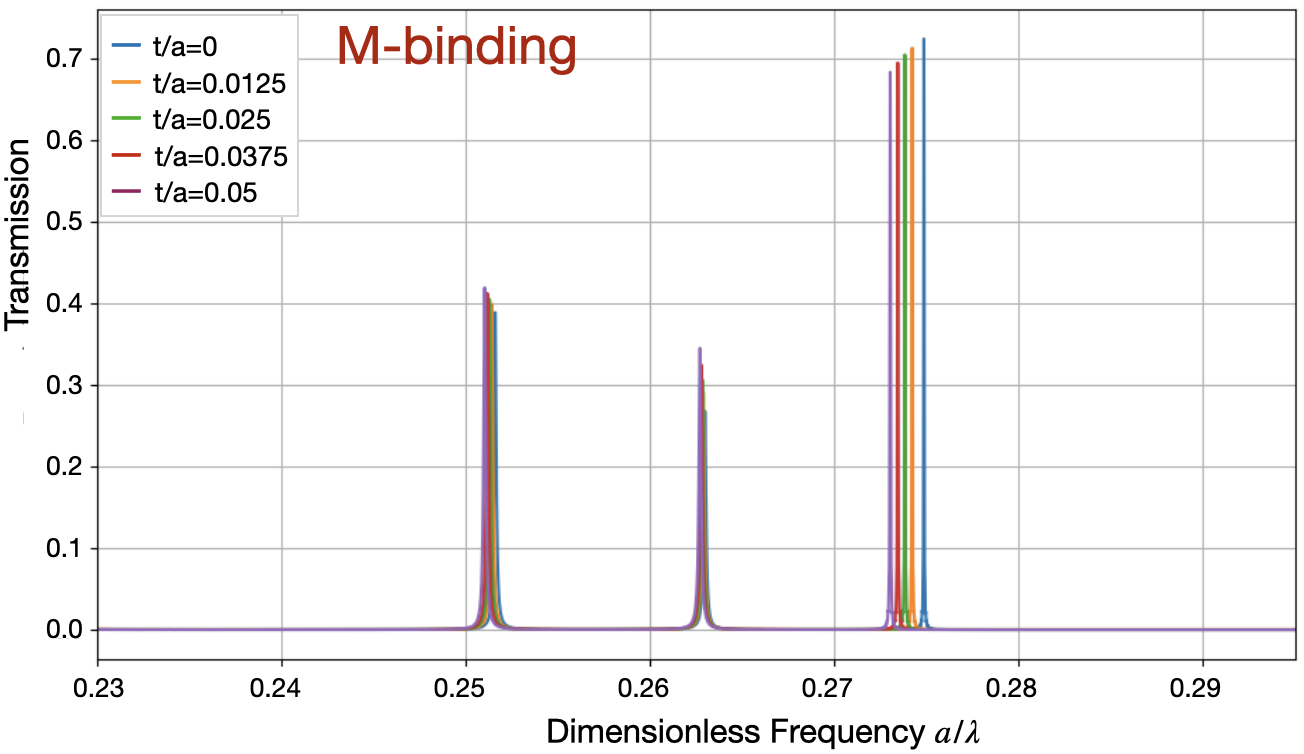}}
   \subfigure[]{ \includegraphics[width=0.43\textwidth]{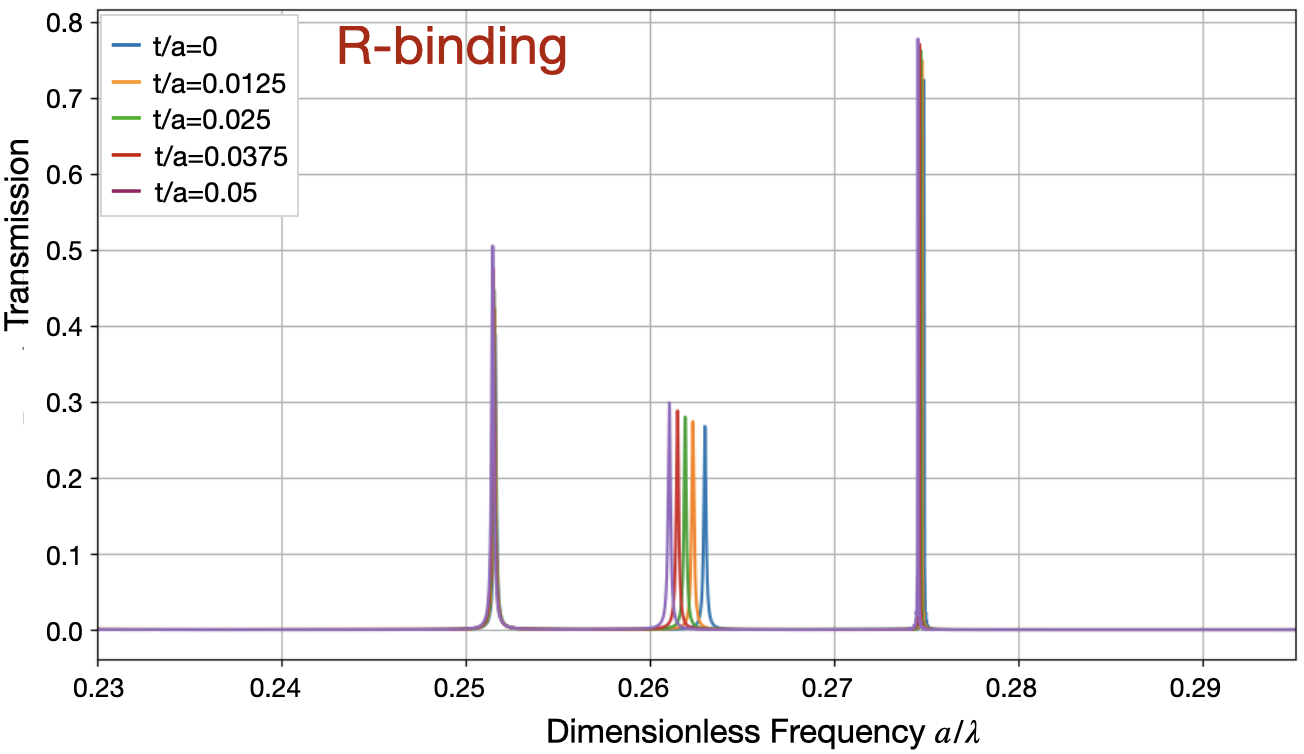}}
     \hspace{0.4cm}
  \subfigure[]{\includegraphics[width=0.43\textwidth]{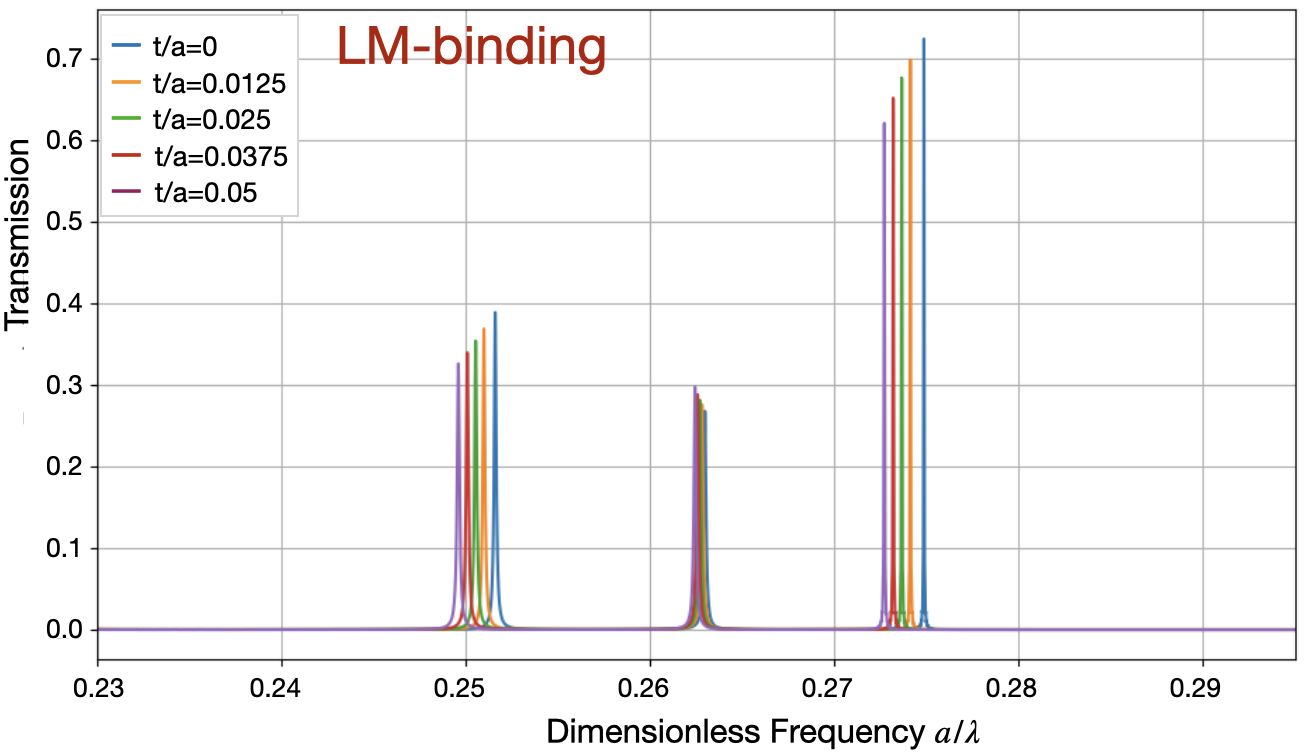}}
     \subfigure[]{\includegraphics[width=0.43\textwidth]{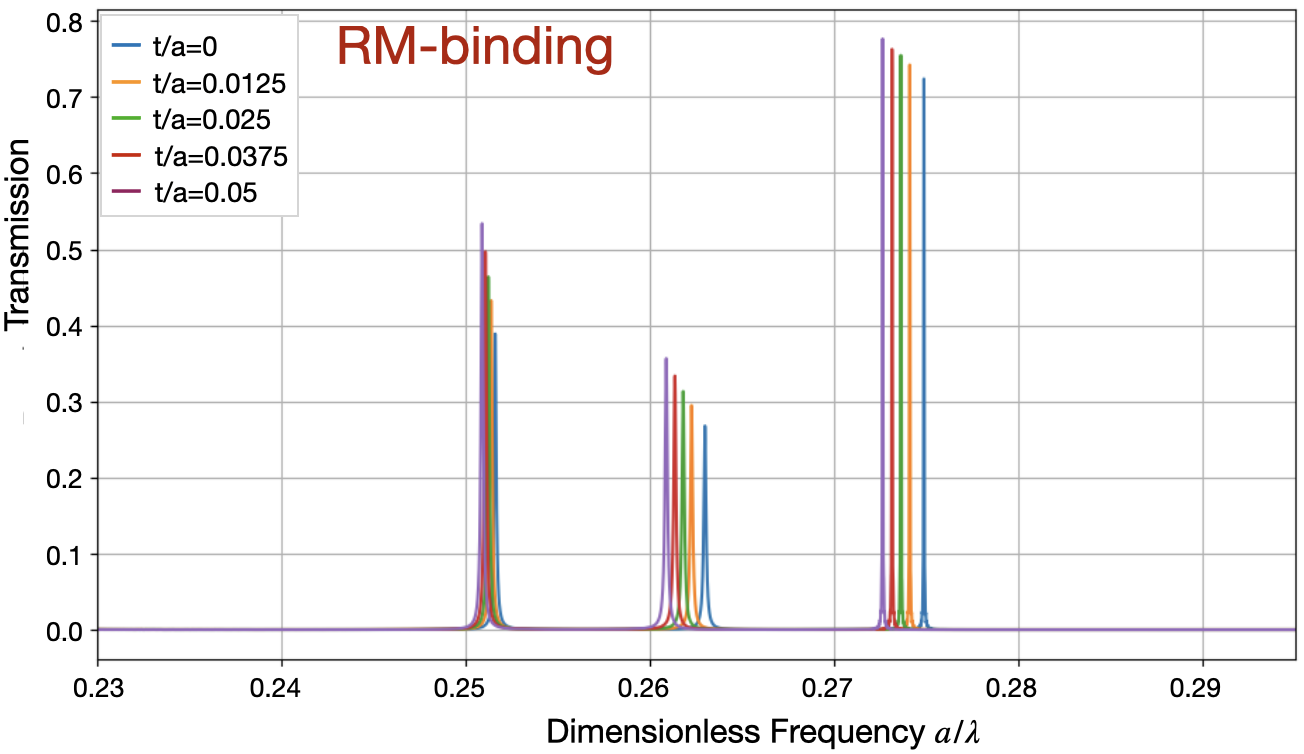}}
    \hspace{0.4cm}
   \subfigure[]{ \includegraphics[width=0.43\textwidth]{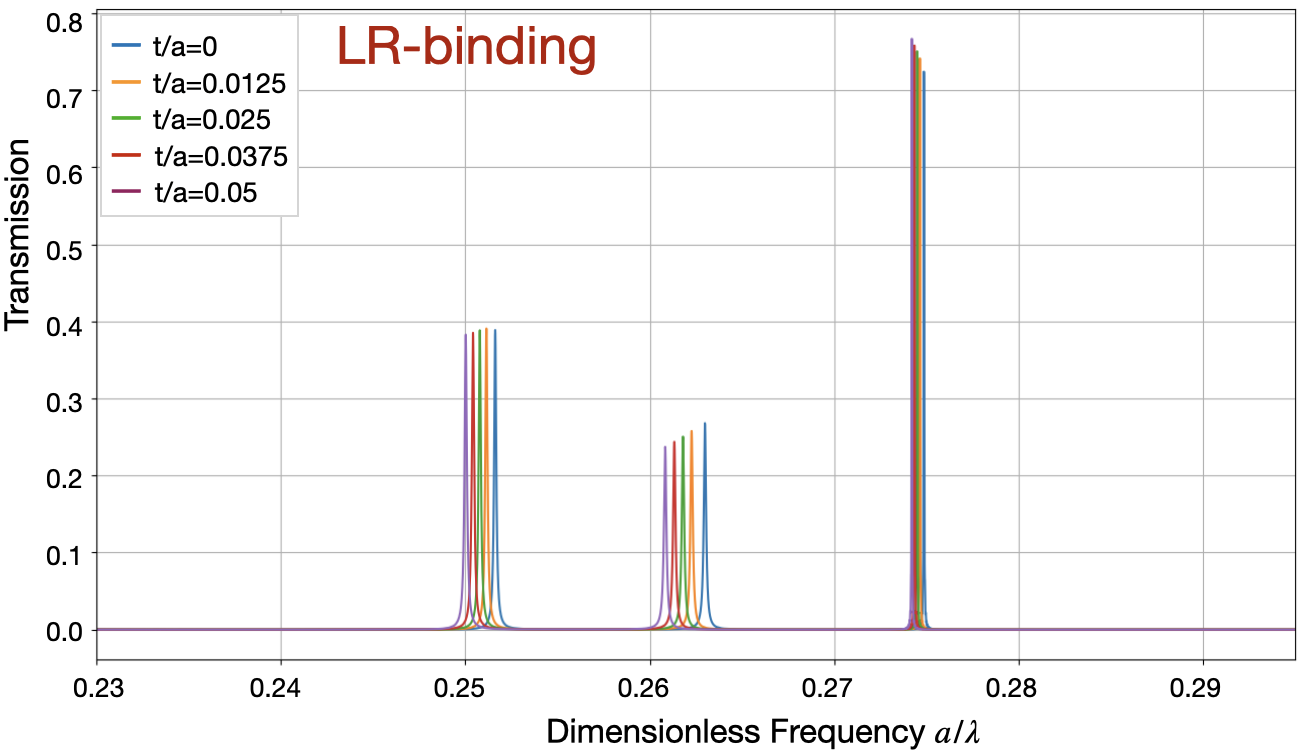}}
   \subfigure[]{\includegraphics[width=0.43\textwidth]{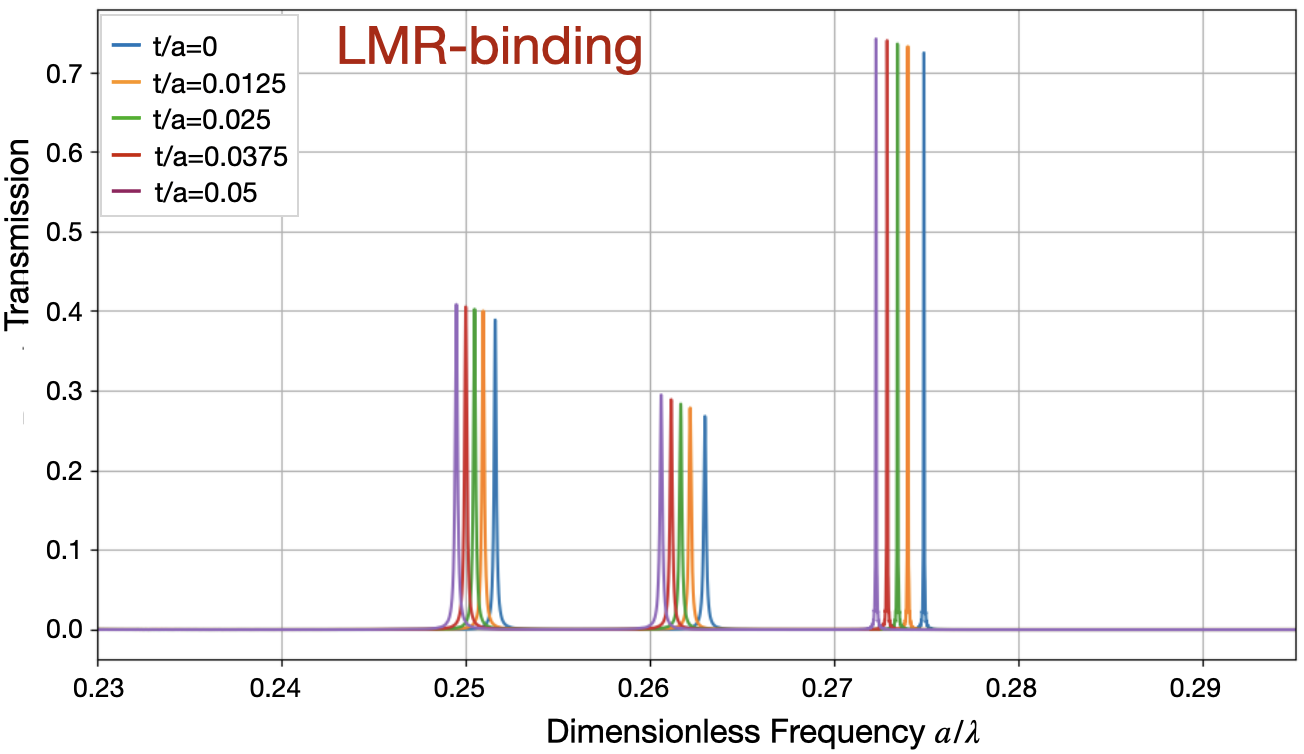}}
    \subfigure[]{\includegraphics[width=0.5\textwidth]{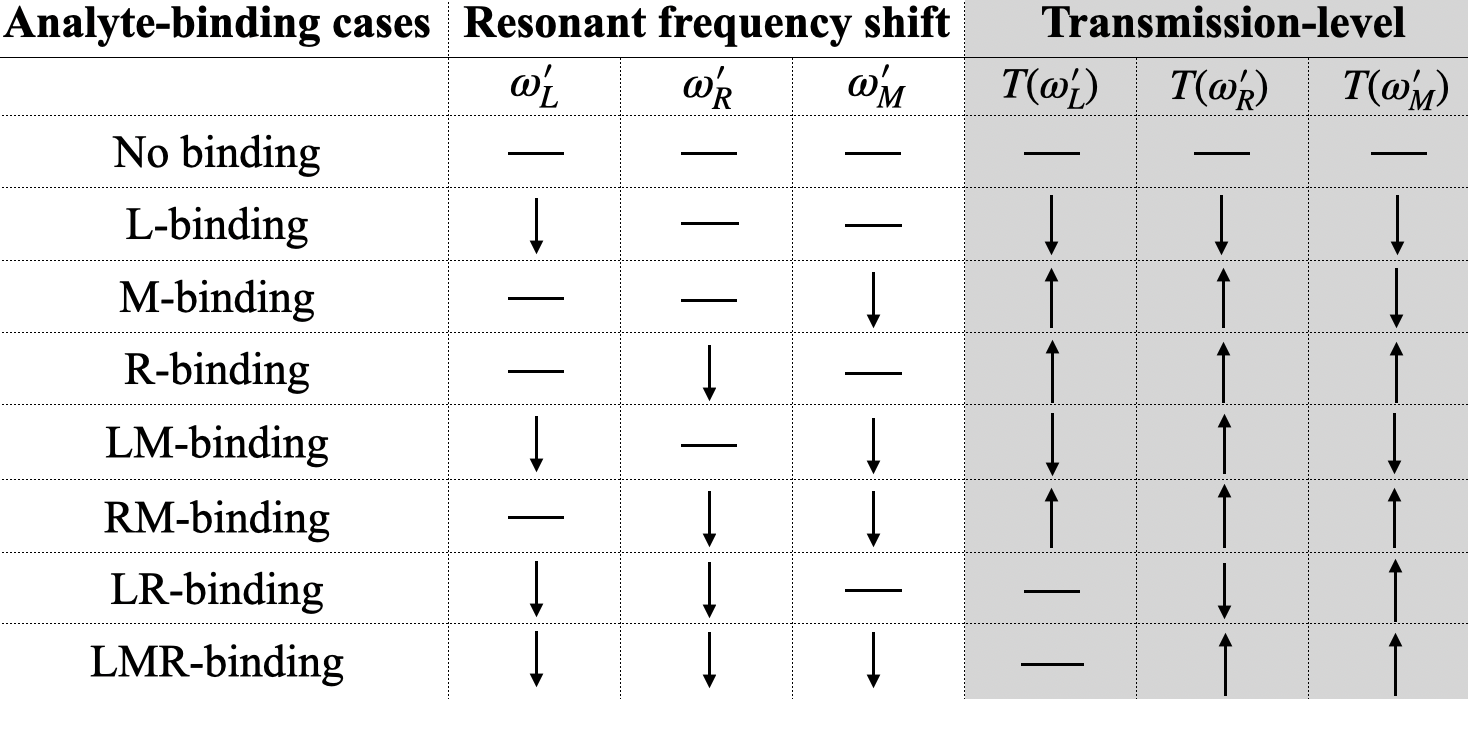}}
    \label{fig: 009-010-0115_Truth Table}
 \caption{Three-defect chip ($s-\bar{s}-\bar{s}$) with three effectively dehybridized resonances and higher transmission-levels. (a) Number of rectangular strips within each unit cell in the defect region is reduced to four, leading to lower the sensitivity. Strip widths within the left-, middle- and right-defect region are $w_l=0.115a$, $w_m=0.09a$ and $w_r=0.1a$, such that the frequency differences between defect modes are reduced to $\Delta \omega_{lm}=0.014097$ $[2\pi c/a]$ and $\Delta \omega_{rm}=0.00533$ $[2\pi c/a]$. The length of each strip is $h=0.4a$. The transmission-levels of $|\text L\rangle'$ and $|\text R\rangle'$ resonances are improved to nearly $0.3\sim 0.4$. Spectral correlations are increased, but minor due to lower sensitivity. Transmission spectra for different analyte-binding cases are shown. (b) L-binding. (c) M-binding. (d) R-binding. (e) LM-binding. (f) RM-binding. (g) LR-binding. (h) LMR-binding. (i) Truth table for all the different analyte-binding cases are completely distinguishable with respect to frequency-shifts alone.}
\label{fig: 3 dehybridized_low sensitivity}
\end{figure}
\end{center}
\end{widetext}

\subsection{Chip 2: Spectrally independent resonances  with high transmission but reduced sensitivity}

To improve the transmission levels, we must enhance mode hybridizations, resulting in higher spectral correlation. If we further lower the sensitivity, the correlation in frequency shifts is not easily discernible. Then, the resonances can still be regarded as effectively dehybridized. To further lower the sensitivity, we reduce the number of rectangular strips within each unit cell in the defect region from eight to four, and choose the widths of strips within the left-, middle- and right-defect region to be $w_l=0.115a, w_m=0.09a$ and $w_r=0.1a$ respectively, as shown in Fig. \ref{fig: 3 dehybridized_low sensitivity}(a). The length of each strip is still $0.4a$. In this case, $\Delta \omega_{lm}=0.014097$ $[2\pi c/a]$ and $\Delta \omega_{rm}=0.00533$ $[2\pi c/a]$, much smaller than in Chip 1. This increases direct mode hybridizations and transmission. The three resonances now have frequencies $\omega'_L=0.251623$ $[2\pi c/a]$, $\omega'_R=0.262994$ $[2\pi c/a]$ and $\omega'_M=0.274866$ $[2\pi c/a]$. Using a strip width $w=0.09a$ for the four rectangular strips within each unit cell, the sensitivity is reduced to $0.00252$ $[2\pi c/a]$ frequency shift per analyte-thickness $t=0.5a$. This corresponds to 1750 nm/RIU with lattice constant $a=5\mu m$, almost half the sensitivity of Chip 1. However, the three resonances respond almost independently to analyte-bindings, as seen in the transmission spectra of Fig. \ref{fig: 3 dehybridized_low sensitivity}(b)-(h). The eight binding cases are completely distinguishable with respect to frequency shifts alone, as shown by the Truth Table (Fig. \ref{fig: 3 dehybridized_low sensitivity}(i)). 

In Ref. 11, a 3D PC short-pillar architecture was introduced with shape-change substitutional defects, leading to dehybridized resonances with high transmissions. These resonances exhibited extremely low sensitivity (below 0.00074 $[2\pi c/a]$ per analyte-thickness $t=0.05a$). This was compensated by a low limit-of-detection, enabled by high quality factors ($\sim 10^4$). In our present chips, the quality factors are $\sim 10^3$. However, this can be enhanced simply by increasing the overall chip width.

\begin{widetext}
\begin{center}
\begin{figure}[htbp]
 \centering
  \subfigure[]{\includegraphics[width=0.95\textwidth]{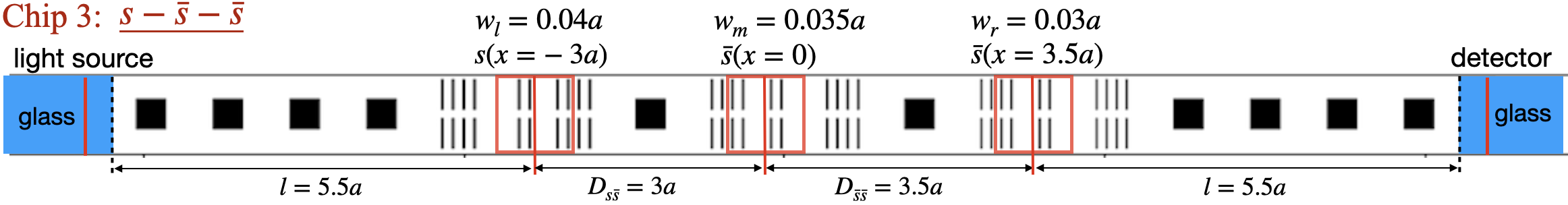}}
    \label{fig: 3 hybridized modes chip}
  \subfigure[]{\includegraphics[width=0.43\textwidth]{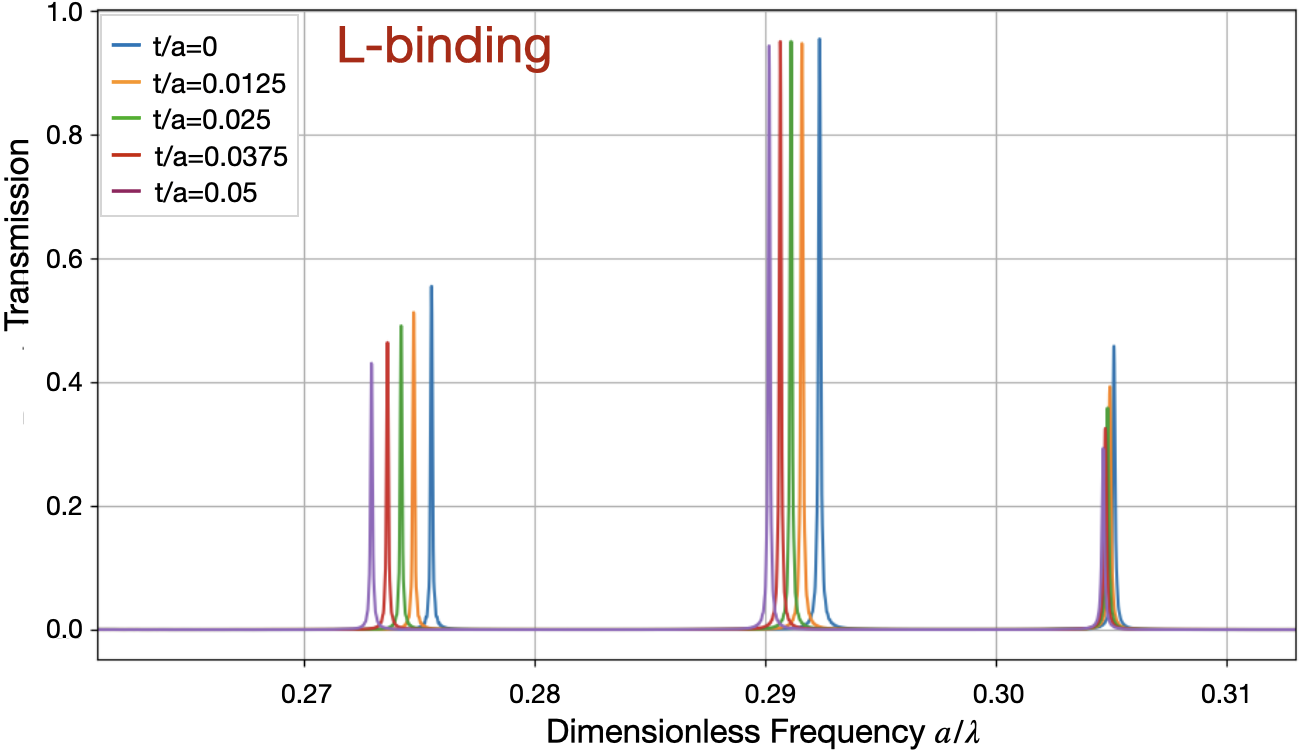}}
    \label{fig: 3 hybrized L-bindings}
    \hspace{0.4cm}
     \subfigure[]{\includegraphics[width=0.43\textwidth]{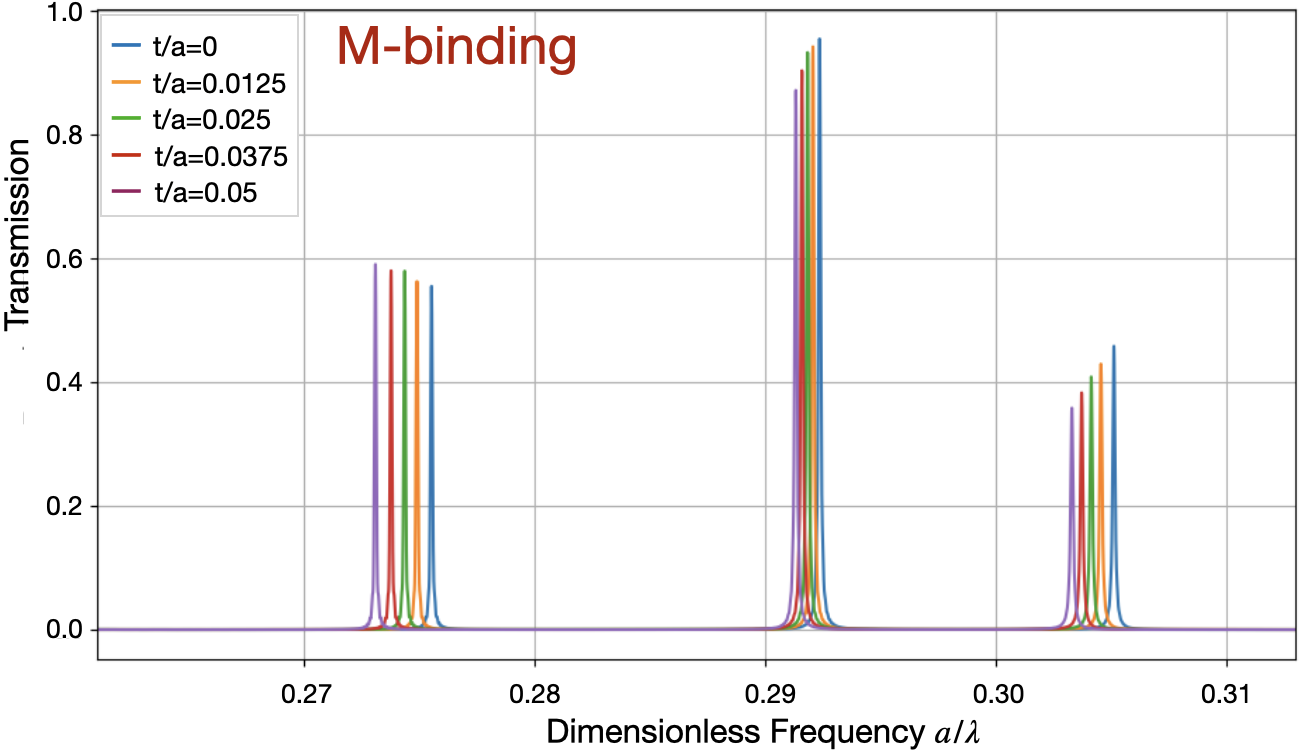}}
    \label{fig: 3 hybrized M-bindings}
   \subfigure[]{ \includegraphics[width=0.43\textwidth]{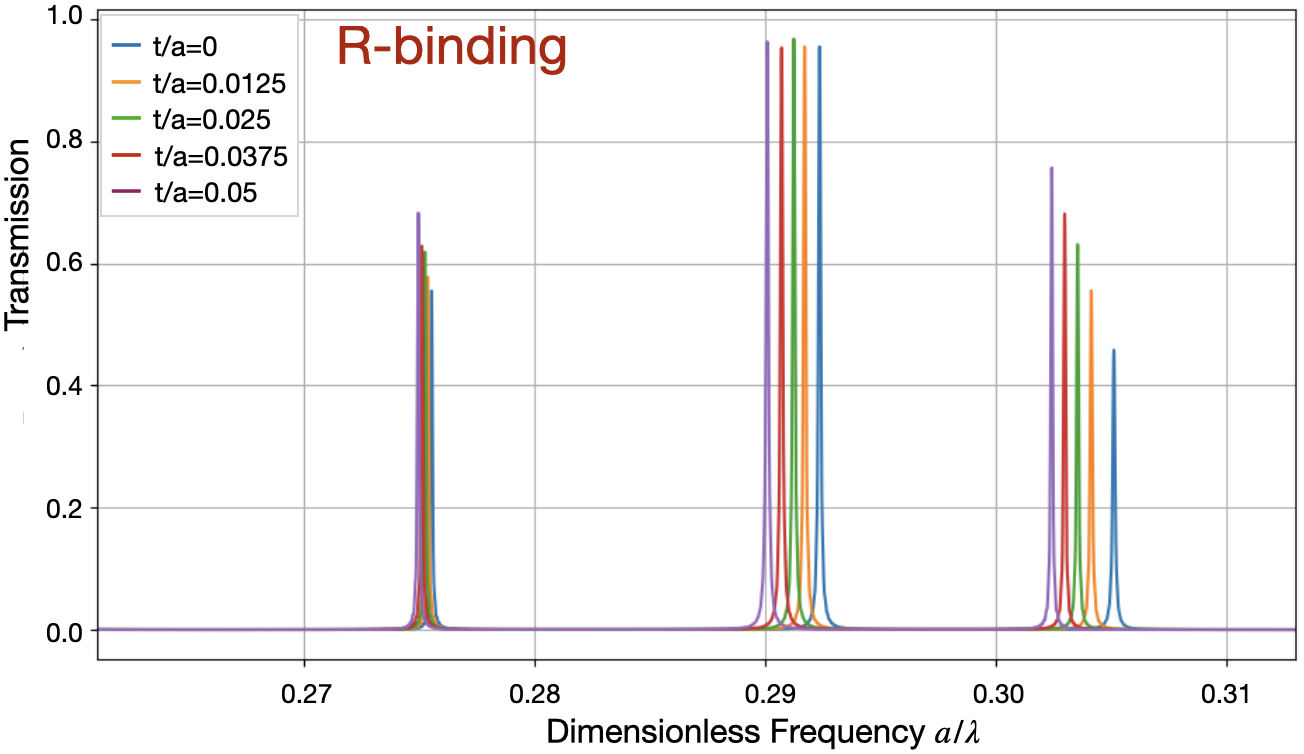}}
    \label{fig: 3 hybrized R-bindings}
     \hspace{0.4cm}
  \subfigure[]{\includegraphics[width=0.43\textwidth]{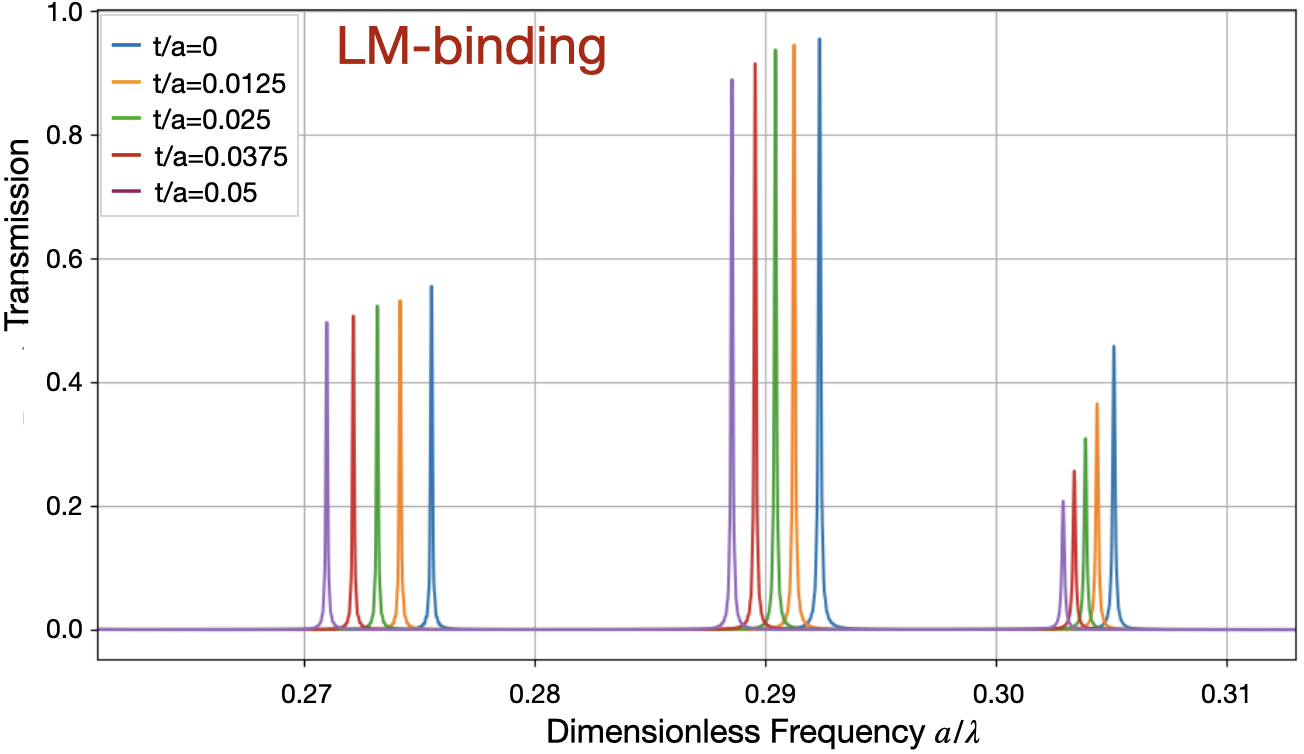}}
    \label{fig: 3 hybrized LM-bindings}
     \subfigure[]{\includegraphics[width=0.43\textwidth]{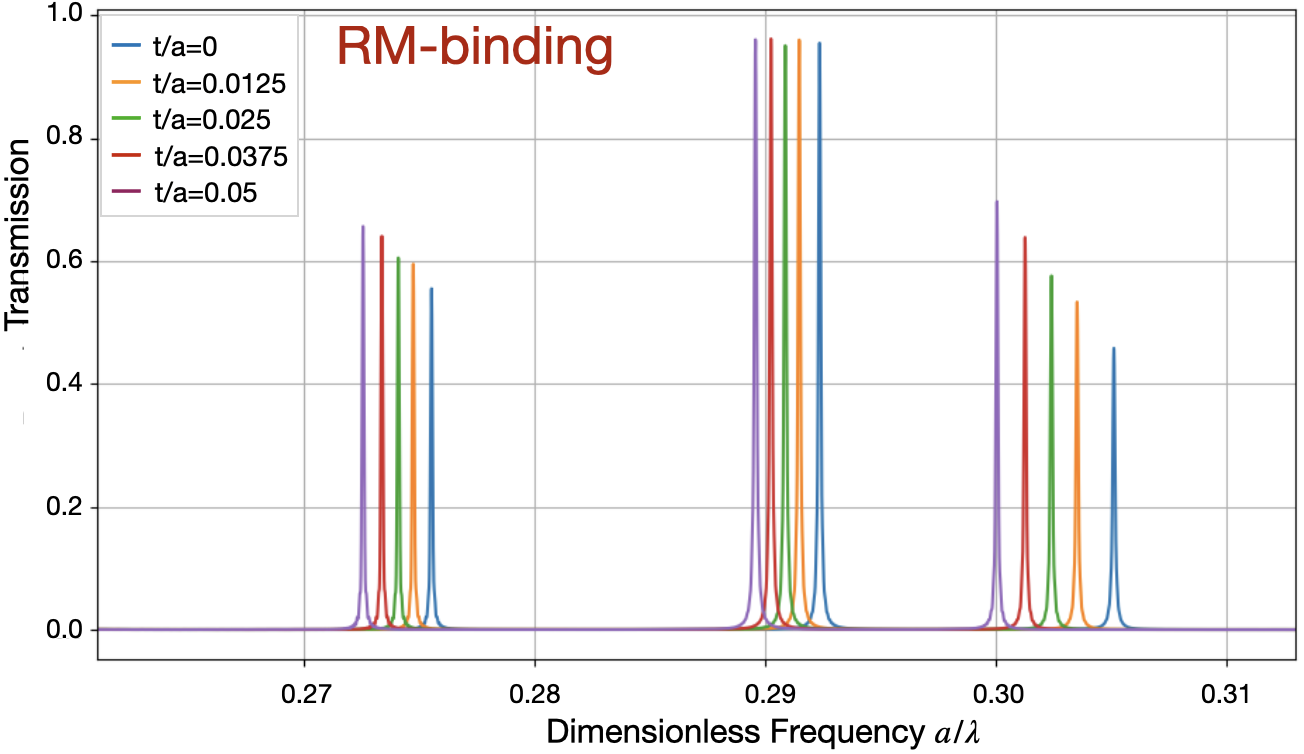}}
    \label{fig: 3 hybrized RM-bindings}
     \hspace{0.4cm}
   \subfigure[]{ \includegraphics[width=0.43\textwidth]{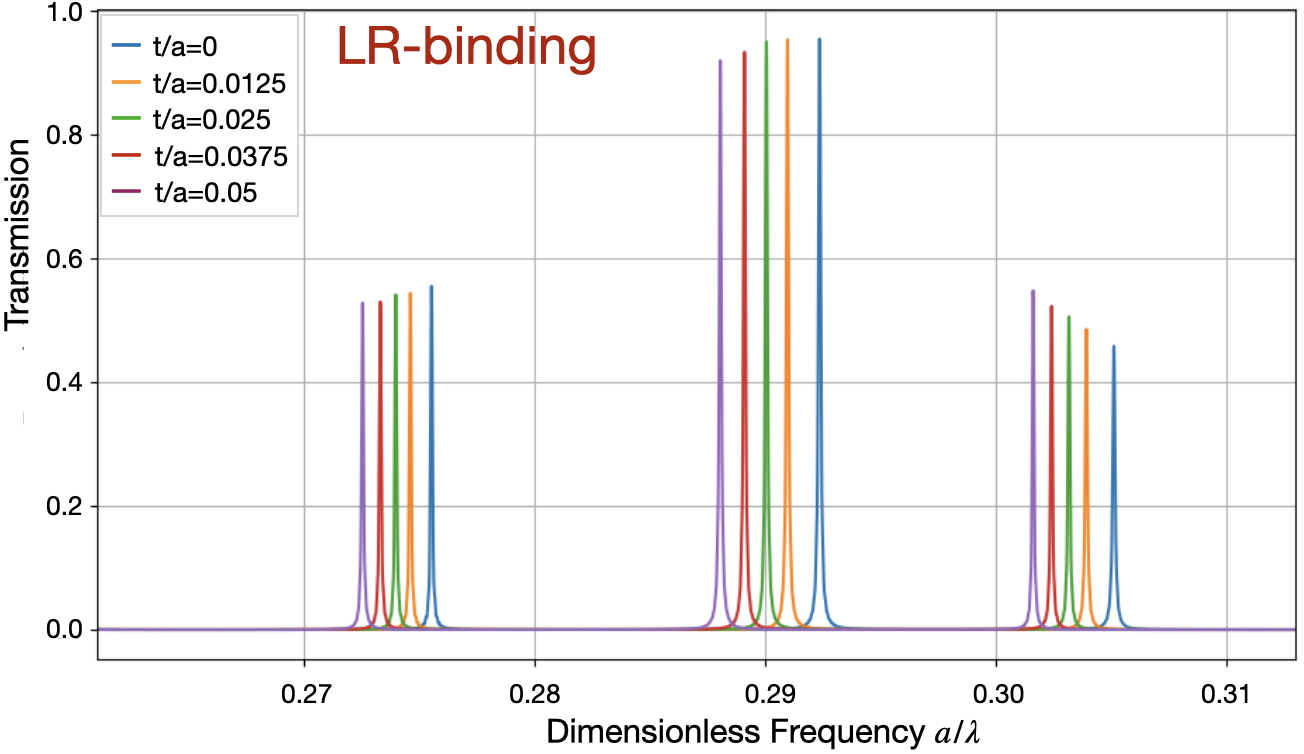}}
    \label{fig: 3 hybrized LR-bindings}
   \subfigure[]{\includegraphics[width=0.43\textwidth]{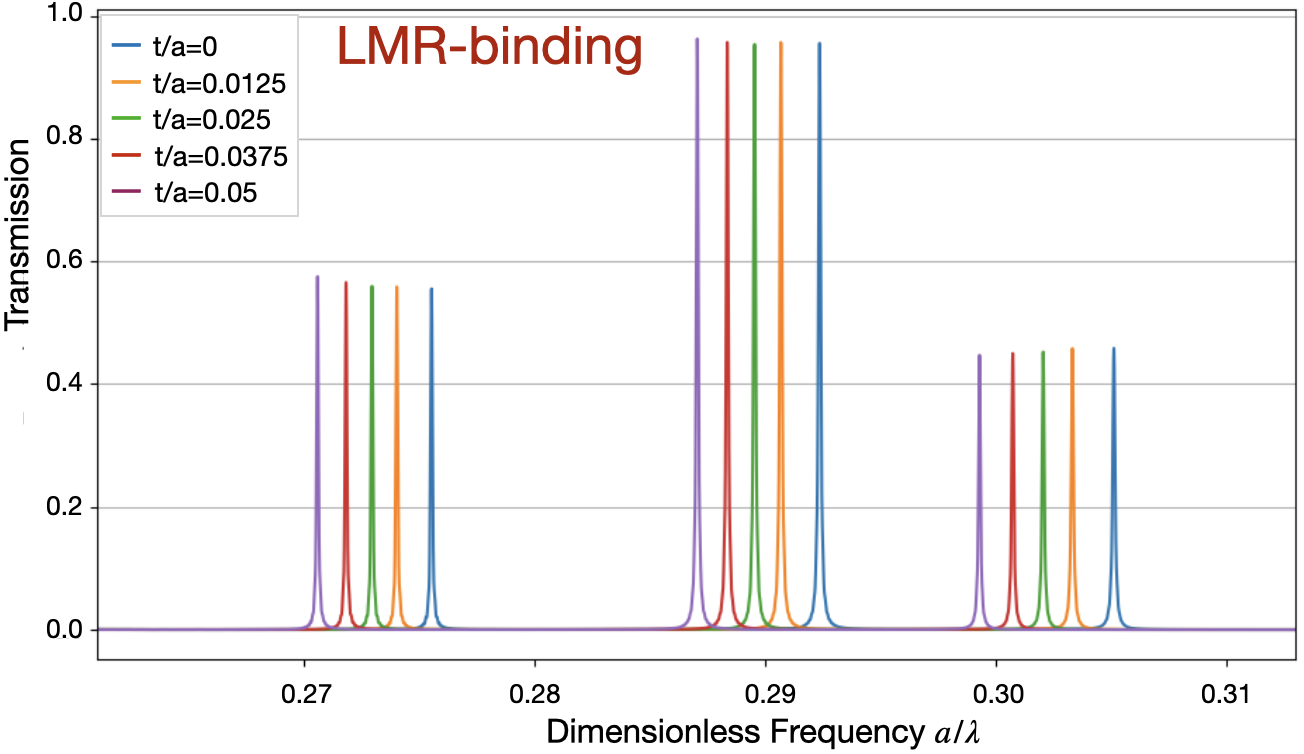}}
    \label{fig: 3 hybridized LMR-bindings}
     \hspace{0cm}
    \subfigure[]{\includegraphics[width=0.5\textwidth]{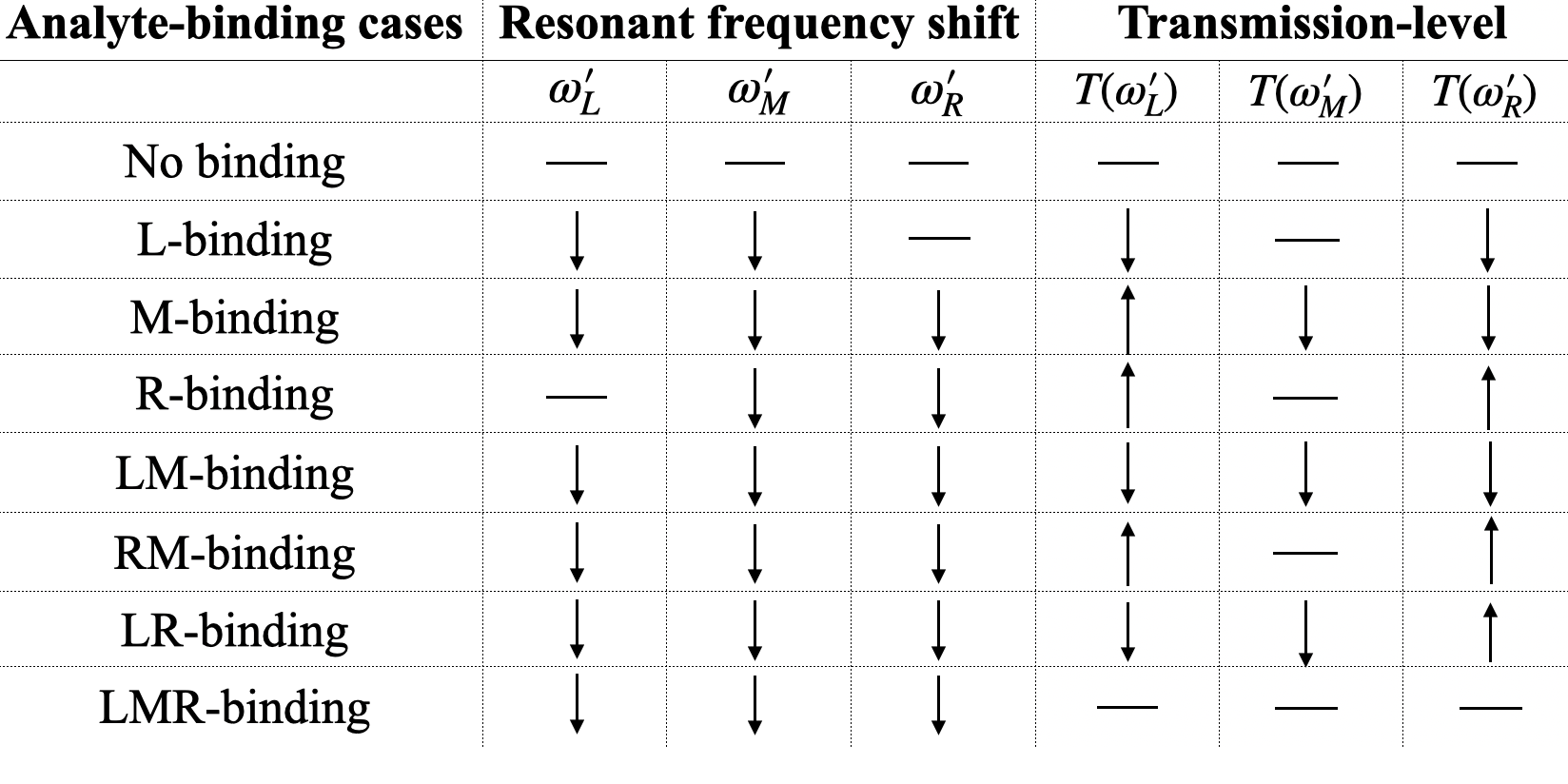}}
    \label{fig: 3 hybridized Truth Table}
 \caption{Three-defect chip ($s-\bar{s}-\bar{s}$) with three highly hybridized resonances. (a) Strip widths within the left domain-wall, the middle anti-domain-wall, and the right anti-domain-wall are $w_l=0.04a$,  $w_m=0.035a$ and $w_r=0.03a$, respectively. The frequency differences between their supported defect modes are $\Delta \omega_{lm}=0.001666$ $[2\pi c/a]$ and $\Delta \omega_{rm}=0.010383$ $[2\pi c/a]$, smaller than Chips 1 and 2. The spatial separations between defects are reduced to $\text D_{lm}=3a$ and $\text D_{rm}=3.5a$. Transmission spectra for different analyte-binding cases are shown. (b) L-binding. (c) M-binding. (d) R-binding. (e) LM-binding. (f) RM-binding. (g) LR-binding. (h) LMR-binding. With higher mode hybridizations, the transmission levels are improved but the spectral responses of the three resonances are more correlated. (i) Truth table for all the different analyte-binding cases. They are distinguishable, but require detailed analysis of both frequency shifts and transmission-level changes.}
\label{fig: 3 hybridized modes bindings}
\end{figure}
\end{center}
\end{widetext}

\subsection{Chip 3: Spectrally correlated resonances with high sensitivity and high transmission}
Finally, we illustrate a chip with three defects that supports three highly hybridized resonances with very high sensitivity. To ensure maximal sensitivity, we fix the strip widths of these three defects to be $w_l=0.04a, w_m=0.035a$ and $w_r=0.03a$, with the resonant frequencies of their supported individual defect modes $\omega_L=0.284683$ $[2\pi c/a]$, $\omega_M=0.286349$ $[2\pi c/a]$ and $\omega_R=0.296732$ $[2\pi c/a]$. The frequency differences of the hybridized modes are reduced to $\Delta \omega_{lm}=0.001666$ $[2\pi c/a]$ and $\Delta \omega_{rm}=0.010383$ $[2\pi c/a]$, (smaller than Chips 1 and 2) to enhance direct mode hybridizations. We identify spatial separations and an arrangement of domain-walls or anti-domain-walls to maximize transmission levels. A smaller spatial separation between any two defects enhances mode hybridization, enabling higher transmissions. The different defect regions are separated by one square dielectric block.

Once the strip widths of the left- and right-defect and their spatial separation are fixed, the middle defect is then chosen to mediate enhanced transmission. We place an anti-domain-wall in the middle with overall chip structure ($s$-$\bar{s}$-$\bar{s}$), as shown in Fig. \ref{fig: 3 hybridized modes bindings}(a). Our simulation reveals higher transmission for this than using ($s$-$s$-$\bar{s}$).

This $s$-$\bar{s}$-$\bar{s}$ chip supports three highly hybridized resonances with frequencies $\omega'_L=0.275535$ $[2\pi c/a]$, $\omega'_M=0.292367$ $[2\pi c/a]$ and $\omega'_R=0.305132$ $[2\pi c/a]$ within the PBG. The frequency differences between these three resonances are reduced compared with Chips 1 and 2. Their mode profiles are shown in Fig. \ref{fig: 3-S-AS-AS_D=1_mode profiles}(b)-(d). Due to strong mode hybridization between $|\text M\rangle$ and $|\text L\rangle$ and between $|\text M\rangle$ and $|\text R\rangle$, the field strength of the $|\text M\rangle'$ resonance in the middle defect region is smaller than that in the left or right defect region. Nevertheless, we still use `M' to denote this resonant mode. On the other hand, the field strengths of the $|\text L\rangle'$ resonance or the $|\text R\rangle'$ resonance in the middle defect region are very strong.

 Chip 3 exhibits both high sensitivity and high transmission. We now test if it can unambiguously distinguish all the eight analyte-binding cases. Fig. \ref{fig: 3 hybridized modes bindings}(b)-(h) show the transmission spectra for all the different analyte-binding cases. The Truth Table (Fig. \ref{fig: 3 hybridized modes bindings}(i)) lists the main features of spectral responses to all eight binding cases. They are distinguishable, but require more detailed analysis of the entire spectral fingerprint. For example, with L-binding, the left-defect mode frequency $\omega_L$ decreases, and thus both the hybridization factors  $\kappa_{lm}/\Delta \omega^2_{lm}$ and $\kappa_{lr}/\Delta \omega^2_{lr}$ decrease. This also shifts the mode profile of the $|\text L\rangle'$ resonance further left of the chip center, leading to lower transmission. Likewise, the mode profile of $|\text R\rangle'$ shifts further to the right of the chip, lowering its transmission. Along with the red-shift of the $|\text L\rangle'$ resonance, there is a correlated red-shift of the $|\text M\rangle'$ resonance, while that of the $|\text R\rangle'$ resonance is very weak.  
 
 As a second example, consider M-binding. Here all three resonances red-shift noticeably, since both the left-defect mode $|\text L\rangle$ and the right-defect mode $|\text R\rangle$ are highly hybridized with the middle-defect mode $|\text M\rangle$. The red-shift of the $|\text M\rangle'$ resonance is smaller than that of $|\text L\rangle'$ and $|\text R\rangle'$. This is because the field strength of the $|\text M\rangle'$ resonance in the middle defect region is weaker than that of $|\text L\rangle'$ and $|\text R\rangle'$ in the middle defect region, as seen in Fig. \ref{fig: 3-S-AS-AS_D=1_mode profiles} in Appendix C. As the middle-defect mode frequency $\omega_M$ decreases, the hybridization factor  $\kappa_{lm}/\Delta \omega^2_{lm}$ increases while the hybridization factor $\kappa_{rm}/\Delta \omega^2_{rm}$ decreases. The $|\text L\rangle'$ mode profile shifts toward to the chip center and its transmission slightly rises. The $|\text R\rangle'$ mode profile shifts away from the chip center and its transmission declines. The decline in transmission of the $|\text M\rangle'$ resonance can be understood as follows. The mode $|\text M\rangle$ is located slightly left of the chip center. For M-binding, the hybridization factor $\kappa_{rm}/\Delta \omega^2_{rm}$ is decreased and the hybridization factor $\kappa_{lm}/\Delta \omega^2_{lm}$ is increased. This causes the mode profile of $|\text M\rangle'$ resonance to move even further to the left of the chip center, reducing its transmission. Spectral responses of all the other binding cases can be explained in a similar way.
 
 A drawback of the strongly hybridized Chip 3 is that spectral fingerprints, such as LR-binding, are not easily distinguishable without detailed analysis. With LR-binding, all three resonances red-shift. However, the decline in transmission of the $|\text L\rangle'$ resonance and the rise in transmission of the $|\text R\rangle'$ resonance are very small. The transmission-level changes, while detectable, can be masked by noise in the biofluid. This implies a lack of robustness of the highly hybridized Chip 3. For the effectively dehybridized Chip 2, distinguishability is much more resilient to random fluctuations in the biofluid. In Appendix B, we present another three-defect chip ($\bar{s}$-$\bar{s}$-$\bar{s}$) with adjusted defect strip widths ($w_l=0.04a, w_m=0.05a, w_r=0.06a$) with correlated spectral peaks, but with more distinctive transmission-level changes. The spatial separation $\text D_{lr}=7a$ between the left- and right-defect is slightly larger and mode hybridizations are smaller. In this case, transmission levels are reduced but still detectable.
\\

\section{Conclusion and Outlook}
We have presented a conceptual 2D design of PBG-based optical biosensors with light trapped by topological domain-wall defects. A very high sensitivity ($0.0052$ $[2\pi c/a]$ frequency shift per analyte-thicknesss $t=0.05a$, or nearly 3000 nm/RIU with size of unit cell $a=5\mu m$) is achieved by replacing the simple square dielectric blocks with fragmented dielectric rectangular strips in the defect region. We give a detailed analysis of how transmission levels and spectral correlation in responses to analyte-binding are closely related to optical mode hybridization. We find that the correlation in frequency shifts decreases more rapidly than transmission levels as mode hybridization is reduced. This fact makes it possible to achieve effective dehybridization (spectral independence) in the two-defect case. However, in the three-defect case, spectral independence achieved solely by dehybridization leads to extremely low transmission levels. Alternatively, by retaining mode hybridizations while lowering sensitivity, spectral correlation may be barely detectable. These  effectively dehybridized resonances exhibit relatively high transmission. So there are fundamental trade-offs among three biosensor features: (1) effective dehybridization (spectral independence), (2) high transmission, and (3) high sensitivity. Three distinct three-defect chips were presented that can distinguish all different analyte-binding cases through a single spectroscopic measurement. One with high sensitivity but reduced transmission and the second with high transmission but reduced sensitivity, support three effectively dehybridized resonances that have independent spectral responses to analyte-binding. These chips are preferred because different analyte-binding cases are completely distinguishable by frequency shifts alone.  The third chip supports three highly hybridized resonances with high sensitivity and high transmission, but their spectral responses to analyte-binding are strongly correlated. In this case, frequency-shifts alone are insufficient to distinguish all different analyte-bindings. Here, more detailed analysis of the entire spectral fingerprints is required. One issue is that the transmission-level changes for some analye-binding cases may be small and could be masked by noise in the biofluid.

Even higher sensitivity can be achieved with more elaborate designs or other types of defects. For example, increasing the number of rectangular strips within one unit cell from eight to twelve, with properly chosen strip width, such as $w_{def}=0.02a$, will improve the sensitivity further to 0.00758 $[2\pi c/a]$ frequency shift per analyte-thickness $t=0.05a$, which is $46\%$ higher than our optimal sensitivity here. However, such a chip may be more challenging to fabricate. The chips given here all involve topological defects that involve a disruption of lattice periodicity. These were combined with substitutional defects (involving a shape change within a unit cell) to enhance sensitivity and adjust mode frequencies. Purely substitutional defect with fragmented rectangular strips can also achieve a high sensitivity. However, such chips typically require period doubling along the direction of fluid flow to enable optical coupling by an external plane wave.\cite{Abdullah15, Shuai16, Abdullah19, Dragan21} Combining topological domain-walls with substitutional defects enables additional design flexibility, not possible with purely substitutional defects. This may be useful in the design of 3D chips with finite pillar height. Nevertheless, other issues will arise. For example, an improperly fragmented unit cell may radiate more light into the 3rd dimension, thereby lowering the quality factor.  Accordingly, further design considerations may be required to retain high sensitivity while not not losing light vertically. 

There are specific types of fabrication errors that may be consequential to device functionality. One type is surface roughness or random shape fluctuations of the silicon structures (depicted in our manuscript as black squares and rectangles). The PBG of our structures provides considerable robustness against such fabrication errors. In our 2D simulation, these simply cause some small shift in the frequencies of the optical transmission resonances. This is a matter of calibration of each fabricated device. Although the transmission resonances may vary slightly from one device to another, the shift in resonance frequencies due to analyte binding is the same from one device to another. In other words, the user should simply calibrate the analyte-induced spectral fingerprint with respect to the spectral fingerprint in the absence of analyte binding.   

In a real-world, 3D biosensor, the black squares and rectangles representing silicon must be vertically extruded to describe finite-height silicon pillars. A very consequential fabrication error is the possibility of shape fluctuations of the pillars in the vertical direction.  This can lead to scattering of light (loss) into the vertical direction and consequent reduction of the in-plane transmission resonance quality factors.  This is particularly important if there are very thin rectangular strips of silicon as seen in our fragmented defects (Fig. \ref{fig:improve sensitivity}).  This is an important topic for future research. 

We hope that our design guidelines described herein will aid the development of real-world 3D photonic crystal biosensors, requiring minimal biofluid volume, while achieving low limit-of-detection, high sensitivity, and spectral independence, with efficacious transmission quality.

\acknowledgments
This work was supported by the Natural Sciences and Engineering Research Council of Canada.

\appendix
\section{Spectrally independent resonances with high sensitivity but extremely low transmission}
Here, we illustrate a high-sensitivity three-defect chip with three effectively dehybridized resonances. The architecture is shown in Fig. \ref{fig: 3 dehybridized modes}(a), with defect spatial separations set to be $\text D_{lm}=\text D_{mr}=4a$. The strip widths of the left anti-domain-wall, the middle domain-wall and the right anti-domain-wall are given by $w_l=0.065a$, $w_m=0.03a$ and $w_r=0.05a$, respectively. The frequency differences between the individual defect modes (in the usual notation) are $\Delta \omega_{lm}=0.040359$ $\left[2\pi c/a\right]$ and $\Delta \omega_{rm}=0.023102$ $\left[2\pi c/a\right]$. These relatively large spectral separations suppress mode hybridizations and spectral correlations. The three resonances within the PBG occur at well-separated frequencies $\omega_L'=0.256058$ $[2\pi c/a]$, $\omega_R'=0.273131$ $[2\pi c/a]$, and $\omega_M'=0.299376$ $[2\pi c/a]$, as shown by Fig. \ref{fig: 3 dehybridized modes}(b). (Since we have chosen the smallest strip width for the middle defect, the rightmost transmission peak corresponds to the middle-like resonance $|\text M\rangle'$, and the middle transmission peak corresponds to the right-like resonance $|\text R\rangle'$. ) Their electric field profiles, shown in Fig. \ref{fig: 3 dehybridized modes}(c)-(e), concentrate mostly within the respective defects, with only small overlaps occurring through their tails. 

Fig. \ref{fig: 3 dehybridized modes}(f)-(g) show the transmission spectra for M-binding and LR-binding. For M-binding, the middle-like resonance $|\text M\rangle'$ red-shifts while the induced correlated spectral responses in the left-like resonance $|\text L\rangle'$ and right-like resonance $|\text R\rangle'$ are negligible. Similarly, with LR-binding, the left-like resonance $|\text L\rangle'$ and right-like resonance $|\text R\rangle'$ red-shift while the induced correlated spectral response in $|\text M\rangle'$ is negligible. All three resonances  $|\text L\rangle'$, $|\text M\rangle'$ and $|\text R\rangle'$ are effectively dehybridized and have independent spectral responses to analyte-binding. However, the transmission-levels of $|\text L\rangle'$ and $|\text R\rangle'$ resonances are extremely low, $0.03\sim 0.075$.

This illustrates that with three defects, it is problematic to achieve spectral independence with high enough transmission by only adjusting mode hybridizations.
\begin{widetext}
\begin{center}
\begin{figure}[htbp]
 \centering
 \subfigure[]{\includegraphics[width=0.99\textwidth]{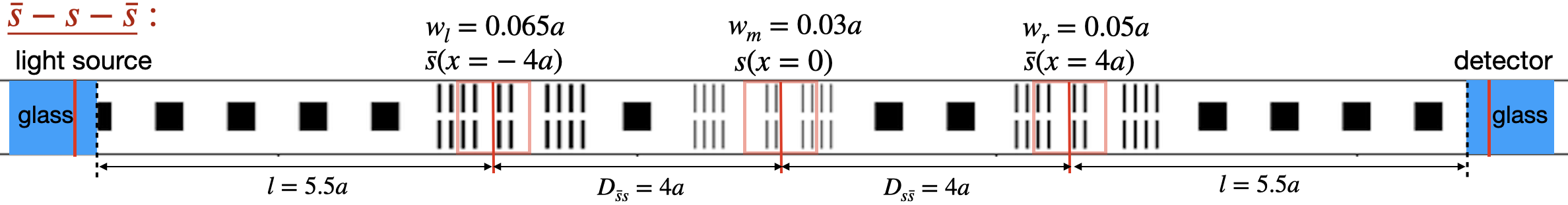}}
    \label{fig: 3-defect dehybridized chip}
  \subfigure[]{\includegraphics[width=0.5\textwidth]{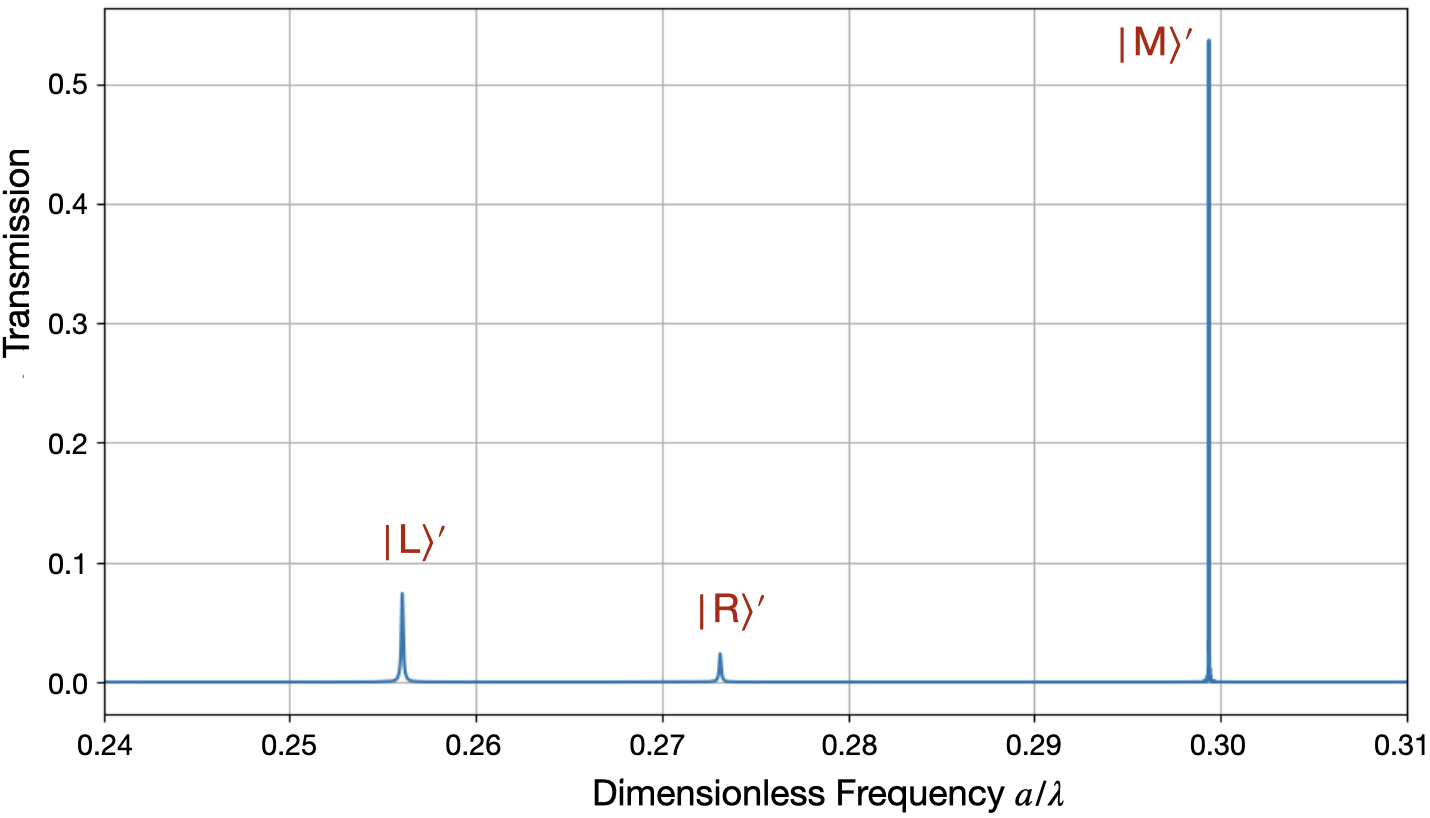}}  
   \hspace{0.3cm}
  \subfigure[]{\includegraphics[width=0.47\textwidth]{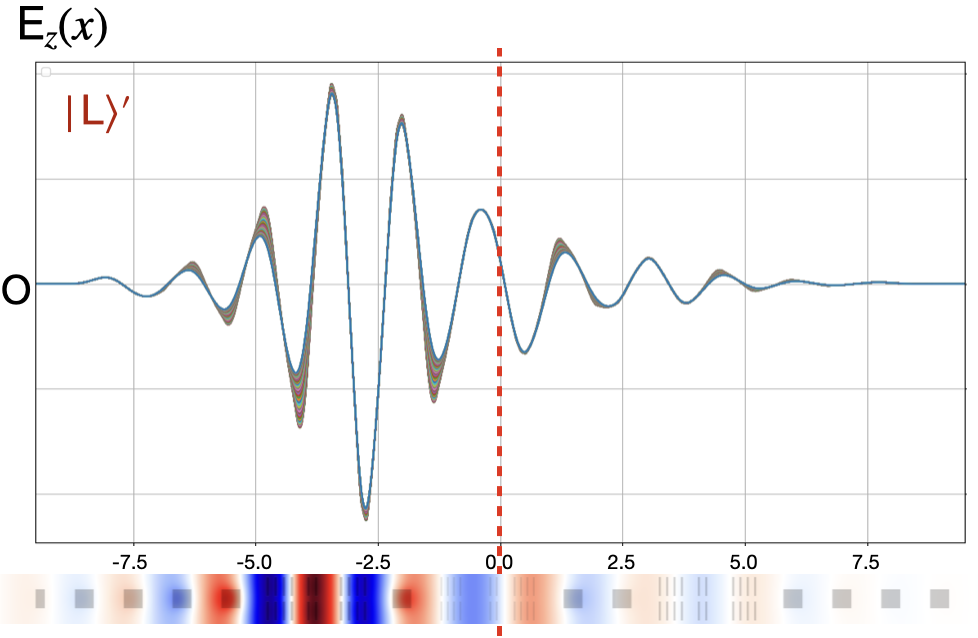}}
  \subfigure[]{\includegraphics[width=0.47\textwidth]{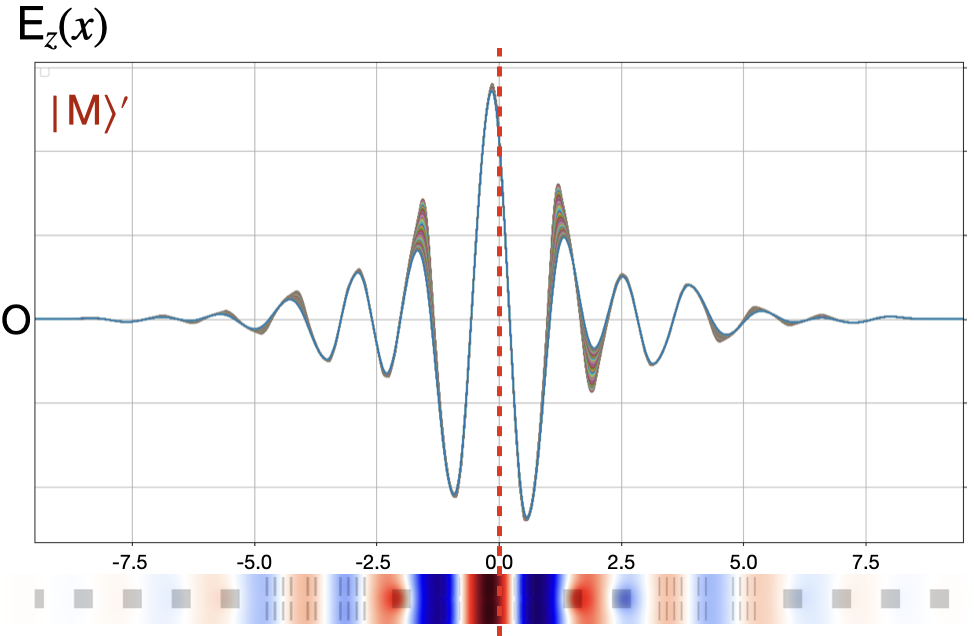}}
   \hspace{0.3cm}
  \subfigure[]{\includegraphics[width=0.47\textwidth]{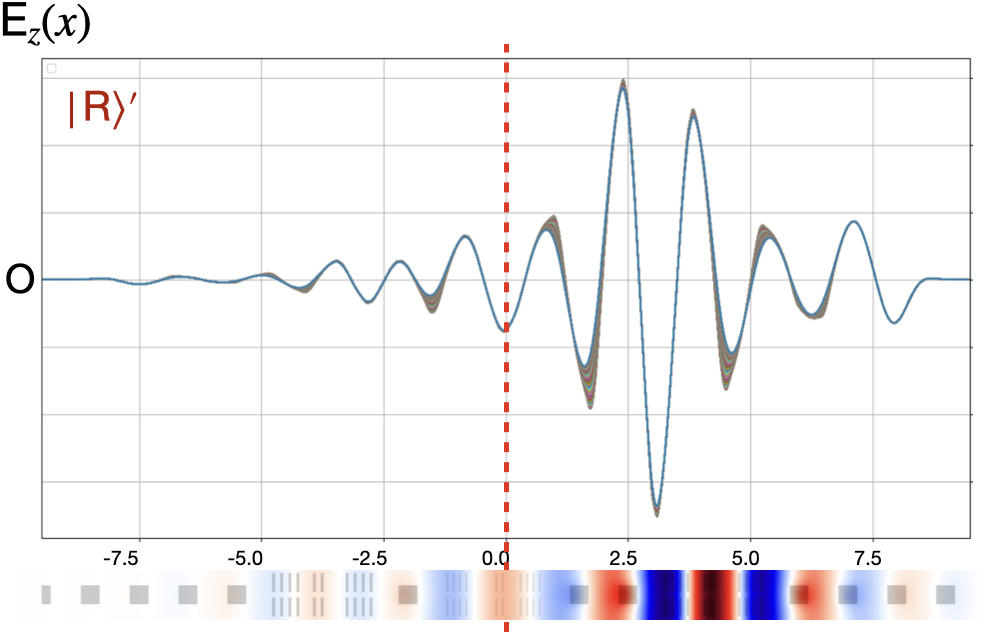}}
    \label{fig: 3 dehybridized mode profiles}
\subfigure[]{\includegraphics[width=0.49\textwidth]{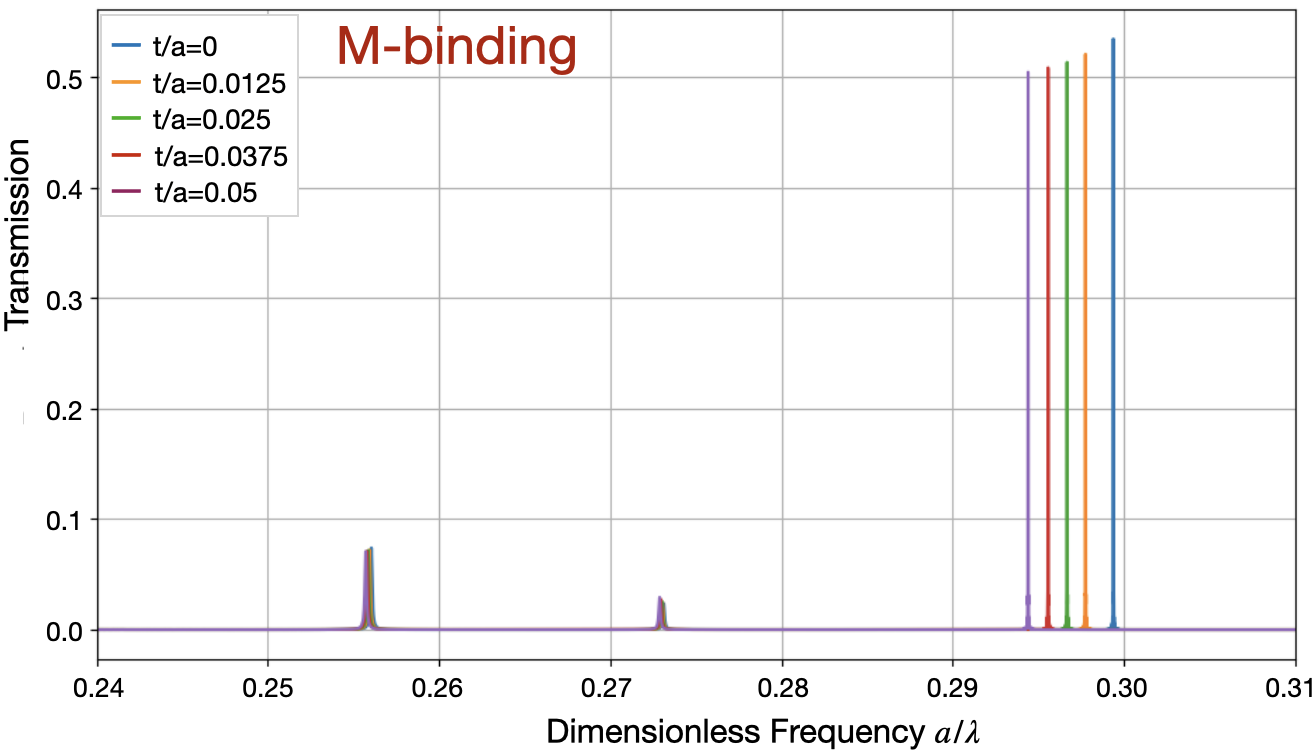}}  
 \hspace{0.1cm}
  \subfigure[]{\includegraphics[width=0.49\textwidth]{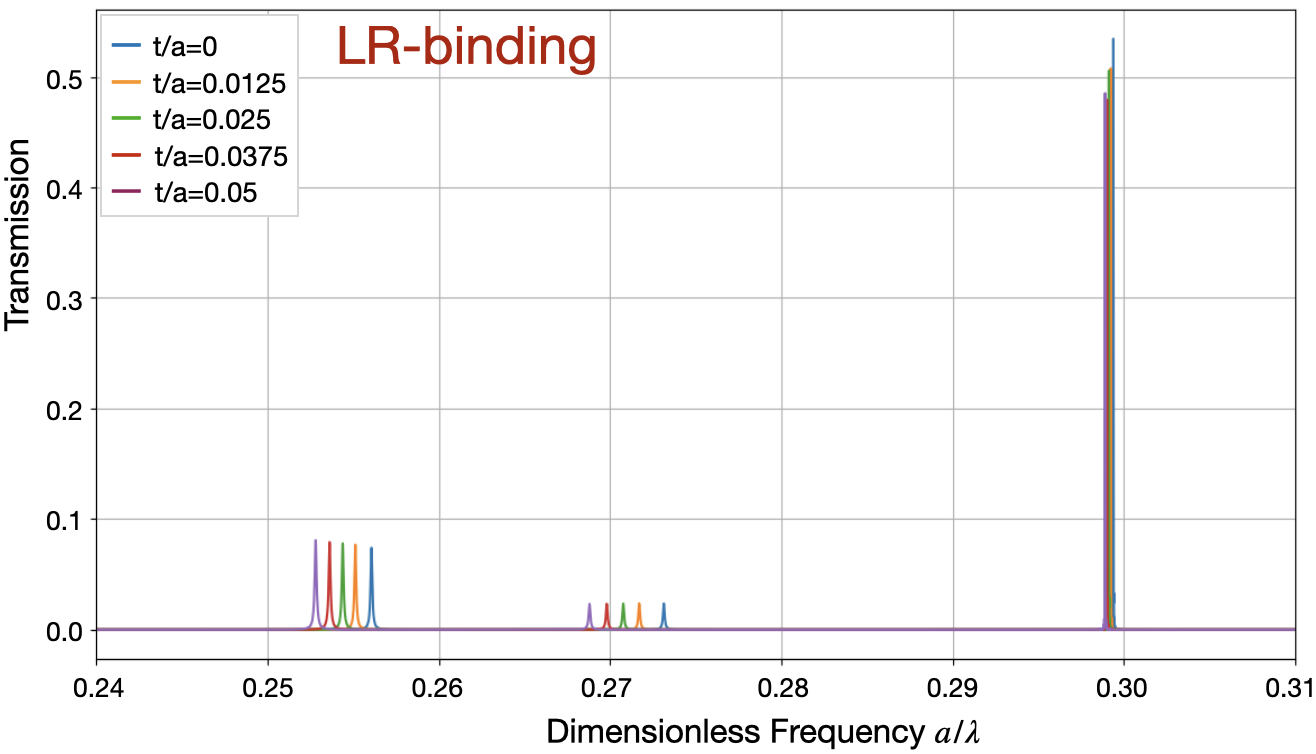}}  
  \caption{Three-defect chip ($\bar{s}-s-\bar{s}$) with three effectively dehybridized resonances. (a) Strip widths within the left anti-domain-wall, the middle domain-wall, and the right anti-domain-wall regions are  $w_l=0.065a$,  $w_m=0.03a$ and $w_r=0.05a$ respectively, leading to relatively large frequency differences, $\Delta \omega_{lm}=0.040359$ $[2\pi c/a]$ and $\Delta \omega_{rm}=0.023102$ $[2\pi c/a]$.  The spatial separations between defects are $\text D_{lm}=\text D_{rm}=4a$. (b) Three resonances within the PBG occur at well-separated frequencies $\omega_L'=0.256058$ $[2\pi c/a]$, $\omega_R'=0.273131$ $[2\pi c/a]$, and $\omega_M'=0.299376$ $[2\pi c/a]$. The rightmost peak with the highest frequency corresponds to the middle-like mode $|\text M\rangle'$, since the strip width of the middle defect is the smallest.  (c), (d), (e) show the electric field profiles of the left-like mode $|\text L\rangle'$, middle-like mode $|\text M\rangle'$, and right-like mode $|\text R\rangle'$ respectively. (f) $\&$ (g) show the spectral responses with respect to M-binding and LR-binding, revealing spectral independence. Since the left- and right-defect have a large spatial separation and their strip widths  are different, $|\text L\rangle'$ and $|\text R\rangle'$ are effectively dehybridized as well. The transmission-levels of $|\text L\rangle'$ and $|\text R\rangle'$ resonances are very low, only about $0.03\sim 0.075$.}
\label{fig: 3 dehybridized modes}
\end{figure}
\end{center}
\end{widetext}

\section{Spectrally correlated resonances with high sensitivity and more distinctive transmission-level changes}

The main drawback of the highly-hybridized, three-defect chip given in Sec. IV C, is that spectral fingerprints for some analyte-binding cases are not easily distinguishable. For example, with LR-binding, transmission-level changes of the resonances are very small and can be masked by noise in the biofluid. 

In Fig. \ref{fig: 3-AS-AS-AS bindings}, we present another three-defect chip ($\bar{s}$-$\bar{s}$-$\bar{s}$) with adjusted strip widths ($w_l=0.04a, w_m=0.05a, w_r=0.06a$). In this chip, the spatial separation between left- and middle-defect is equal to that between right- and middle-defect, $\text D_{lm}=\text D_{rm}=3.5a$, and the middle defect is located at the exact center of the chip. This chip supports three spectrally correlated resonances with high sensitivity, but with more distinctive transmission-level changes. The spatial separation $\text D_{lr}=7a$ between the left- and right-defect is slightly larger and mode hybridizations are smaller. In this case, transmission levels are reduced but still detectable.

\begin{widetext}
\begin{center}
\begin{figure}[htbp]
 \centering
  \subfigure[]{\includegraphics[width=0.95\textwidth]{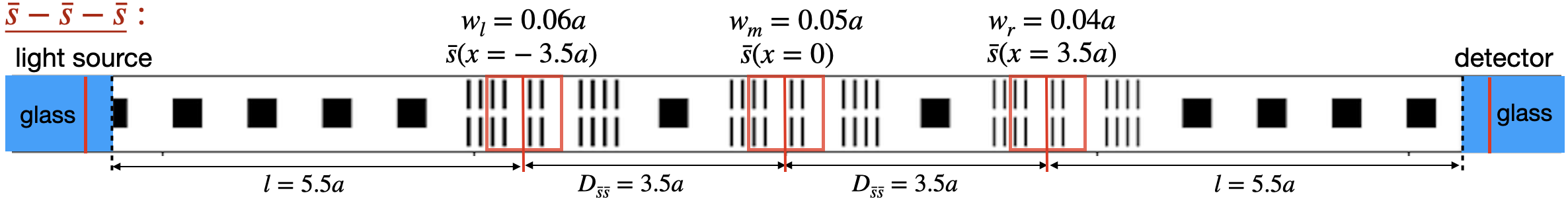}}
    \label{fig: 3 hybridized AS-AS-AS chip}
  \subfigure[]{\includegraphics[width=0.43\textwidth]{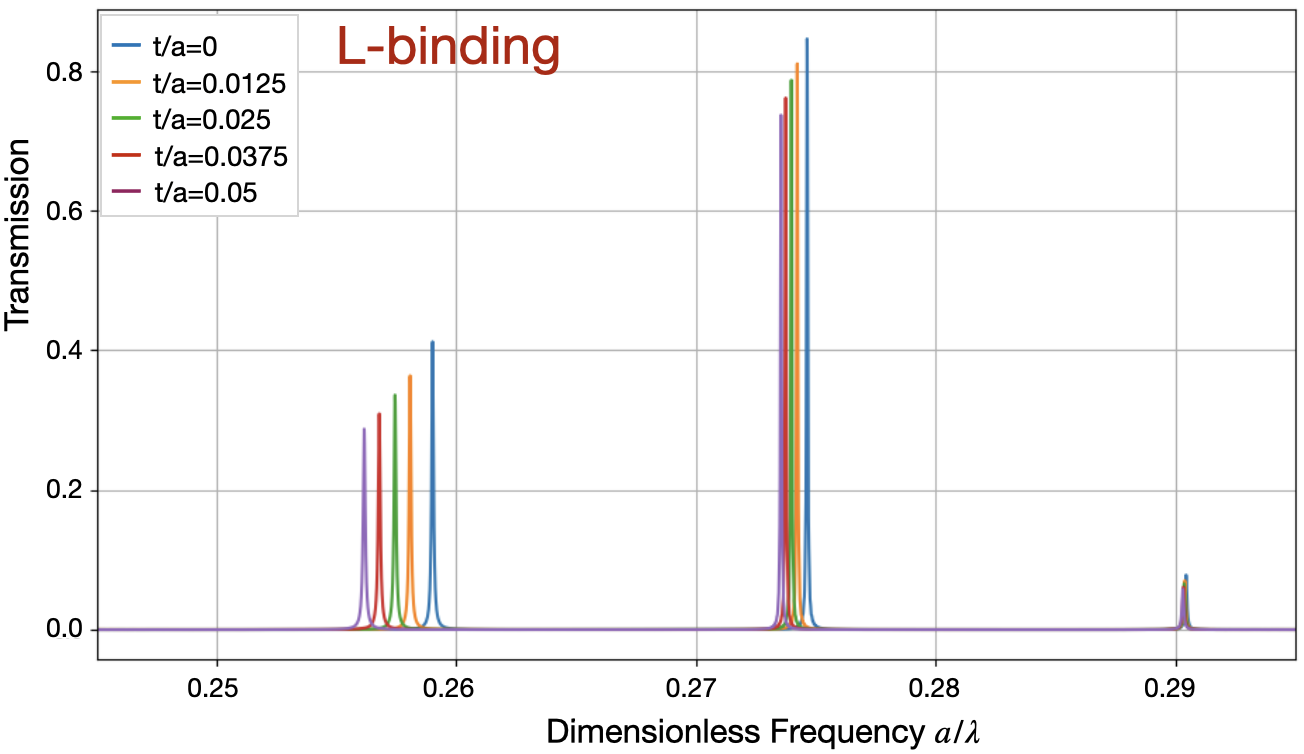}}
   \label{fig: 3 AS hybrized L-bindings}
    \hspace{0.4cm}
  \subfigure[]{\includegraphics[width=0.43\textwidth]{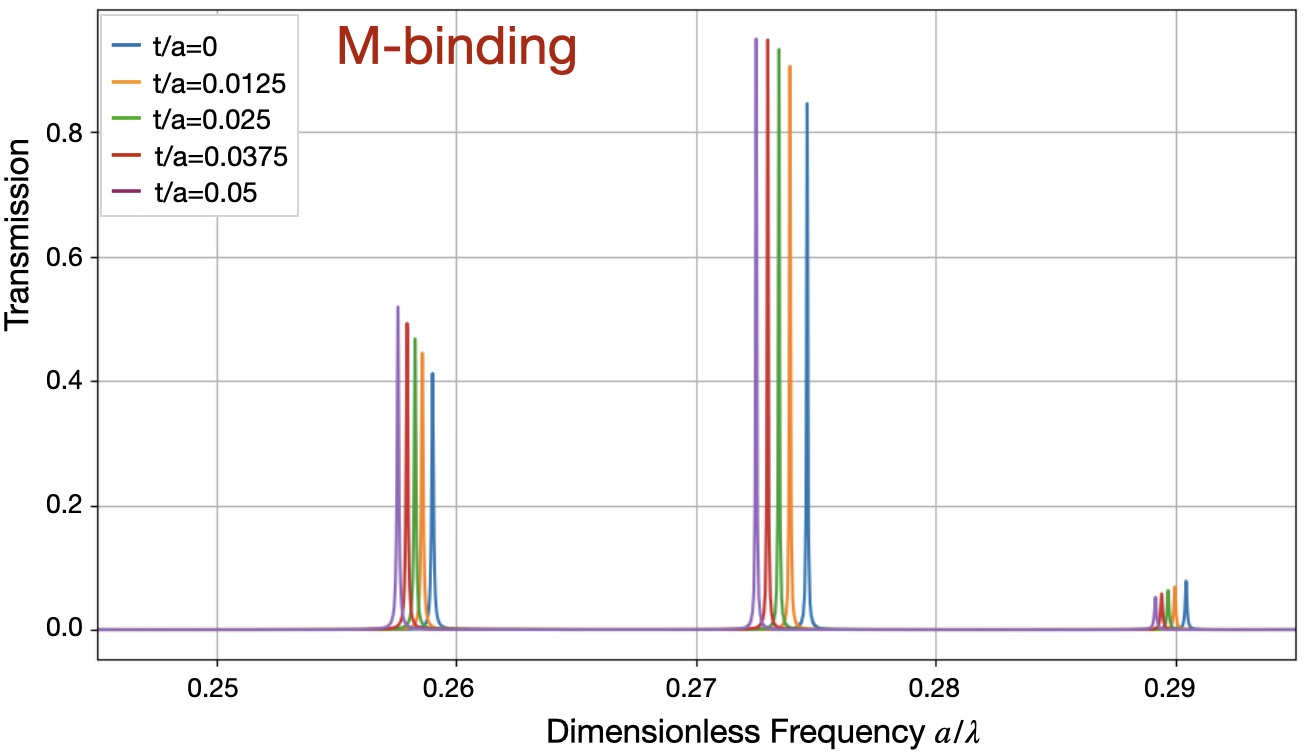}}
    \label{fig: 3 AS hybrized M-bindings}
  \subfigure[]{ \includegraphics[width=0.43\textwidth]{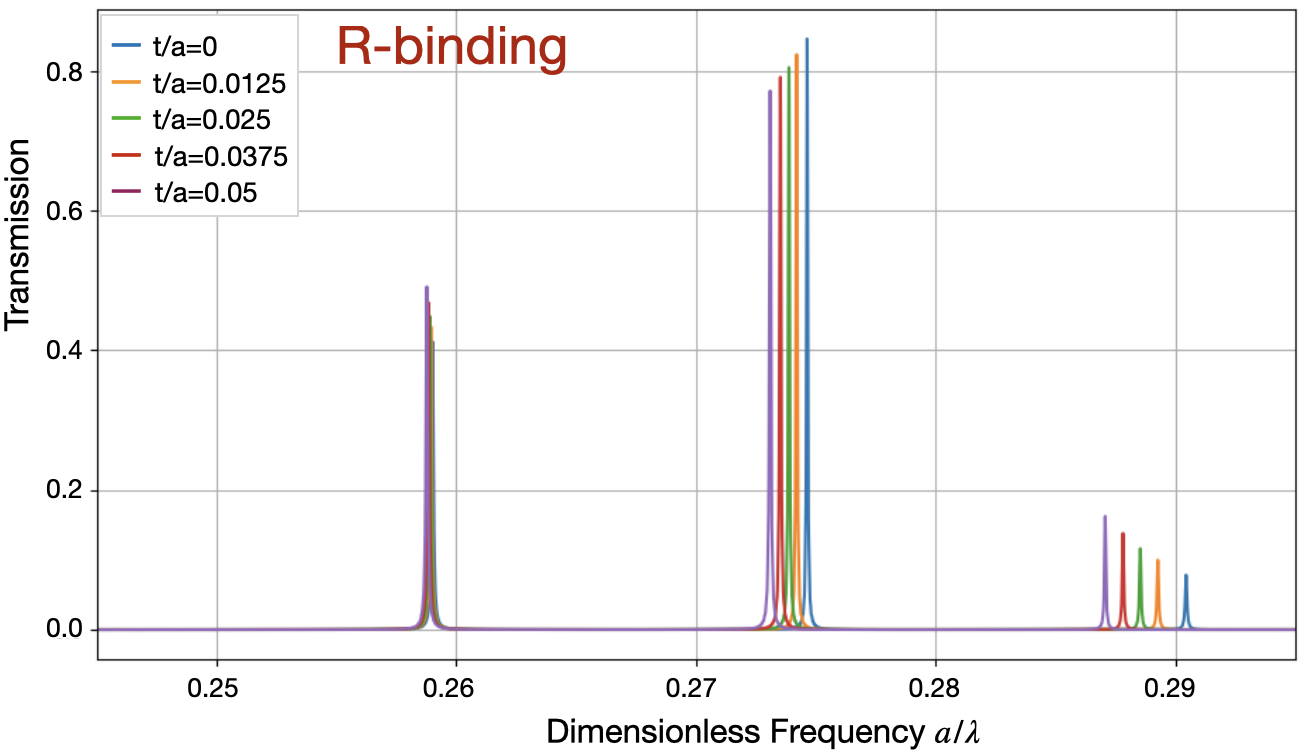}}
    \label{fig: 3 AS hybrized R-bindings}
     \hspace{0.4cm}
  \subfigure[]{\includegraphics[width=0.43\textwidth]{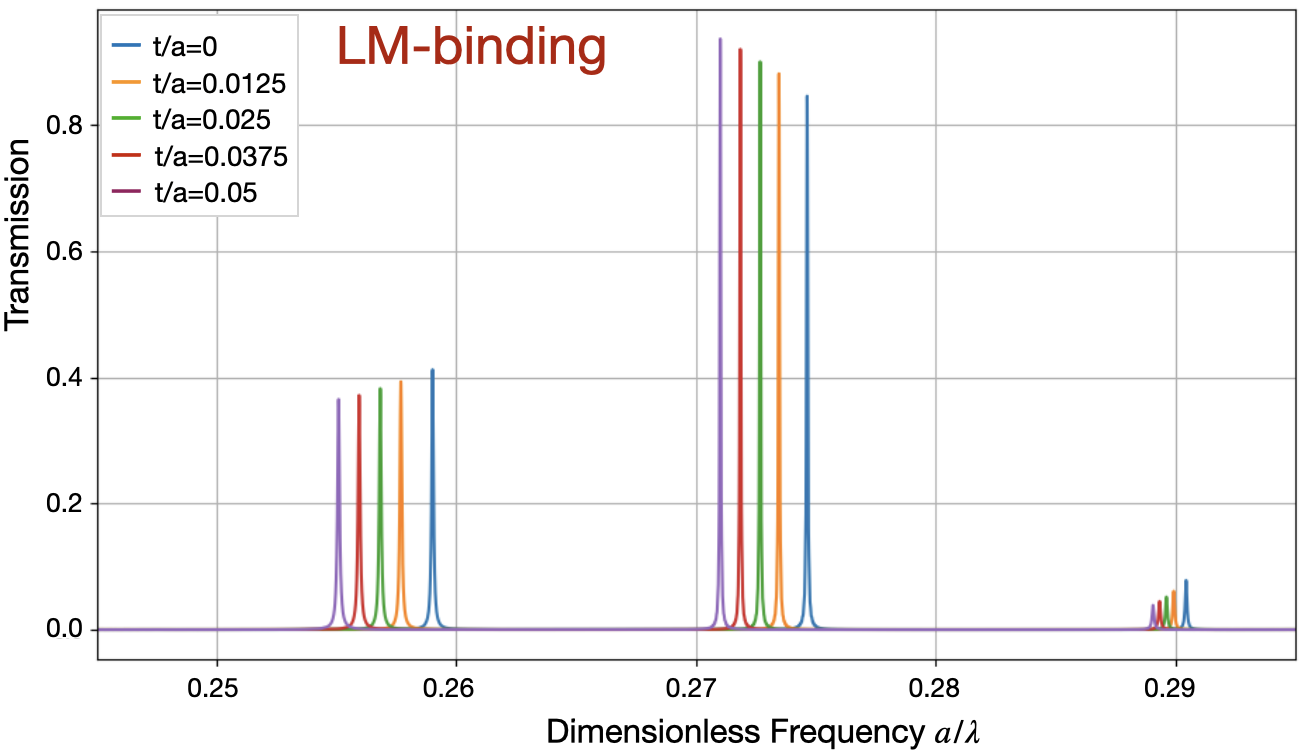}}
    \label{fig: 3 AS hybrized LM-bindings}
  \subfigure[]{\includegraphics[width=0.43\textwidth]{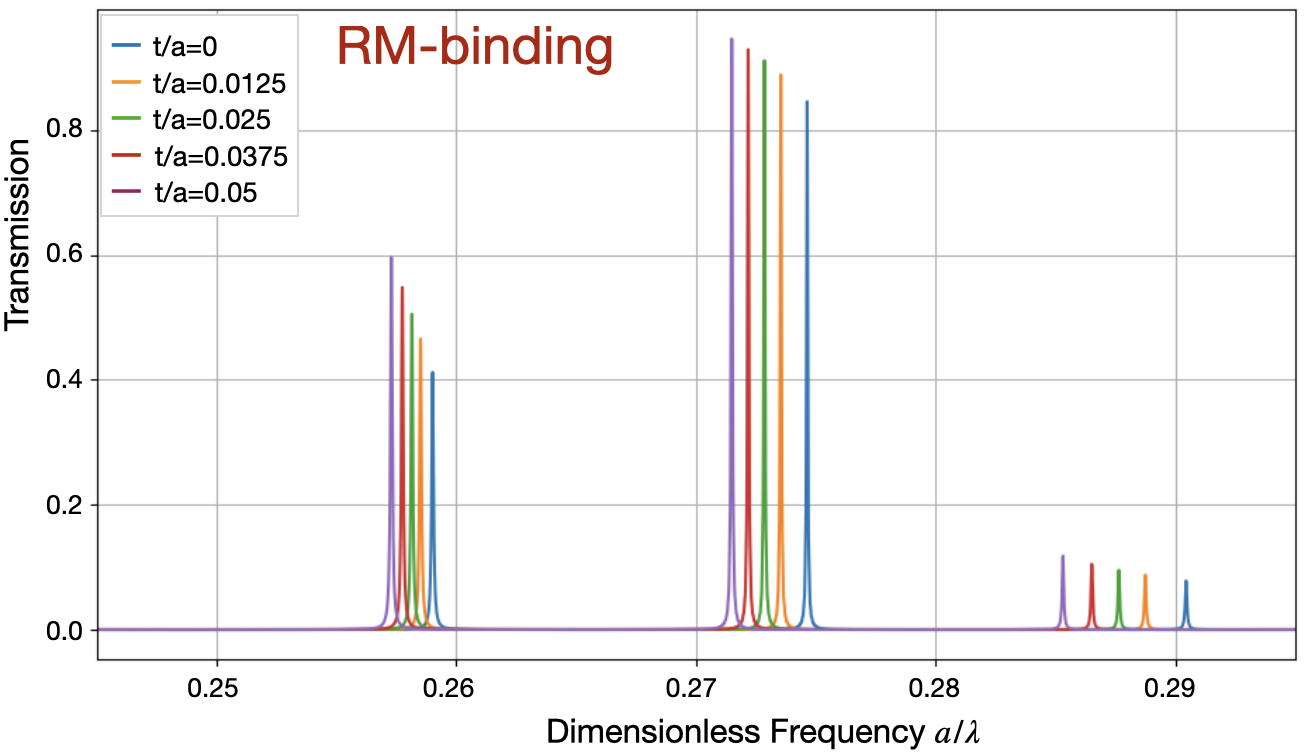}}
    \label{fig: 3 AS hybrized RM-bindings}
     \hspace{0.4cm}
  \subfigure[]{ \includegraphics[width=0.43\textwidth]{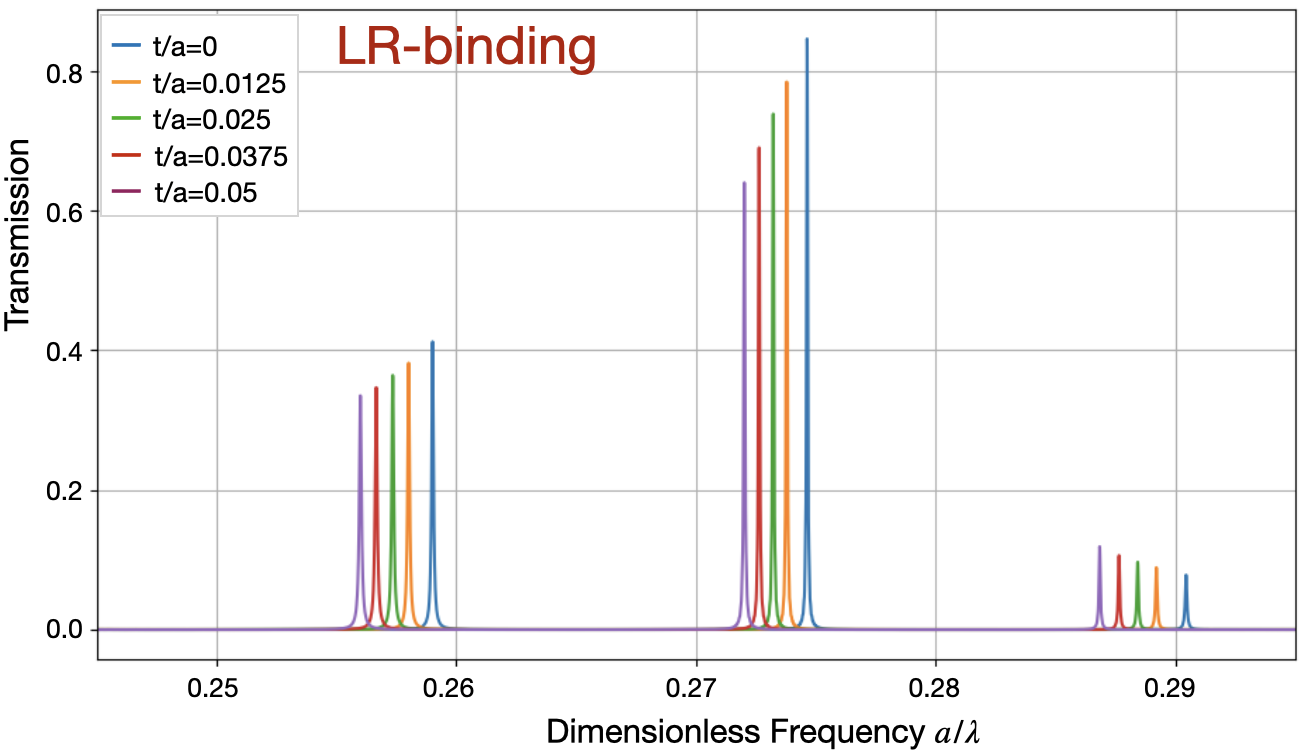}}
    \label{fig: 3 AS hybrized LR-bindings}
   \subfigure[]{\includegraphics[width=0.43\textwidth]{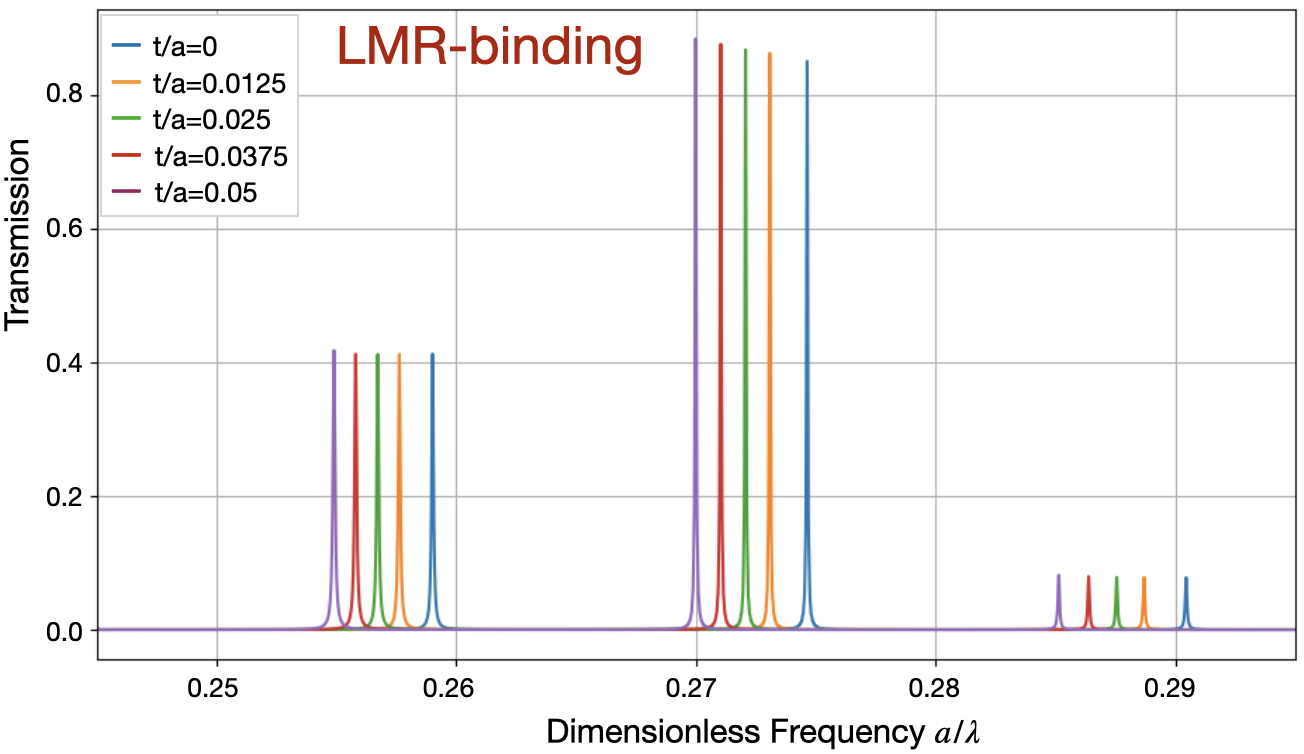}}
    \label{fig: 3 AS hybridized LMR-bindings}
    \subfigure[]{\includegraphics[width=0.5\textwidth]{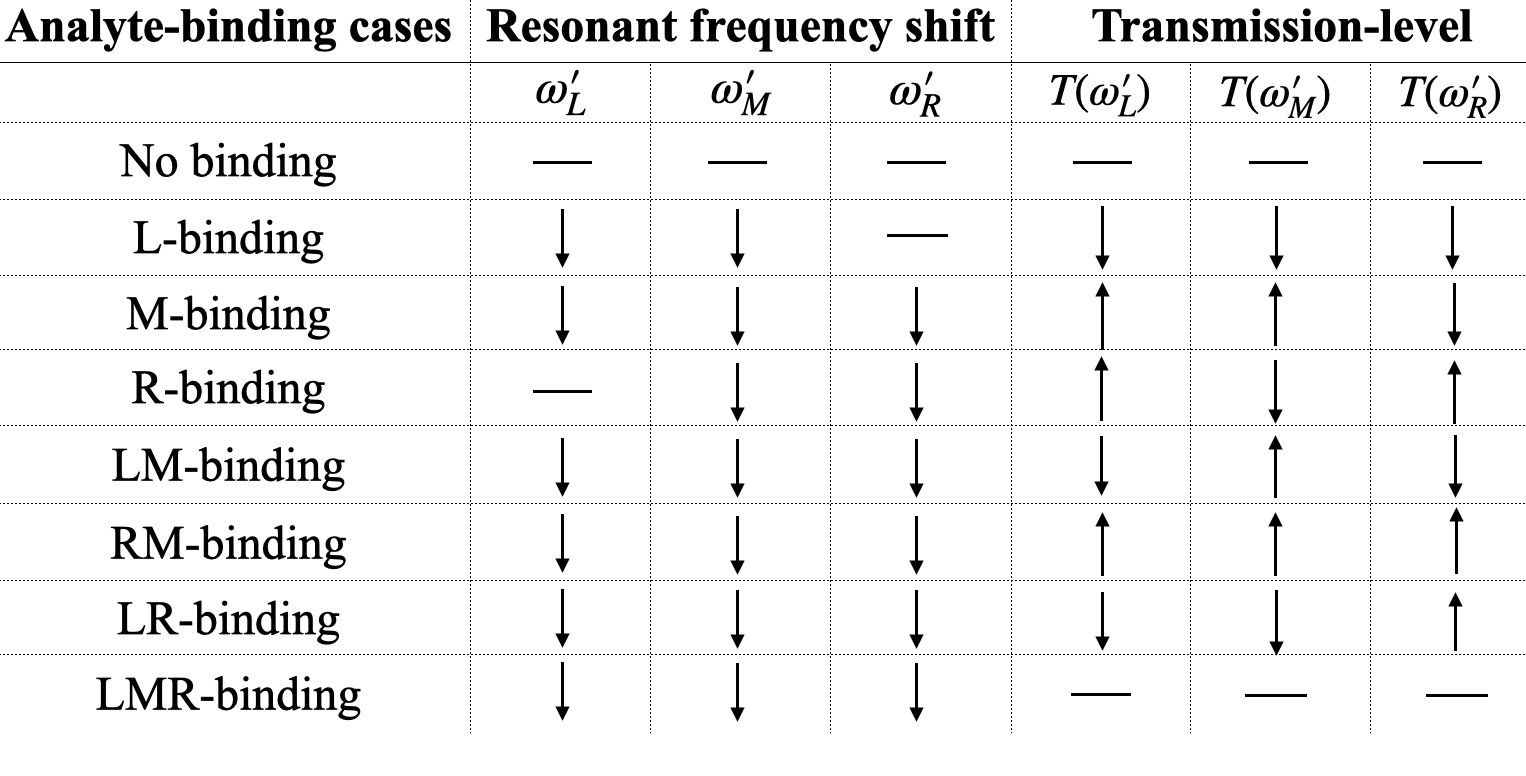}}
    \label{fig: 3 hybridized_as-as_as Truth Table}
 \caption{Three-defect chip ($\bar{s}-\bar{s}-\bar{s}$) with more distinctive transmission-level changes. (a) Strip widths in each defect region are $w_l=0.06a$,  $w_m=0.05a$ and $w_r=0.04a$. The spatial separations between defects are $\text D_{lm}=3.5a$ and $\text D_{rm}=3.5a$. Transmission spectra for different analyte-binding cases are shown: (b) L-binding. (c) M-binding. (d) R-binding. (e) LM-binding. (f) RM-binding. (g) LR-binding. (h) LMR-binding. (i) Truth table for all the analyte-binding cases. Due to high hybridizations, the spectral responses of the three resonances are correlated, but still distinguishable with respect to different analyte-binding cases.}
\label{fig: 3-AS-AS-AS bindings}
\end{figure}
\end{center}
\end{widetext}

\section{Mode profiles of the three resonances in Chip 3}
The three highly hybridized resonances supported by Chip 3 in Sec. IV C, occur at frequencies $\omega'_L=0.275535$ $[2\pi c/a]$, $\omega'_M=0.292367$ $[2\pi c/a]$ and $\omega'_R=0.305132$ $[2\pi c/a]$ within the PBG, as shown in Fig. \ref{fig: 3-S-AS-AS_D=1_mode profiles}(a). The frequency differences between these three resonances are reduced compared with Chips 1 and 2. Their electric field profiles are shown in Fig. \ref{fig: 3-S-AS-AS_D=1_mode profiles}(b)-(d). Due to strong mode hybridizations between $|\text M\rangle$ and $|\text L\rangle$ and between $|\text M\rangle$ and $|\text R\rangle$, the field strength of the $|\text M\rangle'$ resonance in the middle defect region is smaller than that in the left or right defect region. Nevertheless, we use `M' to denote this resonant mode and call it middle-like. On the other hand, the field strengths of the $|\text L\rangle'$ resonance and the $|\text R\rangle'$ resonance in the middle defect region are quite strong. This explains the relatively small red-shift of the $|\text M\rangle'$ peak compared to those of the $|\text L\rangle'$ and $|\text R\rangle'$ peaks, for M-binding shown in Fig. \ref{fig: 3 hybridized modes bindings}(c).

\begin{widetext}
\begin{center}
\begin{figure}[htbp]
 \centering
     \subfigure[]{\includegraphics[width=0.45\textwidth]{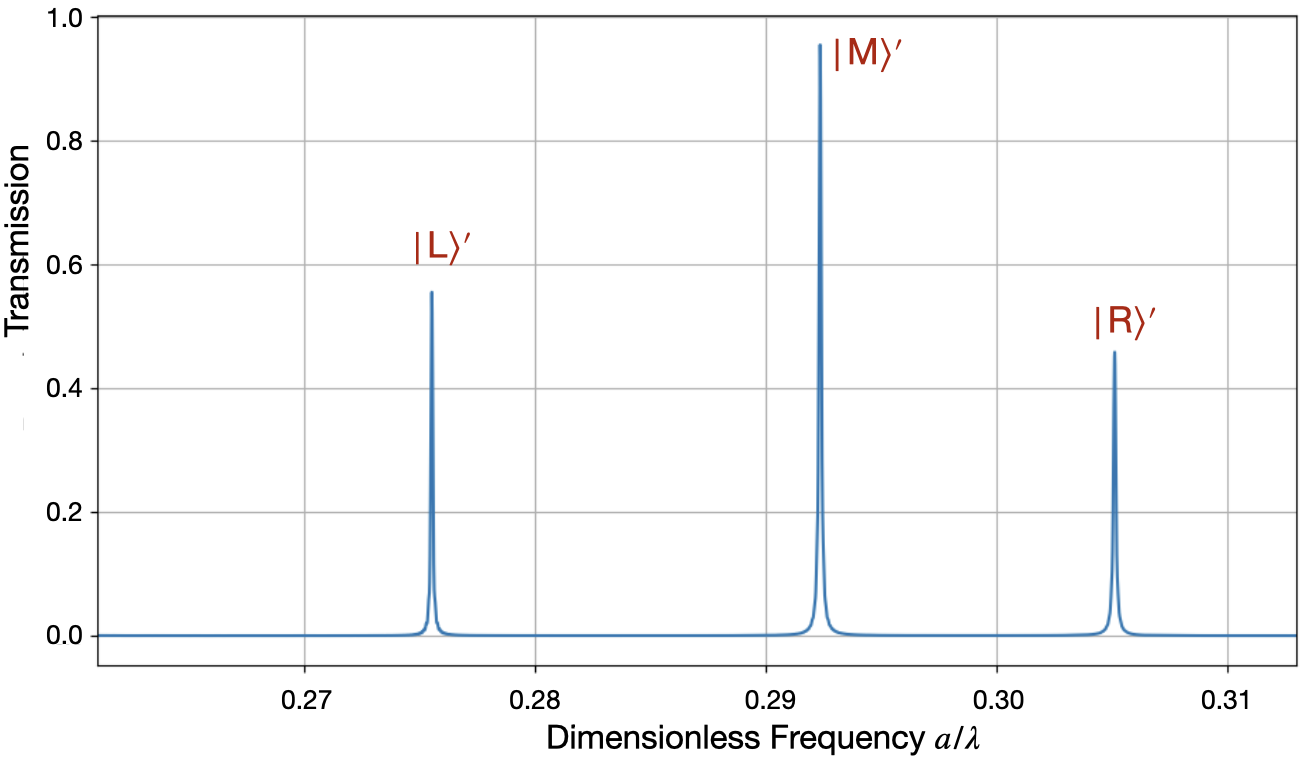}}
     \hspace{0.1cm}
   \subfigure[]{ \includegraphics[width=0.44\textwidth]{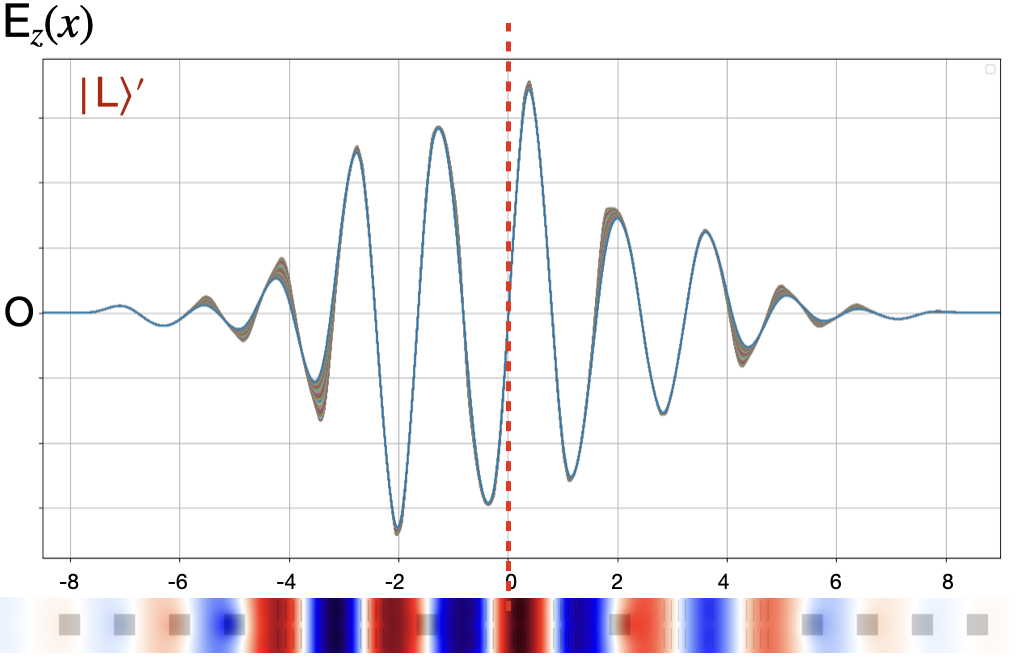}}
   \subfigure[]{\includegraphics[width=0.44\textwidth]{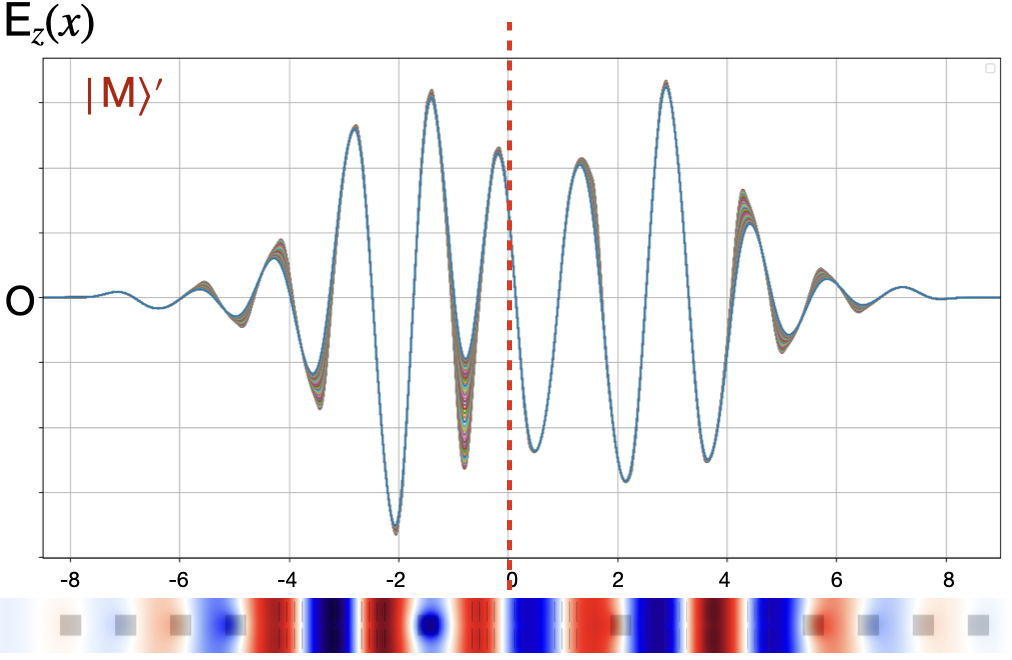}}
     \hspace{0.3cm}
    \subfigure[]{\includegraphics[width=0.44\textwidth]{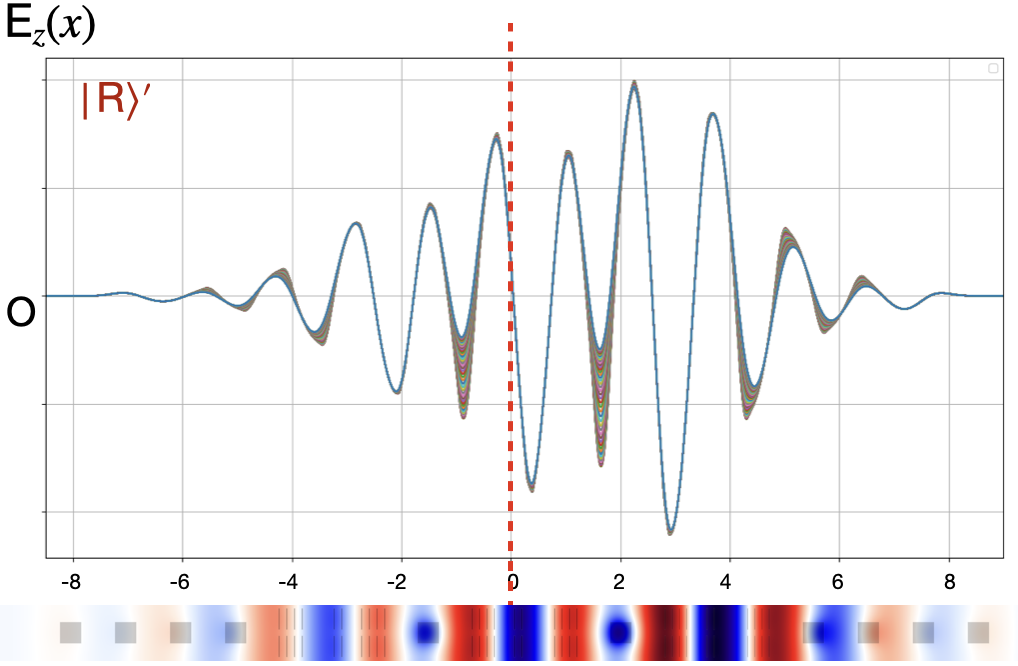}}
 \caption{(a) Three transmission peaks, corresponding to the three highly hybridized resonances $|\text L\rangle', |\text M\rangle', |\text R\rangle'$ in Chip 3 occur at frequencies $\omega'_L=0.275535$ $[2\pi c/a]$, $\omega'_M=0.292367$ $[2\pi c/a]$ and $\omega'_R=0.305132$ $[2\pi c/a]$ within the PBG. Their electric field profiles are shown in (b), (c) $\&$ (d). The field strength of the $|\text M\rangle'$ resonance in the middle defect region is smaller than that in the left or right defect region, while the field strength of the $|\text L\rangle'$ or $|\text R\rangle'$ resonance is quite strong in the middle defect region.}
\label{fig: 3-S-AS-AS_D=1_mode profiles}
\end{figure}
\end{center}
\end{widetext}

\end{document}